\documentclass{aa}  
\usepackage{natbib}

\usepackage{array}
\usepackage{graphicx}
\usepackage{txfonts}
\usepackage[colorlinks=true,allcolors=blue]{hyperref}
\usepackage{placeins}
\usepackage{longtable}
\usepackage{lscape}
\begin{document} 

\title{A comprehensive \textit{Gaia} view of ellipsoidal \\ and rotational red giant binaries}

   \author{Camila Navarrete
          \inst{1}\thanks{CNES Fellow.}
          \and
          Alejandra Recio-Blanco\inst{1}
          \and
          Patrick de Laverny\inst{1}
          \and
          Ana Escorza\inst{2, 3}
          }

   \institute{Universit{\'e} C{\^o}te d'Azur, Observatoire de la C{\^o}te d'Azur, CNRS, Laboratoire Lagrange, Bd de l'Observatoire, CS 34229, 06304, Nice Cedex 4, France\\
              \email{camila.navarrete@oca.eu}
         \and
          Instituto de Astrof\'isica de Canarias, C. V\'ia L\'actea, s/n, 38205 La Laguna, Santa Cruz de Tenerife, Spain
              \and
        Universidad de La Laguna, Dpto. Astrof\'isica, Av. Astrof\'isico Francisco S\'anchez, 38206 La Laguna, Santa Cruz de Tenerife, Spain.
             }

   \date{Received ... ..., 2024; accepted ... ..., 2025}

  \abstract
   {The latest \textit{Gaia} Focused Product Release (FPR) provided variability information for $\sim$1000 long-period red giant binaries, including almost $\sim$700 ellipsoidal binary candidates, the largest sample to date of this binary type having both photometric and spectroscopic time series observations.}
   {We aim to characterise physically (luminosity, mass, radius) and chemo-dynamically (metallicity, [$\alpha$/Fe], Galactic velocities) the population of long-period red giant ellipsoidal binary candidates, and a subsample of rotational variable candidates, from \textit{Gaia} FPR, combining \textit{Gaia} astrometry, photometry and spectroscopy observations.}
   {We cross-matched the \textit{Gaia} DR3 measurements (positions, velocities, atmospheric parameters, chemical abundances) with the catalogue of long-period red giant candidates from the \textit{Gaia} FPR, having photometric and radial velocity variability information. Combined with the photo-geometric distances, the extinction, bolometric magnitude, luminosity, spectroscopic radius and mass were estimated. The accuracy of this method was tested for similar samples in the literature, including red giant binaries having asteroseismic determination of their physical parameters.} 
   {Ellipsoidal variables are characterized to be low to intermediate-mass stars (0.6 $\leq \mathcal{M}_1 \leq 5.0$ M$_{\odot}$), with radii as large as the Roche lobe radius of the binary. Eccentricities tend to be lower for primary stars with smaller radii, as the expected result of tidal circularization. Combined with the orbital properties, estimates for the minimum mass of the companion agree with the scenario of a low-mass compact object as the secondary star. There are at least 14 ellipsoidal binaries with orbital periods and masses of the two stars compatible with model predictions for Type Ia SN progenitors. For the rotational variables, their orbital periods, enhanced chromospheric activity, smaller radii and low mass ($\mathcal{M}_1 \lesssim$ 1.5 M$_{\odot}$) point to a different type of binaries than the original ellipsoidal sample. The Galactic velocities indicate that ellipsoidal variables are found both in the Galactic disk and halo, while rotational variables are predominantly concentrated in the Galactic disk. The velocity dispersion is much higher in ellipsoidal than in rotational binaries, probably indicating older/younger dynamical ages. The enhanced [$\alpha$/Fe] abundances for some of the ellipsoidal binaries, having $\mathcal{M}_{\rm 1} \geq$ 1.0 M$_{\odot}$, resemble the population of young $\alpha$-rich binaries in the thick disk. An episode of mass transfer in those systems may have produced the enhanced $\alpha$ abundances, and the enhanced [Ce/Fe] abundances reported in a few ellipsoidal binaries.}
   {Luminosities, radii and masses were derived for 243 ellipsoidal and 39 rotational binary candidates, the largest Galactic sample of these variables, having chemo-dynamical and physical parameterization. Based on their mean chemo-dynamical properties and stellar parameters, these binaries can be considered as two manifestations of the same phenomena, a close binary with a giant primary, instead of two independent, unrelated binary types.  Detailed future analysis of individual sources will provide insights into the history and future evolution of these binaries.}
   

   \maketitle
%

\section{Introduction}

Binary stars are ubiquitous. Estimates for the stellar multiplicity can go from 20\% up to 100\% depending on the mass of the primary \citep[see e.g.,][and references therein]{Offner23, Moe17, Duchene13, Raghavan10, Bate09}, making the study of the binary systems of paramount importance for stellar evolution models. The binary properties, including the mass ratio, metallicity, and the evolutionary stage of the components, among others, lead to the binary evolution into different possible outcomes, including, for instance, compact binaries with double degenerate components \citep[e.g.,][]{Marsh11, Napiwotzki04}, rapid rotating stars \citep[e.g.,][]{Simonian19, Gaulme20}, semi-detached systems (such as symbiotic binaries, see for example, \citealt{Podsiadlowski07, Mikolajewska12}, cataclysmic variables, see e.g., \citealt{Paczynski85}, X-ray binaries, e.g., \citealt{Bahramian23}), contact binaries \citep[e.g.,][]{Qian20} and chemically peculiar stars like barium and carbon-enhanced metal-poor (CEMP) stars \citep{Jorissen2019, Escorza19, Karinkuzhi21}. Some of these systems can eventually become the progenitors of type Ia supernovae \citep{Maoz14}, hypervelocity stars \citep{Brown15}, stellar black holes or gravitational wave sources \citep[see e.g.,][]{Amaro12}, having a strong impact on the chemical enrichment and evolution of galaxies. 

The \textit{Gaia} space mission \citep{GaiaDR3} is revolutionizing the field of binary stars mostly since its third data release \citep[DR; see the review of][and references therein]{ElBadry24}. Over $\sim$800\,000 non-single stars and their orbital solutions were published for the first time in DR3 as part of the \texttt{non\_single\_star} catalogue \citep[NSS;][]{GaiaNSS}, being the largest all-sky collection of binaries compared to all the previous compilations \citep{ElBadry24}, despite being only a fraction of the genuine binaries that could be observed by \textit{Gaia}. Unlike previous observations, binary systems observed by \textit{Gaia} also have precise astrometric observations (including parallaxes and proper motions), $G$, $G_{\rm BP}$, $G_{\rm RP}$-band photometry, and for the brightest (G$_{\rm RVS} <$ 12 mag) spectroscopic observations. From the epoch photometry and spectroscopy observations, the variability of the systems can be analyzed, making this dataset a unique opportunity to study different types of binary systems in a holistic approach which was not available before \textit{Gaia}. A probe of this is the increasing number of studies using the \textit{Gaia} NSS catalogue to explore, for example, the wide-binary population \citep{Hernandez23, Hollands24}; to search for massive unseen compact companions \citep{ElBadry22, Fu22, Gomel23, Jayasinghe23}, including the discovery of the most massive stellar black hole, \textit{Gaia} BH3 \citep{GaiaBH3}; and population studies of main-sequence and brown dwarf binaries \citep[e.g.,][]{Bashi23, Stevenson23}. Given the importance of radial velocity (RV) and time series epoch observations, the \textit{Gaia} mission has published the variability information for classical pulsators, such as Cepheids and RR Lyrae in its DR3 \citep{Ripepi23, Clementini23} and, more recently, as part of the \textit{Gaia} Focused Product Release \citep[FPR;][]{GaiaFPR}, for long-period variables (LPVs, having photometric periods\footnote{In the context of the Gaia FPR, the ``photometric period'' is based on the periodogram and may not represent the true orbital period in the case of binary systems, particularly for ellipsoidal binaries in which the photometric light curve shows two minima per cycle. The true period of a binary is therefore obtained unequivocally based on the RV time series.} $\gtrsim$ 10 d to 1000 d), including Mira pulsators, semi-regular variables, and red giant binary systems. 

Among red giant binaries, ellipsoidal variables are close binaries in which the primary star (usually a red giant) is tidally distorted because of the gravitational attraction of the companion (typically a main sequence star). This distortion produces an elongated, ellipsoidal shape of the primary, substantially filling the Roche lobe of the system, which leads to ellipsoidal light variations as the primary orbits around the secondary. Therefore, the velocity variation (orbital period) is twice the light variability (photometric period). Ellipsoidal variables are known to define a linear sequence in the period-luminosity diagram of LPVs in the Large Magellanic Cloud \citep[LMC,][]{Soszynski04, Nie12, Pawlak}. Close red giant binaries can become tidally synchronized, being one of the mechanisms to explain the existence of rapidly rotating giants \citep{Phillips24, Leiner22}. Due to the tidal influence, the giants can also present enhanced chromospheric activity, due to the formation of starspots. This type of binary is commonly associated with the RS Canum Venaticorum-type \citep[RS CVn;][]{Hall76, Montesinos88, deGrijs}. Although ellipsoidal variables belong to the Rotational variability class, their brightness variations are due to the changes in the orientation of the tidally distorted primary while in the case of rotational variables such as RS CVn stars, the brightness variations are due to the rotation and starspots in the stellar surface.

In this work, we focus on the sub-sample of red giant binaries among the LPVs which are part of the latest \textit{Gaia} FPR. Aiming to characterize chemo-dynamically this sample, we cross-match the binary candidates from the \textit{Gaia} FPR with DR3 astrometry, photometry (\texttt{gaia\_source} table), spectroscopic atmospheric parameters (\texttt{astrophysical\_parameters} table) and orbital (\texttt{nss\_two\_body\_orbit} table) information. This paper is organized as follows. Section~\ref{sec:binary_candidates} presents the selection of the binary sample in terms of variability cuts, and the properties of the sample (periods, effective temperature, surface gravity, activity level). In Section~\ref{sec:LMR} the procedure to derive the stellar parameters including luminosities, masses and radii of the primary stars, and the validation of the method with external catalogues is presented. Section~\ref{sec:binary_proper} presents the binary population properties of the sample, including the mass function distribution, period-radius-eccentricity relations and mass estimates for the secondary component. Section~\ref{sec:chemo-dynamical} explores the chemical and dynamical properties of the sample including metallicity distribution function, Galactic velocities, velocity dispersions and s-process [Ce/Fe] abundances. Section~\ref{sec:PL_relations} presents the period-luminosity (PL) diagram for our sample and explores correlations between the dispersion of the PL and the mass, metallicity and radii of the primary star. We discuss the intrinsic differences among ellipsoidal and rotational variables and summarize our results in Section~\ref{sec:conclusions}.

\section{Long-period binary candidates in \textit{Gaia} DR3}

\label{sec:binary_candidates}

\subsection{Photometric and radial velocity variability}

Long-period variable candidates were selected from 	
the \texttt{gaiafpr.vari\_long\_period\_variable} table, following the criteria in \cite{GaiaFPR}. In particular, non-pulsating, binary sources were isolated making cuts in the photometric $G$-band amplitude, Amp$_G$, and the RV amplitude, A$_{V_R}$:
\begin{equation}\label{eq:amp_condition_ell}
    {\rm Amp}_G < 0.3\,{\rm mag}\;\;\;{\rm and}\;\;\;\frac{{\rm Amp}_G}{{\rm mag}} < 2.5\cdot10^{-3}\cdot\left(\frac{A_{V_R}}{{\rm km\,s}^{-1}}\right)^2 \,\text{.} 
\end{equation}
From this sample, only those stars in the ``top-quality sample" (TQS, \texttt{flag\_rv} = True) were selected. The TQS consists of stars with compatible RV and photometric variability, having (P$_{V_R} \simeq$ P$_{\rm ph}$ or P$_{V_R} \simeq$ 2 P$_{\rm ph}$) for all P$_{\rm ph}$ $\in$ \{P$_{\rm G}$, P$_{G_{\rm BP}}$, P$_{G_{\rm RP}}$\}. There are 881 sources that pass these cuts. 

\begin{figure}
   \centering
\includegraphics[width=0.5\textwidth]{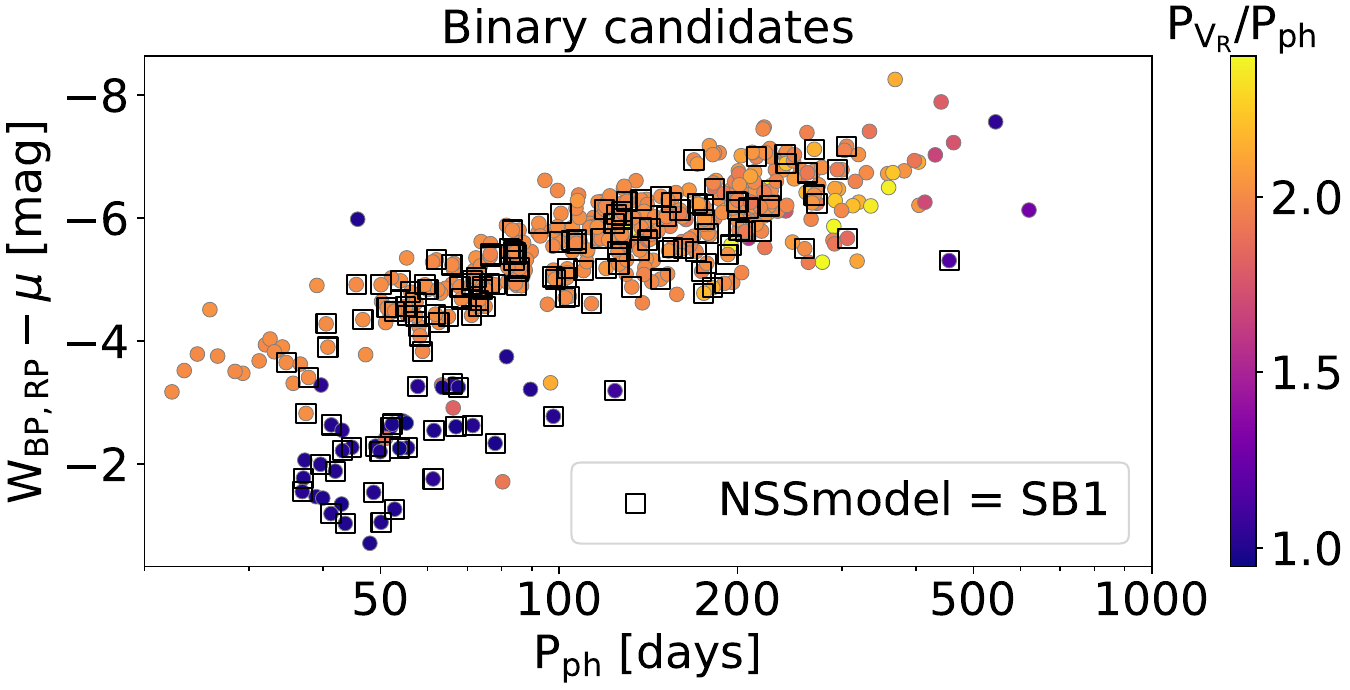}
 \caption{Photometric period versus the absolute Wesenheit magnitude, W$_{\rm BP, RP}$ - $\mu$, for the long-period binary candidates. The color code corresponds to the ratio between the period derived based on the RV curve and the photometric period. Stars having binary solutions in the \textit{Gaia} NSS catalogue are marked with empty squares. Two sequences are recovered: a sequence composed of brighter stars with P$_{V_R}$/P$_{\rm ph}$ $\simeq$ 2, and a tidally-locked population having lower luminosities and shorter periods.} \label{PL_NSS}
\end{figure}

To characterize the physical properties of the sample, a good estimate of their distance and atmospheric parameters is needed. Therefore, we also imposed a cut in parallax to select sources with reliable astrometry: 0.0 $<\epsilon_{\varpi}/\varpi <$ 0.15, and Renormalized Unit Weight Error (RUWE) $<$ 1.4 (see e.g., \citealt{Lindegren21}, and the discussion on its impact on the analysis of single-line spectroscopic binaries in \citealt{Gosset24}), obtaining 636 sources. The cut in RUWE aims to remove partially resolved binaries in which the astrometric solution has large uncertainties. This cut therefore removes true binaries from our sample (28 sources) for which the RV determination and photogeometric distances can be highly affected by their binary nature. The 636 sources were cross-matched to have an entry in the \texttt{gaiadr3.astrophysical\_parameters} table, leading to an initial sample of 434 sources (hereinafter, the input sample). Their distribution in the period-luminosity diagram is shown in Figure~\ref{PL_NSS}, in which the photometric period based on the $G$-band time series, P$_{\rm ph}$, and the \textit{Gaia} Wesenheit index \citep[W$_{\rm BP,RP}$, see][]{Lebzelter18, Lebzelter19} are used. The distance modulus $\mu$ of each star was estimated using the parallax. The stars are color-coded according to the ratio of the RV and photometric variability (P$_{V_R}$/P$_{\rm ph}$), and those sources included in the NSS catalogue \citep{GaiaNSS}, having a binary solution SB1 (single line spectroscopic binary, 154 sources) are marked with empty black squares. In this diagram, the brightest sequence is composed of stars with W$_{\rm BP, RP}$ - $\mu \geq$ --3.0 mag, 20.0 $\leq$ P$_{\rm ph}$ (days) $\leq$ 400.0 days, and P$_{V_R}$/P$_{\rm ph}$ $\simeq$ 2, consistent with the sequence of ellipsoidal binaries \citep{Nicholls10,GaiaFPR}, also known as the sequence E for ellipsoidal variables in the LMC \citep[see e.g.,][]{Soszynski07}.

The second sequence is restricted to shorter periods, P$_{\rm ph} \leq$ 150.0 days, having P$_{V_R}$ $\simeq$ P$_{\rm ph}$. The nature of this sequence is uncertain, although it is mostly composed of binaries according to the NSS solutions, which does not favour the scenario of pulsating stars. Their period ratio is consistent with (almost) tidally-locked binaries, in which the orbital period is the same as the photometric period. This sequence resembles the secondary, although $\sim$1 mag fainter, sequence found for contact binaries (including ellipsoidal binaries) for the LMC reported in \cite{Muraveva2014}. It is worth mentioning that, when using the RV period P$_{V_R}$, the sequence of ellipsoidal variables is shifted and both sequences merge into one. Most of these stars have a variability type “Rotational” in the compilation of \cite{Gaia_var}; therefore, we would refer to them as “rotational” binaries from now on. In this variability class, the rotation of a star with a spotted surface produces the observed photometric variability, which is enhanced in close binary systems compared to a single red giant star with a similar rotation period \citep{Yadav15, Gehan22}.

\begin{figure}
   \centering  
   \includegraphics[width=0.5\textwidth]{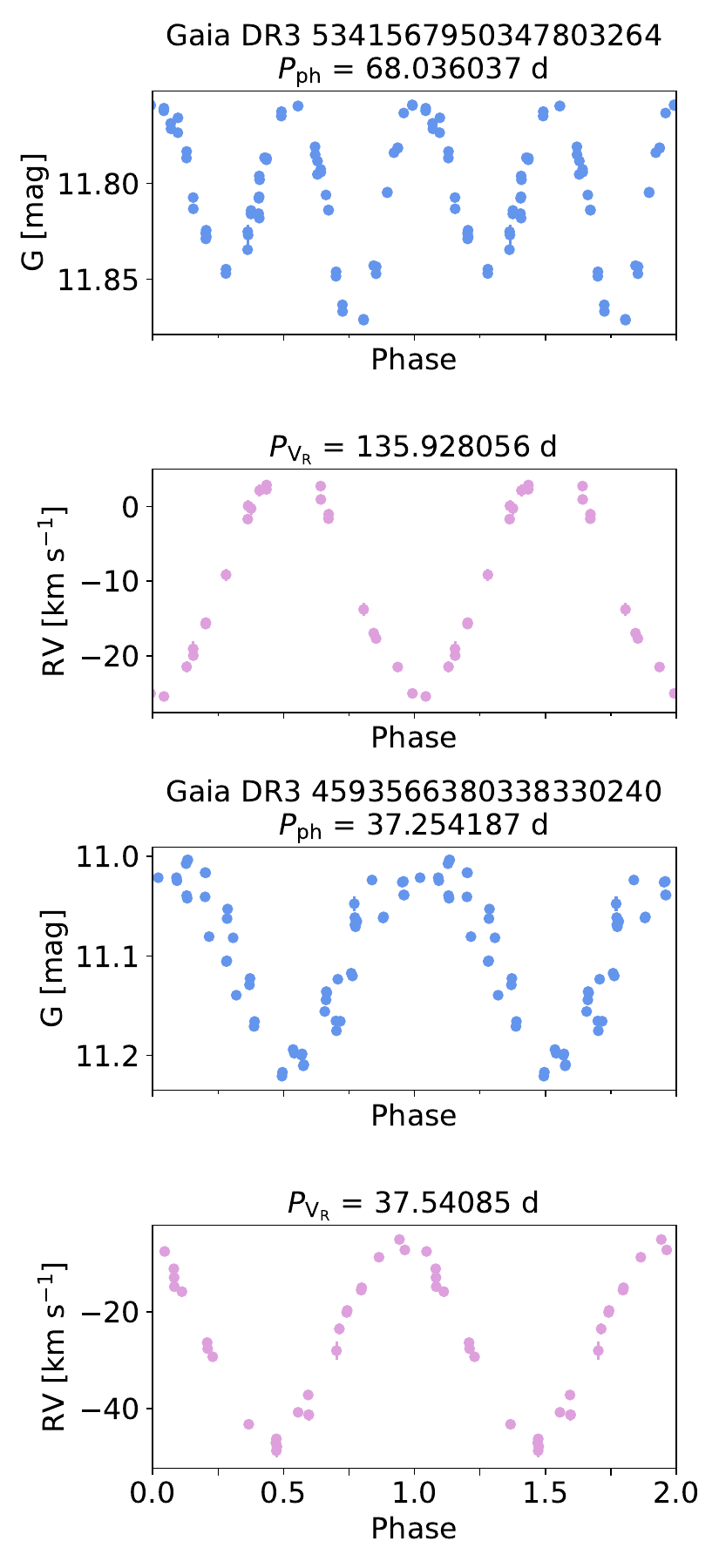}
 \caption{$G$-band and RV curves for one ellipsoidal binary candidate (the two top panels) and one rotational candidate (the two bottom panels). The $G$-band light curve was phase-folded using the period from the RV curve in the case of the ellipsoidal variable, recovering the two minima per orbital cycle.} \label{fig:light_curves_2p}
\end{figure}

To confirm the difference between the two groups, the \textit{Gaia} photometric periods were compared to those from the literature (see Appendix~\ref{app:periods}). Based on the overall good agreement, we can conclude that the two sequences seen in the period-luminosity diagram in Figure~\ref{PL_NSS} are not due to an incorrect photometric period estimate, but rather to an intrinsically different variability phenomenon. We visually inspected the phase-folded $G$-band light curves using the period obtained based on the RV curve. For ellipsoidal binary candidates, two minima of different depth are recovered, while for the rotational candidates, only one minima is obtained. Figure~\ref{fig:light_curves_2p} shows two photometric and RV light curves for one ellipsoidal (left) and one rotational (right) star candidate. From this inspection, we discarded 19 ellipsoidal candidates (P$_{V_R}$/P$_{\rm ph} \lesssim$ 2.0) having light curves with one single minimum or being highly irregular. For the rest of the analysis, we will analyse separately ellipsoidal (being restricted to P$_{V_R}$/P$_{\rm ph} \geq$ 1.5) and rotational (P$_{V_R}$/P$_{\rm ph} <$ 1.5) binary candidates.

In the following sections, four samples will be analyzed, based on independent input catalogues (i) the input sample, having atmospheric parameters (effective temperature and surface gravity) for 372 ellipsoidal and 43 rotational candidates from either the \textit{GSP-Spec} module \citep{GaiaRVS} or XGBoost \citep{xgboost}, 9 and 32 of them, respectively, also having an estimate for the \textit{Gaia} stellar activity index parameters \citep[only considering those with positive values for the \texttt{activityindex\_espcs} parameter, see][]{Gaia_activity}; from the spectroscopic atmospheric parameters, the physical parameters such as the luminosity, radius and mass will be derived. (ii) Orbital properties, including orbital periods and semi-amplitudes (for the full input sample) and eccentricities (available only for 154 stars from  \texttt{gaiadr3.nss\_two\_body\_orbit}); (iii) kinematic properties, including position, distances, radial velocity and proper motions, available for all the sources in the input sample, from \textit{Gaia} \texttt{gaiadr3.gaia\_source}; (iv) chemical properties, including [M/H], [$\alpha$/Fe], [Ce/Fe], based on \textit{Gaia} \textit{GSP-Spec} data, only for ellipsoidal variables, available for 125, 43 and 8 binaries, after applying quality cuts, respectively (see Section~\ref{sec:chemo-dynamical}).

\subsection{Atmospheric parameters: T$_{\rm eff}$, $\log{g}$, [M/H]}\label{sec:atm}

The input sample of binaries has been observed with either the \textit{Gaia} Radial Velocity Spectrometer \citep[RVS, $\mathcal{R}$ $\sim$ 11500, ][]{Sartoretti23} and/or with the BP/RP low-resolution specto-photometers \citep[XP spectra, $\mathcal{R}$ $\sim$25 -- 100,][]{Gaia_deAngeli}. Atmospheric parameters from RVS spectra were derived and validated by the \textit{Gaia} \textit{GSP-Spec} module, available in the \texttt{gaiadr3.astrophysical\_parameters} table, see \citet[hereinafter the ``\textit{GSP-Spec} sample/parameters'']{GaiaRVS}. Atmospheric parameters derived from low-resolution \textit{Gaia} XP spectra using a data-driven approach from the XGBoost algorithm \citep[see][]{xgboost} are also available for most of the sources (hereinafter, the ``XGBoost sample/parameters'').

In the case of ellipsoidal variable candidates, for those with the \textit{GSP-Spec} flag \texttt{KMgiantPar = 0} \citep[i.e., KM-giants without issues in their parametrization, see Table C.8 from][]{GaiaRVS}, the \textit{GSP-Spec} surface gravity and metallicity were calibrated based on their effective temperature through the relations presented in Table A.1 in \cite{ARB24}. The [$\alpha$/Fe] abundances were calibrated using the coefficients in Table 4 of \cite{GaiaRVS}. In the rest of this analysis, only the calibrated atmospheric parameters will be used. More than half of our sample has quality flag \texttt{KMgiantPar = 1, 2}, corresponding to cool giant stars (T$_{\rm eff} <$ 4000 K, $\log{g} <$ 3.5), for which the raw logg estimation had a significant bias and was replaced by a constant value \citep[see the full definition of this flag in Table C.8 of][]{GaiaRVS}. For those stars, when available, we adopted the atmospheric parameters from XGBoost. In this case, we set the [$\alpha$/Fe] abundance (necessary to derive the bolometric correction, see Section~\ref{sec:LMR}), according to the standard relation for the Galactic disk:
\begin{align}
[\alpha/\text{Fe}] &=  +0.0\text{ dex, for [M/H] $\geq$ 0.0 dex;} \nonumber \\
[\alpha/\text{Fe}] &= -0.4 \text{ dex}\times \text{[M/H]}\text{, for $-$1.0 dex $\leq$ [M/H] $<$ 0.0 dex;} \nonumber\\
[\alpha/\text{Fe}] &= +0.4\text{ dex, for [M/H] $<$ $-$1.0 dex.}
\label{eq:alpha}
\end{align}

\begin{figure}
   \centering  \includegraphics[width=0.5\textwidth]{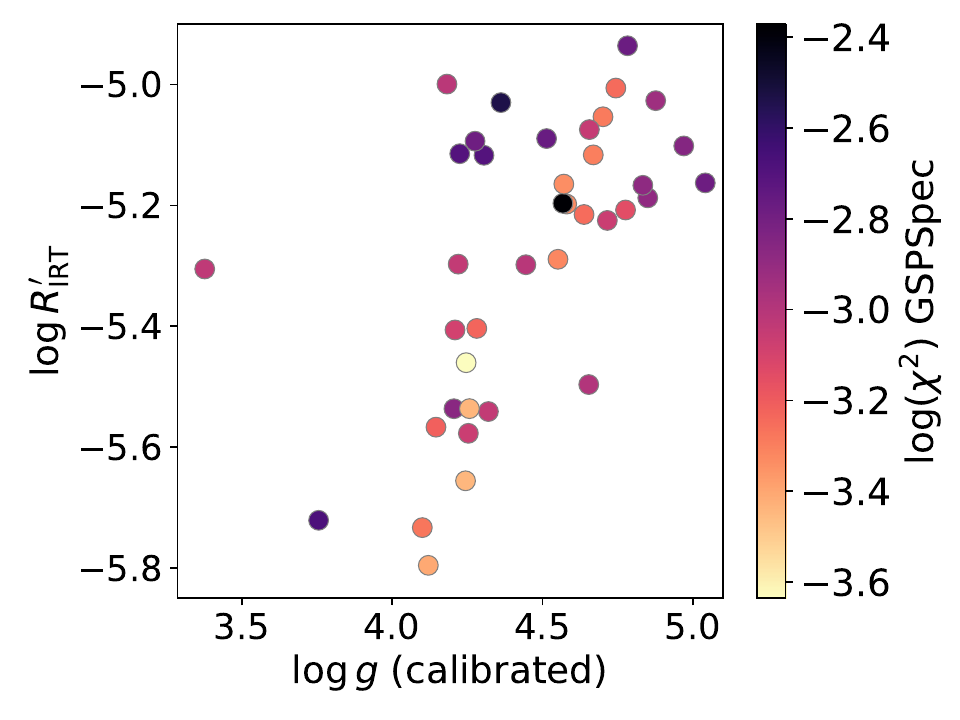}
 \caption{Surface gravity and the activity index $\log{R^{\prime}_{\rm IRT}}$ for stars with available \texttt{activityindex\_espcs} measurements in \textit{Gaia} DR3. The stars are color-coded according to $\log{\chi^2}$, the goodness-of-fit of the RVS spectra from the \textit{Gaia} \textit{GSP-Spec} module.} \label{fig:logg_activity}
\end{figure}

In tidally-locked rotational binaries, the induced rotation is generally associated with an enhanced chromospheric activity, which can bias the atmospheric parameter measurements \citep{Morel04, Cao23}. To test the potential effect of stellar activity in the \textit{GSP-Spec} atmospheric parameters, we compared the $\log{g}$ values with the activity index $\log{R^{\prime}_{\rm IRT}}$ \citep[following Eq. (6), (7) and Table 1 in][]{Gaia_activity} derived using T$_{\rm eff}$ and [M/H] from \textit{GSP-Spec}, and the activity index from \textit{Gaia} DR3 \texttt{activityindex\_espcs} which is available for sources with photometric $\log{g} \geq$ 3.0 dex. Figure~\ref{fig:logg_activity} shows the $\log{g}$ and $\log{R^{\prime}_{\rm IRT}}$, which is a measure of the flux difference between the observed flux with respect to the radiative equilibrium approximation in the core of the Ca II IRT lines, derived as a synthetic spectrum having the same atmospheric parameters as those derived by the \textit{Gaia GSP-Phot} module \citep{Gaia_activity}. The color-code corresponds to the $\log{\chi^2}$ of \textit{GSP-Spec}, which represents the goodness-of-fit of the atmospheric parameters. There is a trend, although with significant scatter, between both quantities, which could be explained considering that active stars have an excess in the Ca II equivalent width, with respect to an inactive star, which could impact the fit in the \textit{GSP-Spec} module leading to potentially overestimated $\log{g}$ values and worse (higher) values of $\log{\chi^2}$. Stars having $\log{R^{\prime}_{\rm IRT}} \geq$ -- 5.3 are considered as highly active in \cite{Gaia_activity}, and are exactly those with the largest surface gravities and worse goodness-of-fit, indicating that the surface gravity estimates for these stars are strongly affected by the level of activity. As the surface gravity and also the metallicity can be biased due to activity, the atmospheric parameters for those stars with \textit{GSP-Spec} $\log{g}_{\rm calibrated} \geq$ 2.2 dex were not considered and instead, the atmospheric parameters from XGBoost were adopted (see the Appendix~\ref{app:logg} for a discussion about the selection of this threshold).
 
\begin{figure}
   \centering
\includegraphics[width=0.5\textwidth]{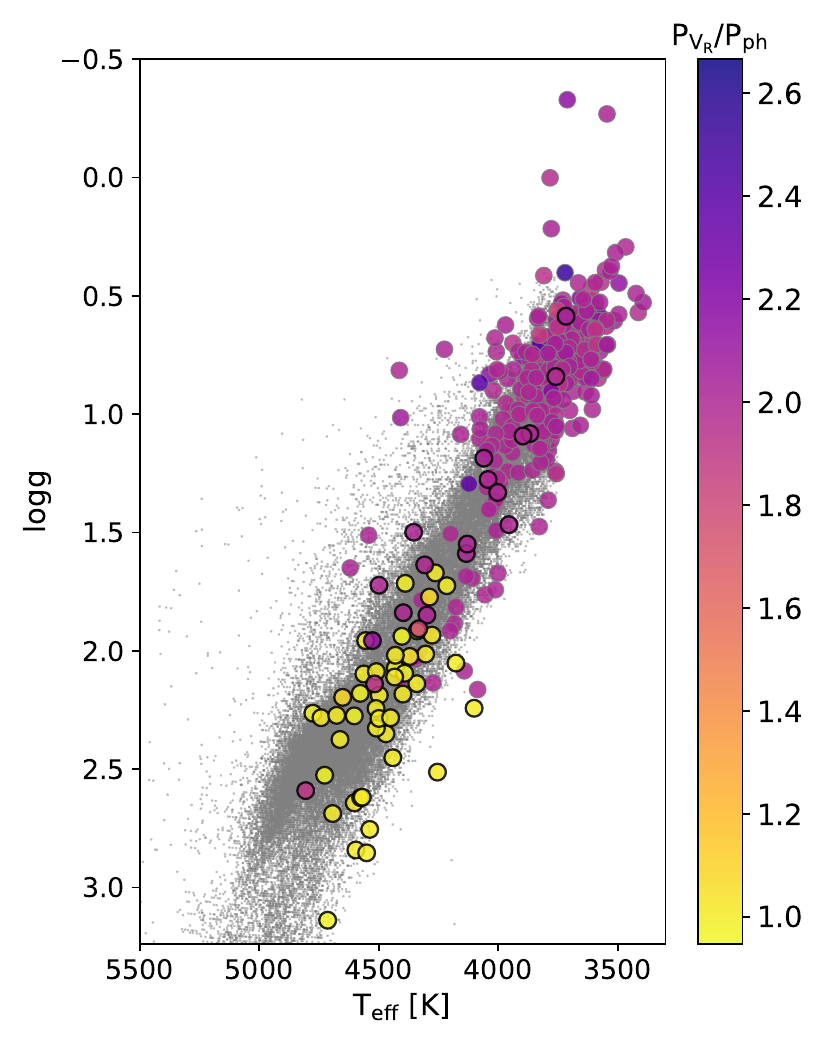}
 \caption{Kiel diagram for the complete sample of red giant binaries, having spectroscopic effective temperature and surface gravity (either from \textit{GSP-Spec} or XGBoost), color-coded based on the ratio between the orbital and photometric period. The sources for which the parameters from XGBoost were preferred against the parameters from \textit{GSP-Spec} are shown as circles with black edges. The background, grey distribution corresponds to a precise \textit{Gaia} \textit{GSP-Spec} sample, selected as in \cite{ARB24}.}
 \label{CMD_period}
\end{figure}

Figure~\ref{CMD_period} shows the Kiel diagram for the sample. Stars for which the parameters adopted are from XGBoost due to an overestimated surface gravity measurement in \textit{GSP-Spec} are shown with black edges, being mostly those rotational binaries with $\log{g} >$ 2.0 dex. The binary stars of our sample are color-coded according to the ratio between the RV and photometric period, to separate the sample into those that are most likely of ellipsoidal type (P$_{V_R}$/P$_{\rm ph} \simeq$ 2.0) to the ones that are most likely tidally locked rotational variables. A sample of high-precision, accurate \textit{Gaia} \textit{GSP-Spec} data set \citep[having median uncertainties of 10 K, 0.03 and 0.01 dex in T$_{\rm eff}$, $\log{g}$ and {[M/H]}, respectively; following the selection cuts as in][]{ARB24} is included as grey dots. The Kiel diagram is consistent with cool red giants, being the ellipsoidal binary candidates more evolved than the rotational variables, although they have $\log{g}$ values above the position of the red clump \citep[see][]{ARB24}. The different atmospheric parameters for the two groups are consistent with the presence of two sequences in Figure~\ref{PL_NSS}, where the rotational candidates are hotter and less evolved, with shorter orbital periods, compared to the bulk of ellipsoidal candidates.

In summary, from the input sample of 434 red giant binaries, atmospheric parameters from  \textit{Gaia} \textit{GSP-Spec}, having \texttt{KMgiantPar = 0} and calibrated $\log{g} <$ 2.2 dex, were adopted for 86 ellipsoidal candidates and 2 rotational candidates, while XGBoost atmospheric parameters were adopted, if available, for the rest of the cases, accounting for 292 ellipsoidal and 41 rotational candidates. From the sample, one possible contaminant is \textit{Gaia} DR3 517611939345861120, which is the brightest star among the non-ellipsoidal sample, even brighter than the sequence of ellipsoidal variables in the PL diagram for its period, P$_{\rm ph}$ = 45.7 d (see Fig.~\ref{PL_NSS}). Its atmospheric parameters both from \textit{GSP-Spec} and XGBoost place it far from the red giant sample, being much hotter than the rest of the sample. We detail the peculiarities of this system in Appendix~\ref{app:comments} and we discard it from the rest of the analysis. In total, there are reliable atmospheric parameters for 424 red giant binaries.

\section{Physical stellar parameters}

\subsection{Luminosities, radii and spectroscopic masses}\label{sec:LMR}
The stellar parameters of the primary star in each system, e.g., luminosity, mass and radius, were derived using both \textit{Gaia} photometry and spectroscopy, following de Laverny et al. (in prep.). In the following, we shortly enumerate the steps to derive those quantities. 

The color-excess E(BP$-$RP) was first estimated considering the effective temperature, surface gravity and metallicity and the observed \textit{Gaia} (BP-RP) color. The true color was obtained using the colour-T$_{\rm eff}$ relations for \textit{Gaia} DR3 bandpasses\footnote{\url{https://github.com/casaluca/colte}} \citep{Casagrande21}. To derive the stellar luminosity, the absolute magnitude was obtained considering the photogeometric heliocentric distance derived by \cite{BailerJones} and the A$_G$ extinction, obtained from the E(BP-RP) color-excess. The bolometric correction BC$_{\rm G}$ was obtained using the \texttt{bcutil} package\footnote{\url{https://github.com/casaluca/bolometric-corrections}}, for \textit{Gaia} DR3 magnitudes \citep{Casagrande18}, considering the atmospheric parameters (T$_{\rm eff}$, $\log{g}$, [M/H], [$\alpha$/Fe]) either from \textit{GSP-Spec} or XGBoost. The adopted bolometric magnitude of the Sun was M$_{\rm bol, \odot}$ = 4.74 mag. From the luminosity estimates and the spectroscopic effective temperature, the stellar radius was also derived.

The $\log{g}$ measurements and the derived radii were then used to estimate the mass of the primary star, as 
\begin{equation}
\log{\left( \frac{{\mathcal{M}}}{{M}_{\odot}} \right)} = \log{g} + 2 \log{\left( \frac{\mathcal{R}}{{\rm R}_{\odot}} \right)} - \log{g_{\odot}} \text{,}
\label{eq:Mlogg}
\end{equation}
adopting $\log{g_{\odot}} =$ 4.44.

The associated errors on the distance, apparent $G$ magnitude and atmospheric parameters (T$_{\rm eff}$, $\log{g}$, [M/H], [$\alpha$/Fe]) were propagated through 1000 Monte Carlo realizations. In the case of atmospheric parameters from XGBoost, the errors adopted for the atmospheric parameters were fixed to 50 K in T$_{\rm eff}$, 0.08 dex in $\log{g}$ and 0.1 in [M/H], following \cite{xgboost}. We arbitrarily adopted an error of 0.1 dex in [$\alpha$/Fe]. It is worth mentioning that this analysis does not take into account the error associated with the light contribution of the secondary both in the photometric and spectroscopic measurements. Depending on the brightness ratio among the two components, the associated uncertainties can be non-negligible and hard to estimate. We, therefore, do not take this error into account explicitly in the error determination. For the following analysis, stars for which unreliable extinction and/or bolometric correction estimates, due to unphysical results or values outside the color range covered by the colour-T$_{\rm eff}$ relations from \cite{Casagrande21} were discarded. These are mostly stars with T$_{\rm eff} \leq$ 3700 K. 

As the estimate of the spectroscopic masses depends strongly on the accuracy and precision of the surface gravity measurements \citep[see Eq.~\ref{eq:Mlogg} and the comparison with masses based on asteroseismology in][]{ARB24}, in a few cases, the primary masses recovered had large errors ($>$ 50\%) or a significantly large value. Based on the temperature and luminosity of our sample, it is not expected to have sources with primary masses larger than $\sim$6.0 M$_{\odot}$ (depending on the metallicity). We revised those cases individually (see Appendix~\ref{cleaned_masses}), leading to the rejection of three stars, the validation of their masses (2 stars) or the adoption of a different mass value using the set of atmospheric parameters from XGBoost instead of \textit{GSP-Spec} (6 stars). The median values for the extinction, luminosity, mass and radii, as well as the 84th and 16th percentile values (see de Laverny et al., in prep.) for this revised sample are included in Table~\ref{tab:parameters}. We have included as well a quality flag for the variability classification based on the visual inspection of the light curve. Those having a well-defined phase-folded $G$-band light curve, presenting two clear minima (for the case of ellipsoidal candidates) or one sinusoidal shape (rotational candidates) have a flag "Good" (G), while those presenting more scatter in the light curve or gaps in the phase coverage have marked as "Regular" (R), while for the most irregular cases, we chose to mark them as "Uncertain" (U). As visual inspection may be prone to be subjective, the flags should be considered only as an indication of the quality of the light curve. Our results and conclusions do not change if, for instance, the sources flagged as "U" are removed from the sample.

\begin{figure*}
\centering
\includegraphics[width=\textwidth]{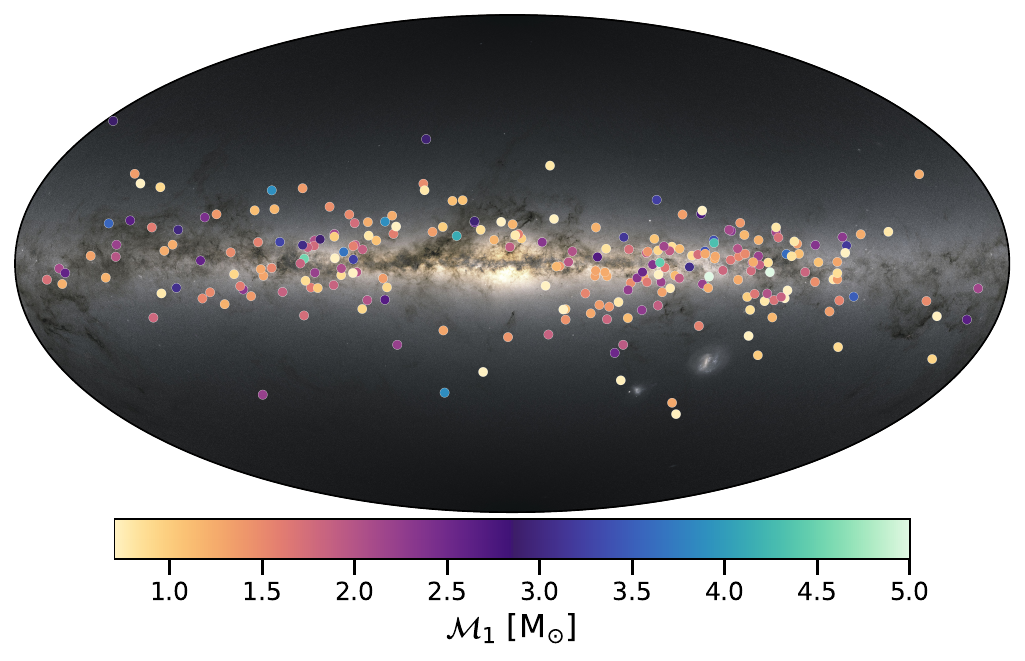}
\caption{All-sky Aitoff projection of the ellipsoidal variable candidates with reliable physical parameters (luminosity, radius, mass) from the primary star. The color-bar corresponds to the mass of the primary $\mathcal{M}_1$. The background image was generated as part of the \textit{Gaia} eDR3, based on more than 1.8 billion sources (Image credits: ESA/\textit{Gaia}/DPAC).}
\label{fig:all_sky}
\end{figure*}

\begin{figure*}[h!]
   \centering
\includegraphics[width=0.99\textwidth]{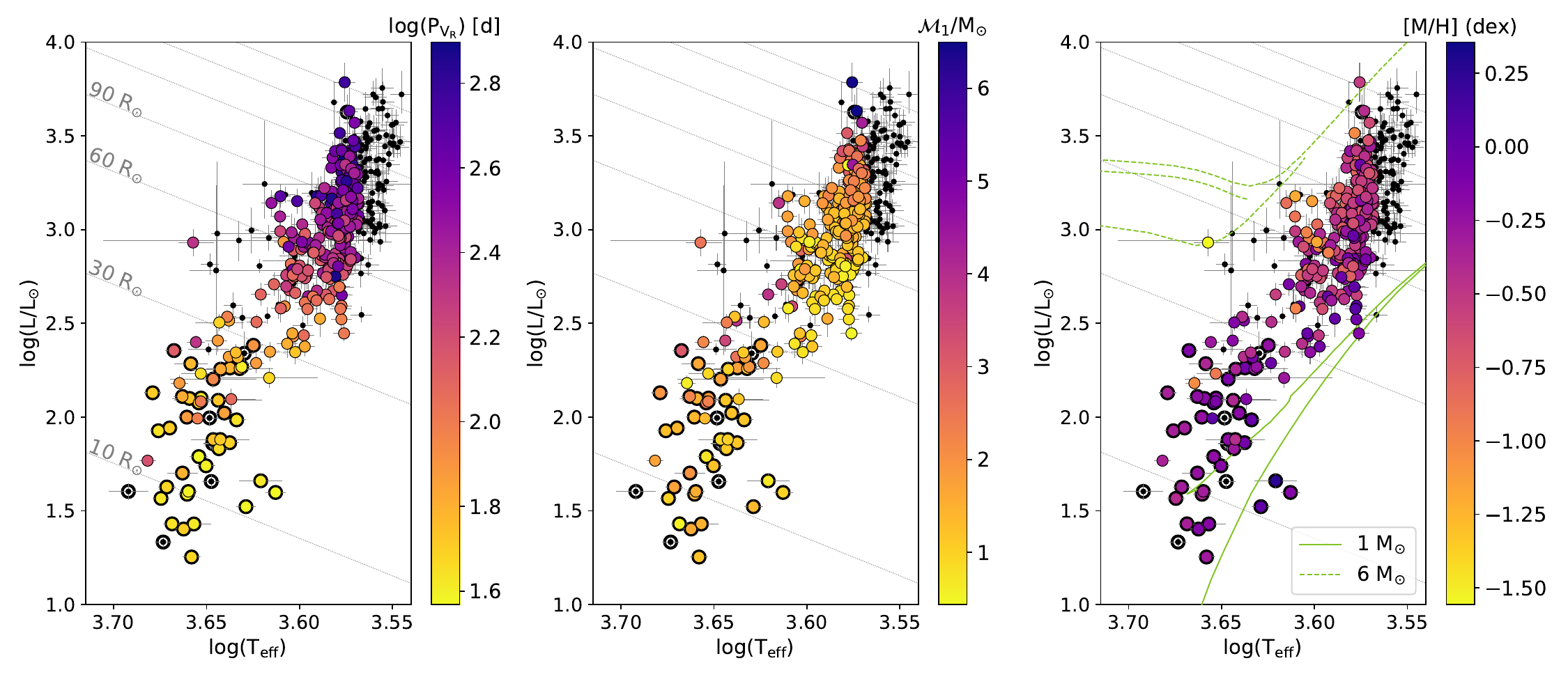}
 \caption{Hertzsprung-Russell diagram for the sample of rotational and ellipsoidal binaries. From left to right, the points are color-coded according to the logarithm of its orbital period, P$_{\rm V_{\rm R}}$, the mass of the primary star and the adopted spectroscopic metallicity [M/H]. Black points correspond to stars with unreliable extinction determination (mostly stars with T$_{\rm eff} \leq$ 3700 K, see Sect.~\ref{sec:LMR}), and therefore, potentially unreliable stellar parameters while those with a thick black edge are the rotational candidates. The grey lines correspond to constant radii of 10, 30, 60, 90, 120, 150 and 180 R$_{\odot}$. In the right panel, the evolutionary tracks for a 1 M$_{\odot}$ (green curve) and a 6 M$_{\odot}$ (green dashed curve) star, having Z=0.004 ([M/H]$\sim-$0.6), from the grid computed by \cite{Lagarde12}, based on the STAREVOL \citep{Siess2000} code, are included.}\label{HR_Radius_Masses}
\end{figure*}

In total, we recovered reliable masses for 243 ellipsoidal variable candidates and 39 rotational candidates. The all-sky Aitoff projection of the ellipsoidal variables is shown in Figure~\ref{fig:all_sky}, where each source is color-coded according to the spectroscopic mass of the primary star. Rotational candidates are not included in this Figure for clarity, since they are less than 40 systems, with similar mass values (their masses can be seen in Figure~\ref{HR_Radius_Masses}).

The Hertszprung-Russell (HR) diagram for the luminosities, radii and masses derived is shown in Figure~\ref{HR_Radius_Masses}. The left panel has a color-bar for the orbital period, in logarithmic scale, also including lines for the stellar radius. For the ellipsoidal candidates, the spectroscopic radii range from $\sim$10 to $>$180 R$_{\odot}$, with a median value of $\sim$60 R$_{\odot}$. The rotational candidates have smaller radii, not larger than $\sim$30 R$_{\odot}$. This is an expected result, since binary systems with longer orbital periods have larger separations, and therefore a larger radius is needed to reach the Roche-lobe and to be detectable as an ellipsoidal variable. The stellar masses for the primary star $\mathcal{M}_1$ (middle panel) show low to intermediate masses for both rotational and ellipsoidal candidates, with a handful of cases for which $\mathcal{M}_1 \geq$ 6.0 M$_{\odot}$ (the more massive stars are compared against evolutionary tracks in Appendix~\ref{cleaned_masses}). 

The right panel of Figure~\ref{HR_Radius_Masses} shows the spectroscopic metallicity [M/H] associated with the adopted atmospheric parameters. The metallicities of rotational candidates are mostly supersolar. Having an enhanced chromospheric activity (see Section~\ref{sec:atm}), their metallicity values could be as well overestimated. In fact, previous studies have found subsolar metallicities for active single-line spectroscopic binaries and RS CVn systems, although the sample sizes were rather small \citep[see e.g.,][]{Morel04, Morel06}. The corresponding adopted metallicities from XGBoost are quite different from those in \textit{GSP-Spec}, probably due to the effect of enhanced activity. For the rest of the analysis, we therefore decided to exclude rotational variables when interpreting chemical content.

Ellipsoidal variables have mainly sub-solar metallicities. For reference, we include the evolutionary tracks of a Z=0.004 ([M/H]$\sim-$0.6), 1.0 M$_{\odot}$ and 6.0 M$_{\odot}$ star as continuous and dashed green lines, respectively. These were computed with the STAREVOL code \citep{Siess2000} by \cite{Lagarde12}. The position in the HR diagram for most of the ellipsoidal variables is compatible with those tracks, although their metallicities may be different from the adopted ones in the evolutionary tracks. Their metallicity distribution function is further explored in Section~\ref{sec:mdf}.

\vfill

\subsection{Validation of the spectroscopic masses}

\begin{figure}
\centering
\includegraphics[width=0.5\textwidth]{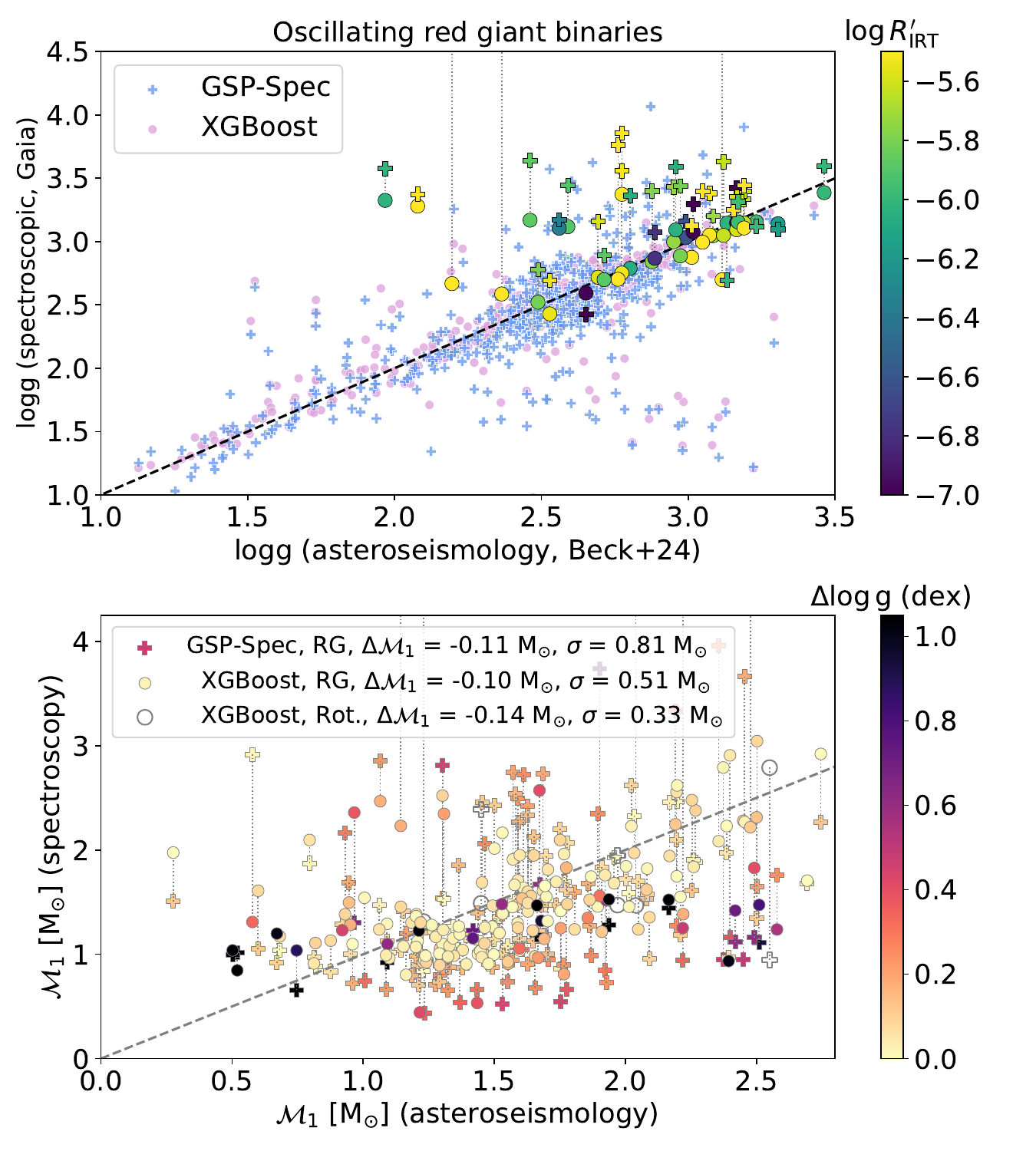}\\
\caption{Spectroscopic \textit{GSP-Spec} (crosses) and XGBoost (circles) parameters and asteroseismic measurements for surface gravity (top panel) and masses (bottom panel) for red giant (RG) binaries \citep{Beck24} and rotational binaries \citep{Gaulme20} having solar-like oscillations. The two spectroscopic values are connected with a vertical dashed line. The color bar shows the activity index and the difference in $\Delta {\rm log\, g}$ in the top and bottom panel, respectively. The median difference and dispersion between spectroscopic and asteroseismic masses are quoted in the bottom panel.}
\label{fig:logg_astero}
\end{figure}

The spectroscopic masses obtained are highly sensitive to the accuracy of the surface gravity, which may be strongly affected by, for example, enhanced chromospheric activity (see Section~\ref{sec:atm}). To test the accuracy of the \textit{Gaia} $\log{\rm g}$ estimates, the catalogue of oscillating red giant binaries from \cite{Beck24} was used as a reference sample. It is based on \textit{Gaia} DR3 orbital solutions and solar-like oscillators having asteroseismology measurements from \textit{Kepler} \citep{Borucki10, Howell14} and TESS \citep[][]{Ricker14} missions. We have cross-match the subsample of giant star binaries in this catalogue (having the same range of orbital periods as our sample) with the atmospheric parameter information from \textit{Gaia} DR3 and XGBoost, to compare surface gravities and activity levels. The top panel in Figure~\ref{fig:logg_astero} shows the $\log{\rm g}$ derived from asteroseismic measurements and the spectroscopic values in \textit{Gaia} DR3, either from the \textit{GSP-Spec} module (crosses) or the XGBoost catalogue (circles). For those sources having an entry for the activity index, the $\log{R^{\prime}_{\rm IRT}}$ activity index was estimated (see Sect.~\ref{sec:atm} and Fig.~\ref{fig:logg_activity}) and it is shown in the color-bar, while the two spectroscopic measurements are connected with a vertical dashed line. There is a good agreement between both spectroscopic and asteroseismic measurements for $\log{\rm g}$ values lower than 2.0 dex, i.e., more evolved stars, which corresponds to our sample of ellipsoidal variables. For larger surface gravities, there are more discrepancies, particularly for the stars having a measured higher activity index. The values of $\log{\rm g}$ larger than $\sim$2.0 dex from \textit{GSP-Spec} tend to be overestimated in most of the cases which is generally not the case for XGBoost (see also the discussion in Appendix~\ref{app:logg} and Fig.~\ref{fig:logg_comp}) since BP/RP spectra have lower sensitivity to chromospheric effects. This comparison supports the selection of XGBoost atmospheric parameters for the less evolved sample, which is mostly composed of rotational variables.

Based on the spectroscopic \textit{Gaia} atmospheric parameters, the luminosity, radius and masses of the oscillating red giant binaries were derived, following the same procedure as for our sample, and the spectroscopic mass estimates compared to the asteroseismic mass, obtained from the scaling relations for $\nu_{\rm max}$, $\Delta \nu$ and effective temperature as in \cite{Beck24}. The bottom panel in Figure~\ref{fig:logg_astero} shows the obtained mass when using the \textit{GSP-Spec} (plus signs) and XGBoost (circles) atmospheric parameters, colored based on the difference in the spectroscopic and asteroseismic $\log{g}$ values. There is an overall good agreement, with a mean bias in the spectroscopic mass of $-$0.1 M$_{\odot}$ for the two sets of atmospheric parameters. The unfilled symbols correspond to rotational binaries from \cite{Gaulme20} which have also indications of enhanced activity. The mean bias in the spectroscopic mass is of $-$0.14 M$_{\odot}$ when using the XGBoost parameters, while in the case of the \textit{GSP-Spec} parameters, some measurements are outside the range of masses shown in the plot, having a bias of more than 1.0 M$_{\odot}$.

Although this literature sample of red giant binaries is not composed of ellipsoidal variables, the atmospheric parameters and orbital period ranges are close to our input sample. This comparison therefore serves as a probe of the method and the effect of accurate surface gravities into reliable physical parameters, particularly the mass. Moreover, the effect of activity into the surface gravity measurements based on \textit{GSP-Spec} and XGBoost samples is evident and supports our decision to adopt XGBoost parameters for stars with enhanced activity (rotational variables, having $\log{g}$ from \textit{GSP-Spec} exceedingly large).

\FloatBarrier

\section{Binary population properties} \label{sec:binary_proper}

\subsection{Period-radius, $e$-period, $f(\mathcal{M})$- period relations}

Orbital period, P$_{\rm nss}$, semi-amplitude of the primary star, $K_1$, and eccentricity, $e$, for 154 binaries in our sample are available in the  \textit{Gaia} NSS \texttt{nss\_two\_body\_orbit} catalogue. All of them have the orbital solution being `SB1'. The orbital period and radial velocity semi-amplitude values from the NSS catalogue are consistent with the radial velocity period P$_{\rm V_{\rm R}}$ and radial velocity amplitude \texttt{amplitude\_rv} reported in the FPR table \citep[see][]{GaiaFPR}, while the eccentricity is only available in the former. Having the orbital parameters, the mass function of the binary system, $f(\mathcal{M})$ can be then estimated as

\begin{equation}
f(\mathcal{M}) = \frac{P{K_1}^3 (1-e^2)^{3/2}}{2\pi G} \label{eq_fm}\text{,}
\end{equation}
for the 154 stars with published eccentricities. For the rest of the sample (i.e., those not included in the NSS catalogue), the FPR values for the period and radial-velocity amplitude are used to get an upper limit for the mass function by adopting an eccentricity of $e=0$. 

\begin{figure*}
    \centering
    \includegraphics[scale=0.5]{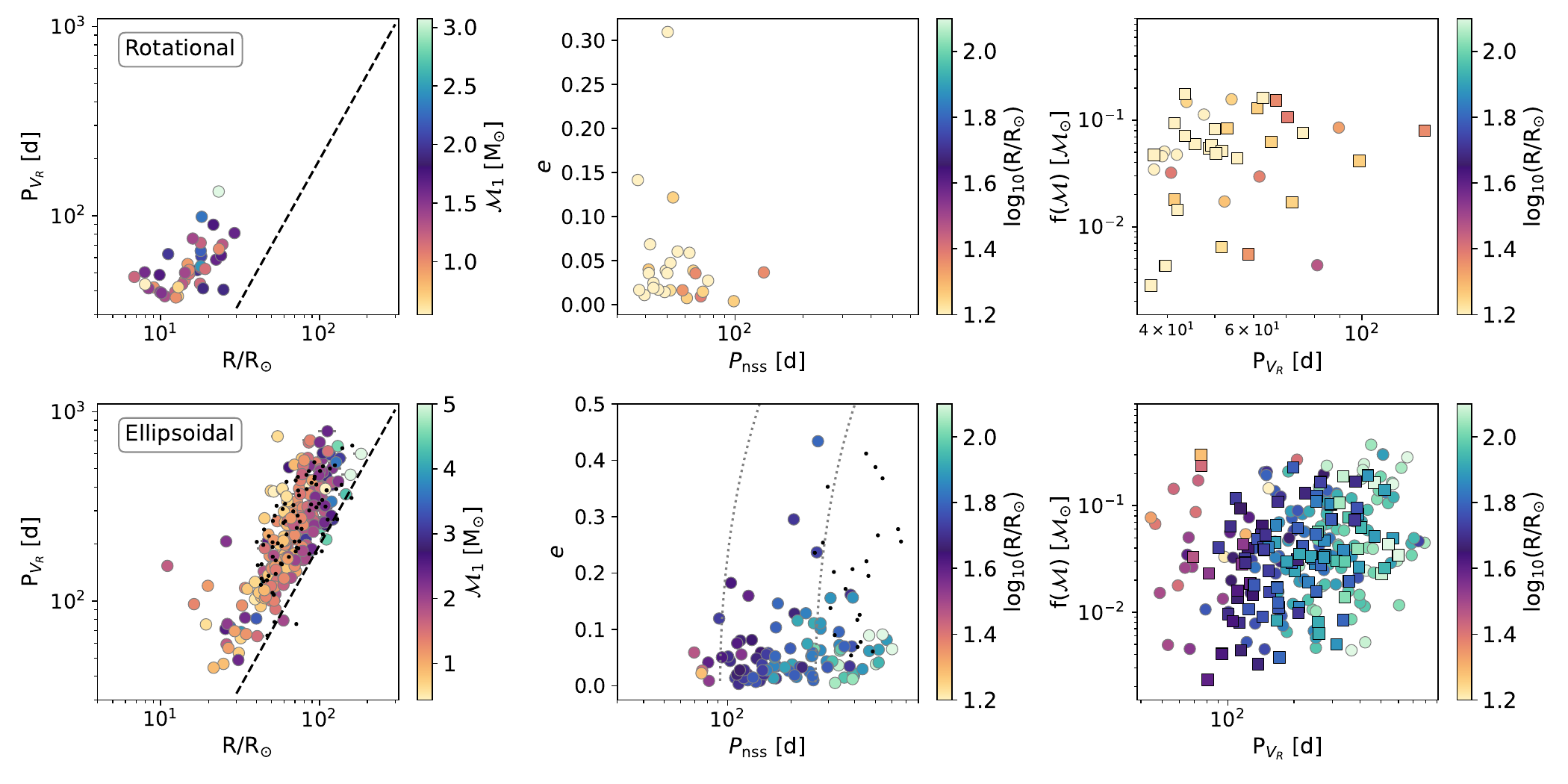}
    \caption{Period versus radius (left), eccentricity (middle) and mass-function (left) for rotational variables (top panels) and ellipsoidal variable candidates (bottom panels). Left panels: Stellar radius of the primary star versus the orbital period, color-coded based on the mass of the primary star. The dashed line corresponds to the threshold period while the black dots are ellipsoidal binaries in the LMC \citep{Nie17}. Middle panels: Orbital period versus the eccentricity for those sources having NSS orbital solution as `SB1', color-coded according to the primary radius. For ellipsoidal variables, the circularization threshold for two values of radius is shown as dotted grey lines and black dots are ellipsoidal binaries in the LMC. Right panels: Orbital period versus mass function $f(\mathcal{M})$, color-coded according to the primary radius. Sources having eccentricity estimates are shown as squares while those without eccentricity measurements are shown as circles.} \label{P_Radius_Masses}
\end{figure*}

Figure~\ref{P_Radius_Masses} shows the period-radius (left panels), eccentricity-period (middle panels) and mass function $f(\mathcal{M})$- period (right panels) distributions for rotational variable candidates (top row) and ellipsoidal variables (bottom row). A threshold or transition period at which the orbit is circular has been found in different binary populations \citep{Jorissen2019, Escorza20}, being this period longer as the radius and luminosity of the star are larger along the red giant branch. An analytical expression for this threshold period can be obtained assuming the radius of the primary star reaches the Roche-lobe in the Kepler's third law, being P$_{\rm threshold}$ a function of the primary radius $R_{\rm 1}$, $\mathcal{M}_1$ and $\mathcal{M}_2$ the masses of the primary and secondary component, respectively. Binary systems having lower periods than this threshold are unlikely to be found as in this case, the primary would have filled up its Roche lobe radius, leading to the subsequent mass-transfer and common-envelope phase, which are short-lived stages. This expression is derived in \cite{GaiaNSS} and corresponds to:

\begin{eqnarray}
\log{(P_{\rm thr}/365.25)} &=& (3/2) \log{(R_1/ {\rm 216} R_{\odot})} \nonumber \\
& & - (1/2) \log{(\mathcal{M}_1 + \mathcal{M}_2)}  \nonumber \\
& & + (3/2) \left( \log{0.38 + 0.2 \log{\mathcal{M}_1/\mathcal{M}_2}} \right)
\end{eqnarray}

The left panels in Figure~\ref{P_Radius_Masses} include this threshold period as a dashed line adopting the mean values $\mathcal{M}_1$ = 1.5, and 1.7 M$_{\odot}$ for rotational and ellipsoidal variables, respectively, and $\mathcal{M}_2$ = 0.7 M$_{\odot}$ (see Fig.~\ref{fig:m2}). The black dots in the bottom left panel are ellipsoidal binaries from the LMC observed from \cite{Nie17}. Almost all the \textit{Gaia} binaries are located below the threshold period, meaning that the radius of the primary, in all the systems, has not reached its Roche-lobe, i.e., the systems are detached (non-contact) binaries. Nonetheless, a large fraction of the ellipsoidal candidates have stellar radii and periods almost at the boundary, consistent with their ellipsoidal nature.

The distribution of orbital periods and eccentricities from the NSS catalogue is shown in the middle panels of Figure~\ref{P_Radius_Masses}, color-coded according to the primary radius. The black dots in the bottom middle panel correspond to measurements for ellipsoidal variables in the LMC \citep{Nie17}. The circularization threshold, i.e., the period at which the binary system should reach zero eccentricity for two values of the radius, $\mathcal{R}_1$ = 50, 100 R$_{\odot}$, and mean values of the masses of the components $\mathcal{M}_1$ = 1.7 M$_{\odot}$, $\mathcal{M}_2$ = 0.7 M$_{\odot}$ (see Fig.~\ref{fig:m2}) is shown as dotted grey lines for the ellipsoidal binaries. The plot for rotational variables (top) and ellipsoidal (bottom) variables show the effects of tidal circularization of the orbits: systems with shorter periods ($P_{\rm nss} \leq$ 100 days) tend to have mostly circular orbits ($e \leq$ 0.05), while for binaries with longer orbital periods, the eccentricities tend to be higher than circular. In fact, tidally locked variables are expected to have circular orbits, as is the case of rotational variables systems. This effect is expected as the timescale for tidal circularization strongly depends on the radius, particularly for giant stars with radiative envelopes \citep{Mazeh08}, as well as on the mass ratio of the system. Based on this distribution, adopting an eccentricity of $e = 0$, for the systems without eccentricities determined in the NSS catalogue, seems reasonable.

The right panels of Figure~\ref{P_Radius_Masses} show the orbital period from the RV curve against the mass function\footnote{It is worth mentioning that none of the eight high-mass $f(\mathcal{M})$ recently reported in \cite{Rowan24} are part of the present analysis as they are not included in the \textit{GSP-Spec} \texttt{astrophysical.parameters} table. Given their magnitude and \texttt{rv\_expected\_sig\_to\_noise}, the RVS spectrum should have been processed by the \textit{GSP-Spec} module \citep{GaiaRVS}, therefore they were most likely removed during the post-processing due to, for instance, unresolved blends affecting the mean spectra \citep{GaiaValidation}.}, color-coded according to the radius of the primary star. To derive the mass function the eccentricity from the NSS orbital solution was adopted, when available (square symbols), while for those systems without eccentricity estimates a circular orbit was adopted (circles). As shown in Eq.~\ref{eq_fm}, the eccentricity factor in the mass function goes as (1-$e^2$)$^{3/2}$, and for eccentricities between zero to 0.3 (middle panels in Figure~\ref{P_Radius_Masses}), the mass function will be reduced by 13\%. The majority of the systems have an eccentricity lower than 0.1 (see also Fig.~\ref{fig:filling_ecc}), which implies a difference of only 0.01\% in the mass function. Therefore, the adopted $e$=0 should have a small impact on the mass function. The mass function for rotational variables is mostly independent of the radius while the ellipsoidal variables have larger radii as larger is the period, as expected. In the case of rotational variables, given their shorter orbital periods than ellipsoidal variables, to get similar values of the mass function they need to have larger radial velocity semi-amplitudes \citep[see Fig. 10 in][]{GaiaFPR}. 
 
\subsection{Minimum mass estimates for the secondary star}\label{sec:min_mass}

The mass of the secondary star in the binary can also be estimated through its mass function (see Eq.~\ref{eq_fm}), expressed as
\begin{equation}
\frac{PK_1^3 (1-e^2)^{3/2}}{2\pi G} = \frac{M_2^3 \sin{i}^3}{(\mathcal{M}_1 + \mathcal{M}_2)^2}\label{eq_fm2}\text{.}
\end{equation}

Using P$_{V_R}$ as the orbital period and the \texttt{amplitude\_rv} as $K_1$, Eq.~\ref{eq_fm2} can be numerically solved to get an estimate of $\mathcal{M}_2 \sin{i}$. The inclination of the binary is not known, therefore an edge-on orientation (i = $\pi$/2) corresponding to the maximum value for $\sin{i}$, is adopted. In this case, the mass function of the system corresponds to the maximum value and the mass of the secondary obtained corresponds to the minimum mass $\mathcal{M}_{\rm 2, min}$. It is worth mentioning that if the inclination is edge-on, the systems would be eclipsing binaries, which is not the case according to the \textit{Gaia} variability classification of all the sources in our sample as LPVs. Although, given the cadence of \textit{Gaia} observations, some eclipsing binaries could have not been detected \citep[see e.g.,][]{Rowan24}. Therefore, the minimum mass should be considered as a strict lower limit. The errors on the secondary minimum mass were obtained through 1000 Monte Carlo simulations sampling the errors in the period and the mass of the primary. As the error for the amplitude of the RV is not reported in the \textit{Gaia} FPR table, the possible contribution of this measurement error is not considered. 

\begin{figure}
   \centering
\includegraphics[width=0.5\textwidth]{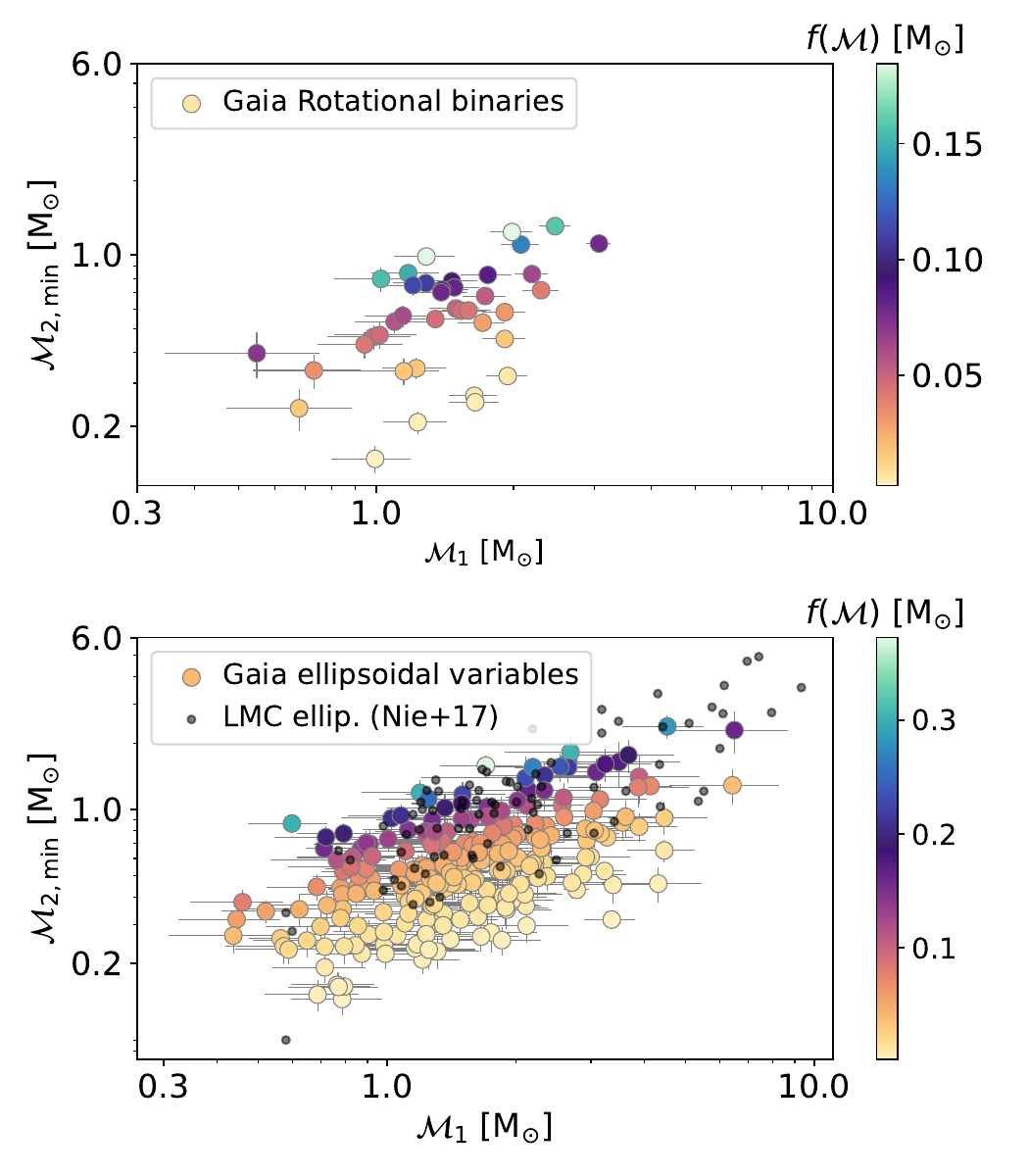} 
 \caption{Masses of the primary $\mathcal{M}_1$ derived using spectroscopic atmospheric parameters and the minimum mass for the secondary star, $\mathcal{M}_{\rm 2, min}$, derived based on the orbital period and RV amplitude. The systems are color-coded according to their mass function $f(\mathcal{M})$.} \label{fig:m2}
\end{figure}

Figure~\ref{fig:m2} shows the mass of the primary $\mathcal{M}_1$ versus the minimum mass of the secondary $\mathcal{M}_{\rm 2, min}$, color-coded according to their mass function $f(\mathcal{M})$. The top panel shows the rotational candidates while the ellipsoidal variables are shown in the bottom panel. For reference, the masses of both components for ellipsoidal variables in the LMC are shown as grey dots \citep{Nie17}. Overall, the masses recovered for the ellipsoidal variables in \textit{Gaia} are similar to those observed in the sample for LMC ellipsoidal variables, despite the fact that the latter has the mass of the companion estimated, including the inclination of the orbit, and not the minimum mass, as it is our case. For a fixed mass function value, the mass of the primary and minimum mass of the secondary have a strong correlation, which is most likely the result of adopting a fixed inclination value. A high fraction of the systems have a minimum mass for the secondary component lower than 1.0 M$_{\odot}$, indicating that possibly the companion could be a stellar remnant (either a white dwarf or a neutron star) or a low-mass main-sequence star. Binaries having a white dwarf and a red giant companion are considered as one possible channel of the single-degenerate scenario to the formation of Type Ia supernovae \citep{Branch95, Liu19, Liu23}. Theoretical predictions for the evolution of red giant + white dwarf binaries \citep{Liu19, Ablimit22} predict possible Type Ia supernova in systems having a secondary mass greater than 0.9 M$_{\odot}$ while having orbital periods shorter than 100 days, as it is the case for several ellipsoidal candidates of the present study.

The masses for the primary and secondary components of the rotational binaries are restricted to $\lesssim$ 3 M$_{\odot}$ and $\lesssim$ 1 M$_{\odot}$, respectively, consistent with a primary star which is a sub-giant while the secondary is a main-sequence star. 

\subsection{Maximum mass estimates for the secondary star}\label{sec:max_mass}

In an attempt to constrain, as much as possible, the mass of the companion, an estimate of the maximum value for the inclination was derived. Following \cite{Mahy22}, we estimated the critical rotational velocity of the primary star (see their Eq. 3) and used the spectral line broadening estimate (\texttt{vbroad} from \textit{Gaia} DR3) as an upper limit for the v$\sin{i}$. This relation is only valid if the orbital and rotational axes are aligned, which is a reasonable assumption since both synchronization and alignment of binary systems are reached in a shorter timescale than circularization \citep[see e.g.,][]{Hut81, Glebocki97, Mazeh08, Daher22}. Based on this, a minimum value for $\sin{i}$ was derived. This implies a minimum value for the mass function and, therefore, an upper limit for the mass of the companion. 

\begin{figure*}
   \centering
   \includegraphics[width=0.45\textwidth]{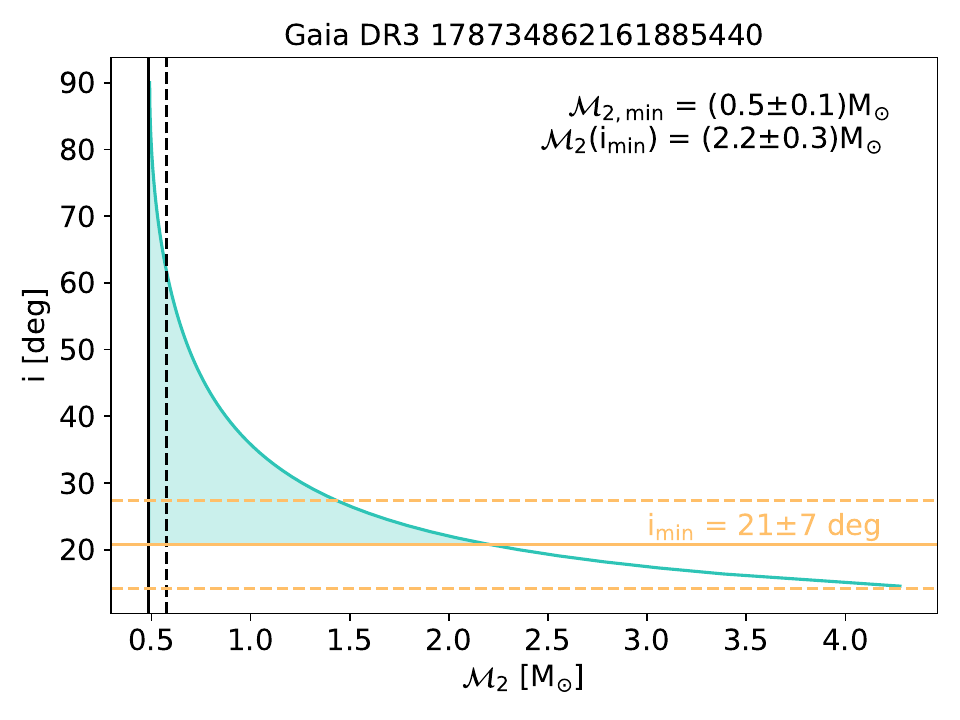}
    \includegraphics[width=0.455\textwidth]{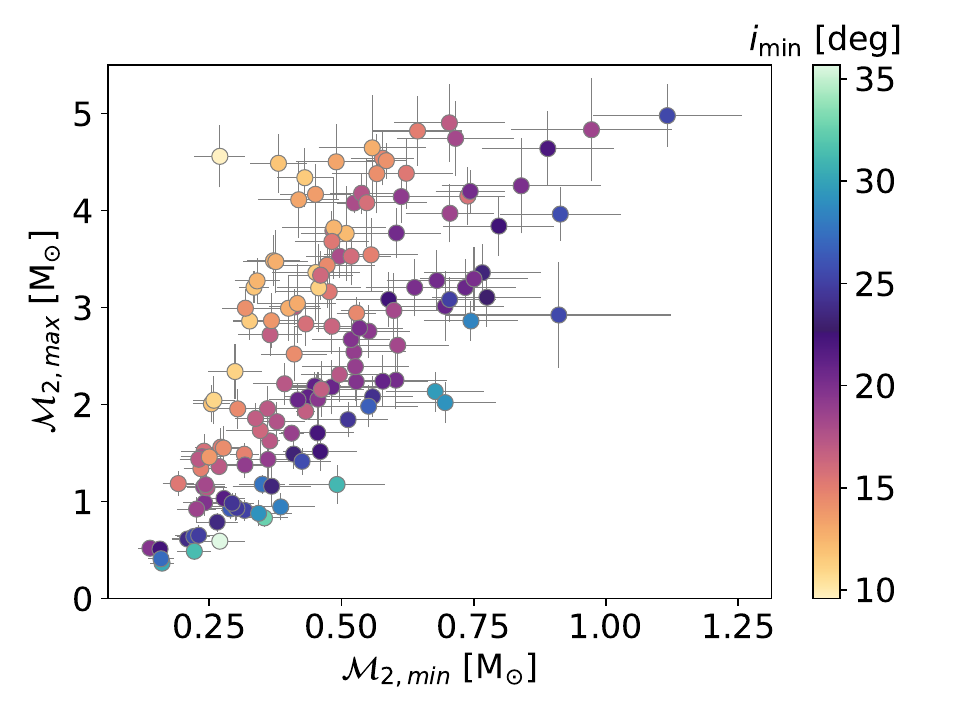}
 \caption{Dependence of the mass of the secondary component as a function of the inclination of the binary. Left panel: secondary mass versus inclination for \textit{Gaia} DR3 178734862161885440 ellipsoidal candidate. The black vertical line marks the minimum mass computed, for $\sin{i}$ = 1, while the dashed vertical line is its associated error. The horizontal orange line is the minimum inclination of the system, while the dashed orange lines are the inclination error. In the top right corner, the minimum mass and the mass obtained from the minimum inclination are reported. The area under the curve corresponds to the range of masses possible for this system. Right panel: minimum secondary mass versus the maximum secondary mass, color-coded by the minimum inclination of the system.} \label{fig:m2_inc}
\end{figure*}

The left panel in Figure~\ref{fig:m2_inc} shows the dependence of the secondary mass on the inclination values for one ellipsoidal candidate, \textit{Gaia} DR3 178734862161885440. The minimum mass corresponds to the value obtained for $\sin{i}$ = 1, while the maximum mass corresponds to the minimum inclination, based on the measured broadening velocity. Both values are reported in the top right corner of the panel. The uncertainty in the minimum inclination (horizontal dashed orange lines), which is obtained through the error propagation of the primary mass and radius, is significant and allows for the mass of the secondary to range from $\sim$1.5 M$_{\odot}$ up to $\sim$4.3 M$_{\odot}$. The inclinations recovered are a lower limit and certainly represent much lower values than inclination values derived for ellipsoidal variables in the literature \citep{Nie17, Wrona22}. Therefore, the obtained maximum mass for the companion could be largely overestimated when the minimum inclination is too low ($i_{\rm min} \leq 30$ deg).

The subsample of stars for which a maximum for the secondary mass can be estimated is shown in the right panel of Figure~\ref{fig:m2_inc}. The minimum and maximum mass for the companion are shown, color-coded based on the minimum value for the inclination. Depending on the minimum inclination, the range of values for the mass of the companion is wide and it is not possible to constrain it. There is a small subsample of systems that have a minimum and maximum mass for the secondary component relatively constrained (i.e., $\mathcal{M}_2$(i$_{\rm min}$) $-$ $\mathcal{M}_{2, \rm min}$ $<$ 1.0 M$_{\odot}$), most of them having a minimum value for the inclination of $i \gtrsim$ 30 deg. 

\begin{figure}
   \centering
\includegraphics[width=0.5\textwidth]{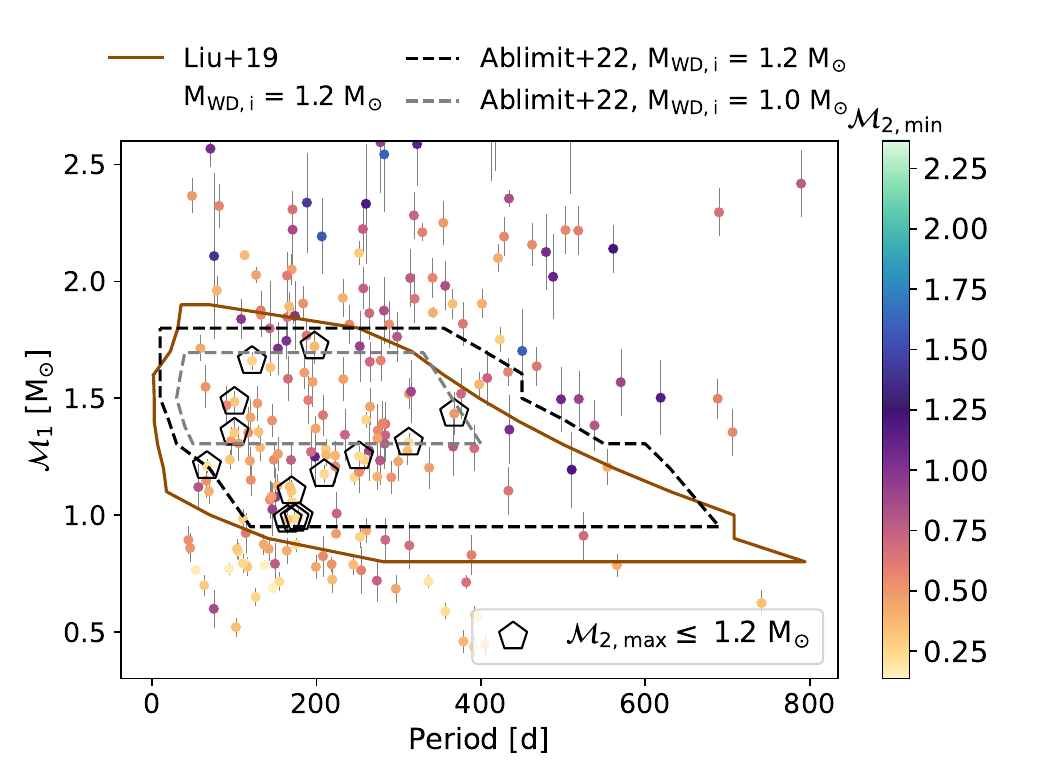} 
 \caption{Orbital period versus the spectroscopic mass of the primary $\mathcal{M}_{\rm 1}$, colored based on the minimum mass of the secondary star $\mathcal{M}_{\rm 2, min}$. The boundaries for potential Type Ia SNe progenitors in red giant stars and white dwarf binaries from \cite{Liu19} and \cite{Ablimit22} are shown as solid and dashed lines. Those ellipsoidal binaries having an $\mathcal{M}_{\rm 2, max} \leq$ 1.2 M$_{\odot}$ are marked with black pentagons.} \label{fig:sne}
\end{figure}

For ellipsoidal binaries in which the maximum mass of the secondary star is $\leq$ 1.2 M$_{\odot}$, some models for the evolution of binaries place them as potential Type Ia SNe progenitors. Figure~\ref{fig:sne} shows the orbital period against the mass of the primary, while the minimum mass for the secondary is shown in the color bar. The parameter space inside which a binary system having a red giant as primary star and white dwarf as a secondary, for different initial masses of the white dwarf is shown as brown solid line \citep[from][]{Liu19}
and black and grey dashed lines \citep[from][]{Ablimit22}. There are 13 ellipsoidal binaries inside the boundaries from the literature and having $\mathcal{M}_{\rm 2, max}$ up to 1.2 M$_{\odot}$. Despite the range of primary masses and periods changes depending on the model, the ellipsoidal binary sample studied in this work may contain promising binary systems to follow-up as potential Type Ia SNe progenitors, particularly those with orbital periods of less than 400 days and $\mathcal{M}_{\rm 1} \leq$ 2.0 M$_{\odot}$.

\subsection{Mass ratio and filling factors}\label{sec:q_f}

\begin{figure}
   \centering 
\includegraphics[width=0.5\textwidth]{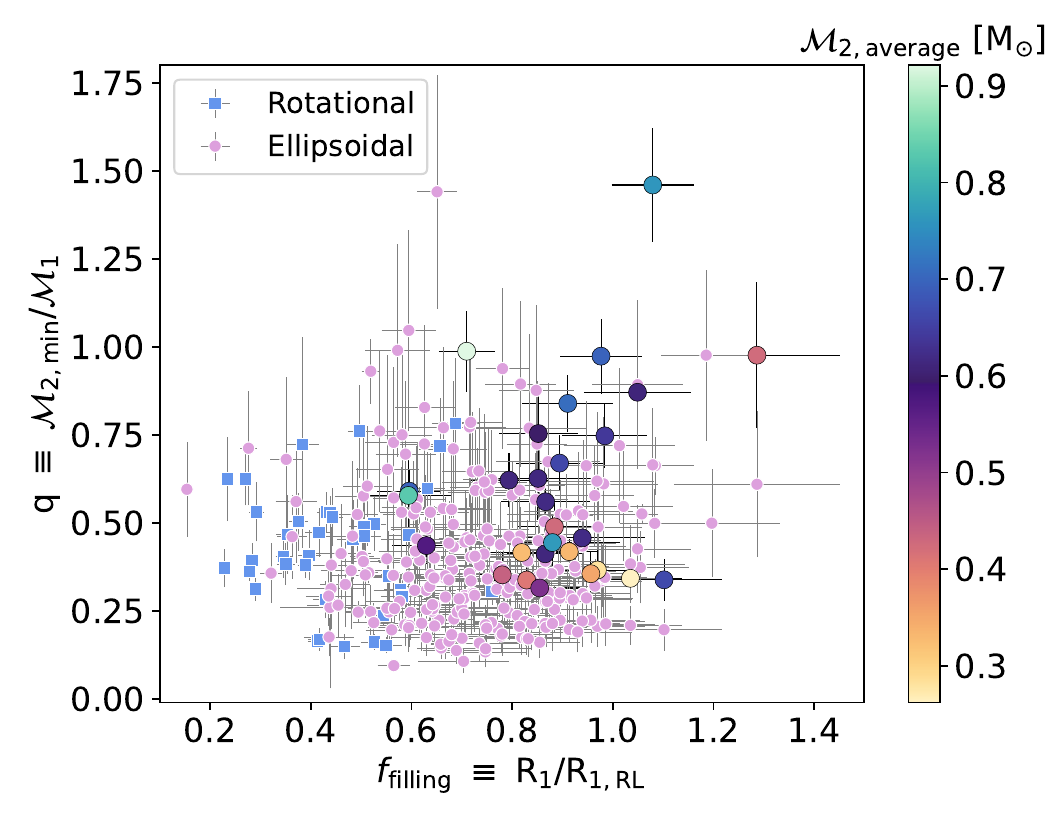}
 \caption{Filling factor versus mass-ratio for rotational candidates (blue squares) and ellipsoidal variable candidates (pale pink circles). For those systems in which the difference between the maximum and minimum mass for the secondary is lower than 1.0 M$_\odot$, the data points are color-coded by the average between the maximum and minimum secondary mass and their mass ratio corresponds to that average mass for the secondary.} \label{fig:filling}
\end{figure}

The ``minimum'' mass ratio $q \equiv \mathcal{M}_{\rm 2, min}$/$\mathcal{M}_1$ as a function of the filling factor, f$_{\rm filling}$ (the ratio between the radius of the primary star and its Roche-lobe radius) is shown in Figure~\ref{fig:filling}, for rotational (blue squares) and ellipsoidal candidates (pink circles). Ellipsoidal binaries having a difference in the maximum and minimum secondary mass being $\leq$ 1 M$_{\odot}$ are shown colored based on the average secondary mass, and their mass fraction was estimated based on this $\mathcal{M}_{\rm 2, average}$. Uncertainties were estimated based on error propagation, including the error on the radius, mass of the primary, minimum mass of the secondary and orbital period. 

The Roche-lobe radius $R_{1, \rm RL}$ was derived based on Eq. 2 from \cite{Eggleton83}, as
\begin{equation}
\frac{R_{\rm 1, RL}}{a} = \frac{0.49 q^{-2/3}}{0.6 q^{-2/3} + \ln{\left(1+q^{-1/3}\right)}}\text{,} \label{eq:ffill}
\end{equation}
being $a$ the semi-major axis of the system. This approximation is valid for circular orbits. In eccentric binaries, the Roche-lobe radius depends on the instantaneous separation between the two stars, being minimum at periastron and maximum at apastron. For the present analysis, we will consider a circular orbit for all the stars, which implies that, for eccentric orbits, the Roche-lobe radius and filling factor estimates will be an average value.

The projected semi-major axis, $a\sin{i}$, can be obtained from the projected semi-major axis of the primary $a_{1}\sin{i}$ as
\begin{align}
a_1\sin{i} &= \frac{K_1 P \sqrt{1-e^2}}{2\pi} \nonumber \\
a\sin{i} &= \left(\frac{\mathcal{M}_1+\mathcal{M}_2}{\mathcal{M}_2}\right) a_{1}\sin{i}
\end{align}

Adopting the maximum value for the $\sin{i} \equiv 1$, an estimate of the maximum semi-major axis, under the assumption of circular orbit, can be obtained and the corresponding maximum Roche-lobe radius R$_{\rm 1, RL}$ and a minimum value for the filling factor f$_{\rm filling} \equiv$ R$_{\rm 1}$/R$_{\rm 1, RL}$. In eccentric orbits, the minimum orbital separation would be $r_{\rm peri} = a(1-e)$, when the Roche-lobe radius will shrink and mass transfer is more likely to occur. In this case, the maximum value for the filling factor would be larger than the value we obtained assuming circular orbits and $\sin{i} \equiv 1$. Adopting a zero eccentricity for eccentric orbits is similar to adopting a median orbital separation $r$ = $(r_{\rm peri} + r_{\rm apo})/2 \sim a$ to estimate the filling factor. 

Figure~\ref{fig:filling} shows that rotational binaries show a cut-off in the mass ratios and filling factors of around $\sim$0.75, while ellipsoidal variables reach higher values, as expected since the former does not show signs of ellipsoidal modulation due to the deformation of the primary star. They all have filling factors smaller than unity, meaning the systems are not in contact. On the contrary, the ellipsoidal candidates span a much larger range of mass ratios and filling factors, even reaching the region in which the secondary star has a larger mass than the primary ($q > 1$, three binaries) or those where the primary star has reached the Roche-lobe radius and therefore the systems could be in contact (f$_{\rm filling} >$ 1). Ellipsoidal binaries having $\mathcal{M}_{\rm 2, min} \geq$ $\mathcal{M}_1$ could be due to secondary stars being either massive main sequence stars or compact objects \citep{Price-Whelan20}, which can be massive but not contributing to the flux in the spectrum (single-line spectroscopic binaries). Another possible scenario would be that these systems are in the process of mass transfer, as is the case of semi-detached binaries. Two of these three systems have filling factors of the order of 0.6, meaning that they are most likely not in contact, while there is one ellipsoidal binary with larger than unity filling factor and mass ratio (Gaia DR3 1107618979843202944). 20 ellipsoidal candidates have f$_{\rm filling} \geq 1$, i.e., the radius of the primary being larger than the Roche-lobe radius and therefore, could be non-detached binaries. A small subsample of potential contact binaries was recovered as well in the analysis of single-line spectroscopic binaries in the APOGEE survey \citep{Price-Whelan20}. Unfortunately, without the orbital inclination, these values are only upper limits; therefore, it could be the case that these systems are detached ellipsoidal variables, as the rest of the sample. 

\begin{figure}
   \centering 
\includegraphics[width=0.5\textwidth]{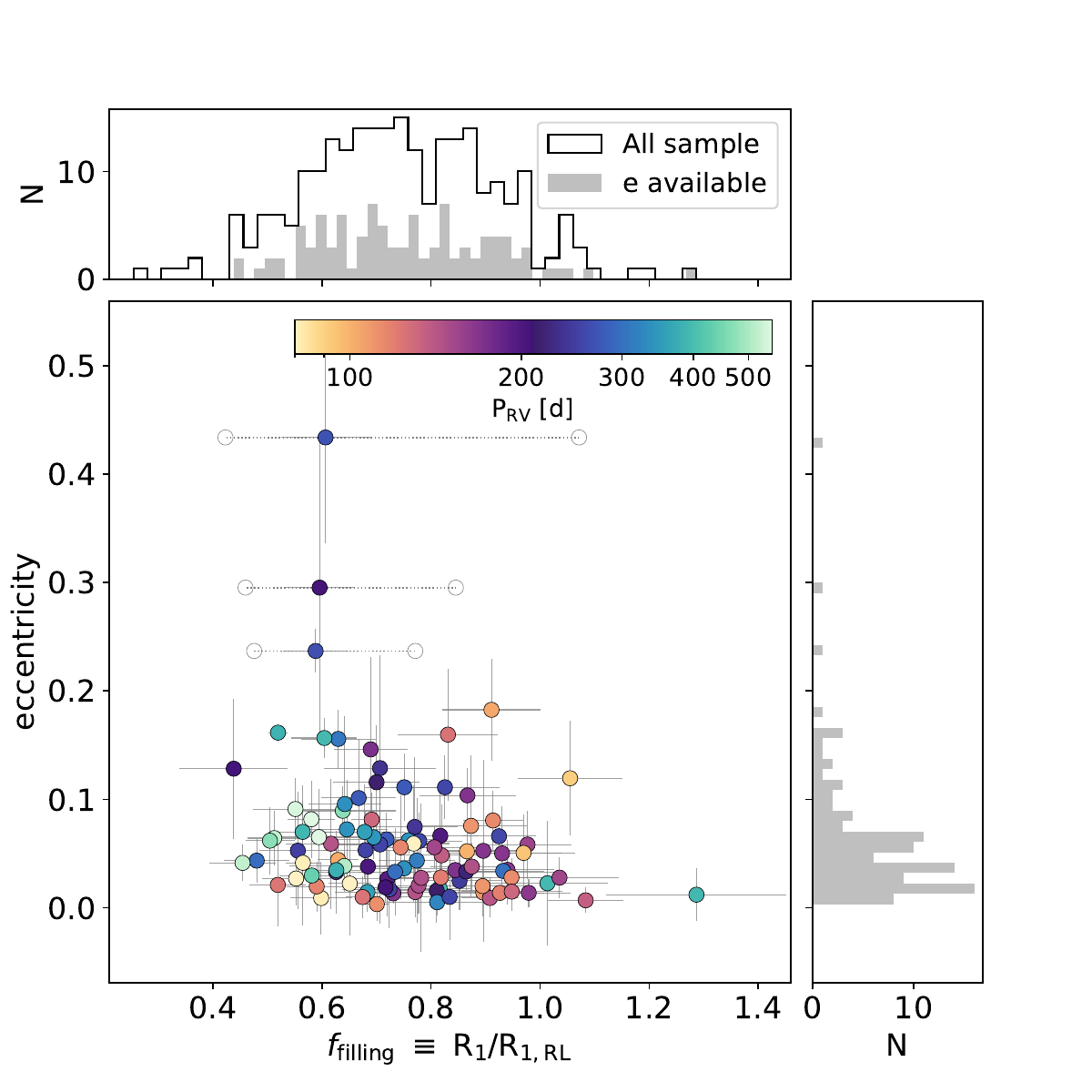} 
 \caption{Filling-factor f$_{\rm filling}$ versus eccentricity for a sub-sample of ellipsoidal variables. The data points are color-coded based on the orbital period. For the most eccentric binaries, the filling factor at apo- and periastron is included as unfilled circles, connected to the average value obtained considered a zero eccentricity. The top and right panel shows the distribution (grey histogram) of filling factors and eccentricities, respectively. The top panel includes the distribution of filling factors for all the ellipsoidals having reliable measurements for their radii (black-edge histogram).} \label{fig:filling_ecc}
\end{figure}

The filling factor should be correlated with the tidal circularization time of the orbit, which goes as (R$_1$/a)$^{-8}$. The orbital separation $a \sin{i}$ can be expressed as a function of the Roche-lobe radius R$_{\rm 1, RL}$ (see Eq.~\ref{eq:ffill}). This implies that systems having larger filling factors are closer to reach the eccentricity zero, while non-circular binaries will tend to have smaller filling factors. In the case of eccentric orbits, the instantaneous filling factor will be larger at periastron and smaller at periastron (see unfilled circles in Fig.~\ref{fig:filling_ecc}) and therefore, adopting an eccentricity of zero in those binaries to derive the filling factor gives a more representative estimate of the average value along the orbit. The orbital period of the system (proportional to the radius of the primary stars) also plays a role in the circularization time. We combined this information in Figure~\ref{fig:filling_ecc} where the eccentricity is shown as a function of the filling factor, and orbital period. The distributions of filling factors (for the full sample for which we have radii and masses) and only those that have eccentricity estimates available are shown as white and grey histograms, respectively, in the top panel. The eccentricity distribution is shown in the right horizontal panel. The dependence on the filling factor to have circular orbits is clearly seen, as the distribution of eccentricities tends to decrease as the filling factor increases. The period dependence is seen as those systems having shorter periods (smaller orbital separations) are those with larger filling factors. 

Table~\ref{tab:binary_ratios} lists the minimum mass for the secondary component, obtained by solving numerically the mass function (Eq.~\ref{eq_fm}, adopting $\sin{i} = 1$) and its standard deviation obtained through Monte Carlo simulations, the maximum value for the mass of the secondary companion, obtained using the broadening velocity (see Sect.~\ref{sec:max_mass}) and its propagated uncertainty, the minimum mass ratio $q$ and the filling factor (adopting a zero eccentricity) and their associated uncertainties. Binary systems for which there is no broadening velocity published in Gaia DR3 or when the estimate is non-physical result (negative masses, relative errors larger than one) do not have a maximum mass estimate in the Table.
 
\section{Chemo-dynamical population properties} \label{sec:chemo-dynamical}

\subsection{Metallicity distribution function}\label{sec:mdf}

\begin{figure}
   \centering
    \includegraphics[width=0.5\textwidth]{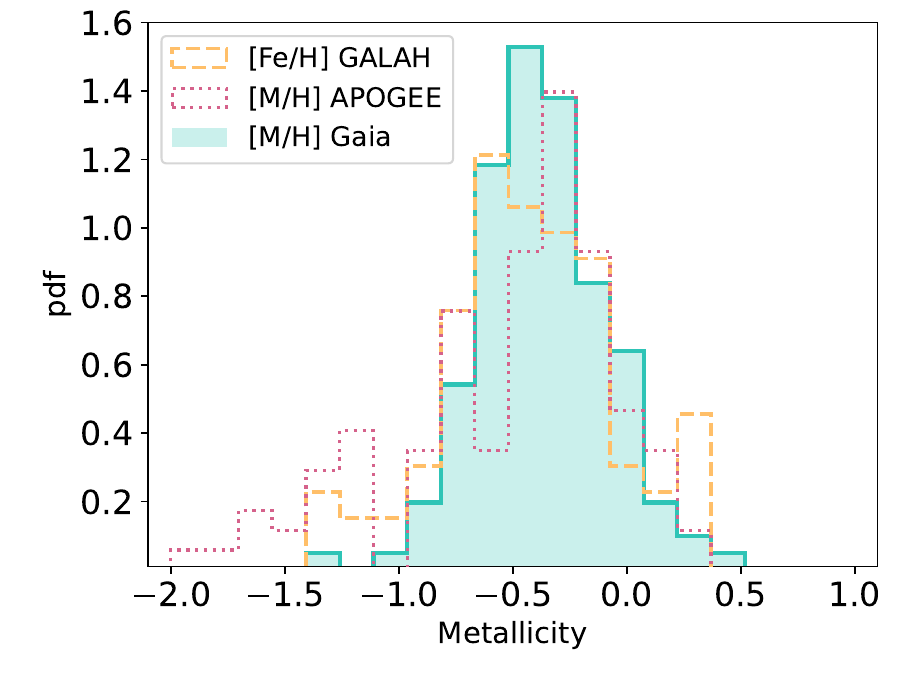} 
 \caption{Metallicity distribution function of ellipsoidal candidates having \texttt{vbroadM} $<$2, \texttt{vradM} $<$2, and $\Delta$[M/H]$\leq$0.1 dex from \textit{Gaia} RVS spectra (aquamarine filled histogram), SB1 stars from the GALAH survey (dashed orange line) and APOGEE close giant binaries (purple dotted line).} \label{fig:mdf}
\end{figure}

The distribution of spectroscopic [M/H] from \textit{Gaia} \textit{GSP-Spec} is shown in Figure~\ref{fig:mdf} for ellipsoidal candidates (aquamarine filled histogram). Only observations having the flag \texttt{vbroadM}$<$2 and \texttt{vradM}$<$2 have been included, meaning that the potential biases due to uncertainties in the rotational and RV shifts account for $\Delta$[M/H]$<$ 0.5 dex \citep[see Tables C.1 and C.2 in][]{GaiaRVS}. The error in metallicity, derived as half the difference between the upper and lower values reported by the \textit{GSP-Spec} module was also restricted to be less than 0.1 dex. The distribution is mostly metal-poor, having a peak at $\sim -$0.35 dex, and a dispersion of 0.4 dex. \textit{Gaia} \textit{GSP-Spec} metallicities, without a restriction in \texttt{KMgiantPar}, were preferred against XGBoost metallicities as the latter is less accurate at [M/H]$<-1$ and cool giants, due to the limitations of the training sample \citep{xgboost}. For a comparison between the metallicities from XGBoost and \textit{GSP-Spec} see Appendix~\ref{app:mh}. 

Figure~\ref{fig:mdf} includes the metallicity distribution of the giant-dwarf and giant-giant SB1 binaries (dashed orange line) from the Galactic Archaeology with HERMES (GALAH) survey \citep{GALAH_Traven} and the close giant binaries from APOGEE \citep[purple dotted line, from][]{Price-Whelan20}. Only those binaries having the same range of effective temperature and surface gravities as our sample of ellipsoidal variables were selected. The metallicity distribution of ellipsoidal candidates based on \textit{Gaia} measurements is similar to the one derived based on GALAH and APOGEE observations, having a broad, metal-poor distribution, closely resembling the MDF measured by APOGEE although without having ellipsoidal binaries as metal-poor as [M/H]$\leq$ -1.5 dex. It is worth mentioning that different surveys have different limiting magnitudes and, therefore, probe up to different Galactocentric radii implying possible metallicity biases. Moreover, some metal-poor stars may not be recovered as such by \textit{Gaia} \textit{GSP-Spec} \citep[see][]{Gaia_cartography, Matsuno24}. Thus, the discrepancies found might not be an intrinsic difference among the binary populations, and, overall, the \textit{Gaia} \textit{GSP-Spec} metallicities for the ellipsoidal samples seem to be in fair agreement with independent MDF for single-line spectroscopic binaries having a giant as a primary star.

\subsection{[$\alpha$/Fe] - [M/H], Galactic velocities}\label{subsec:chemodynamics}

\begin{figure}
   \centering
    \includegraphics[width=0.5\textwidth]{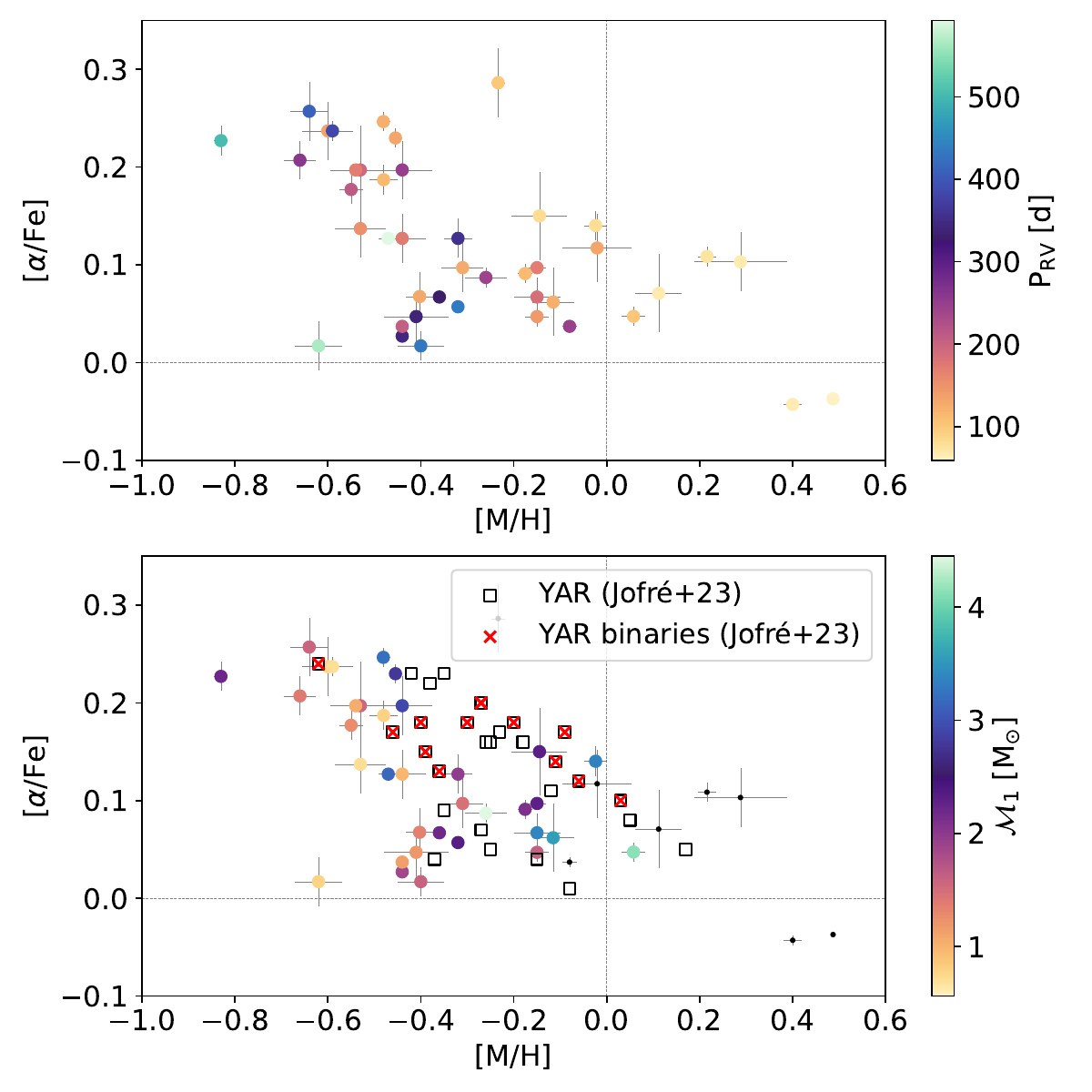} 
 \caption{[$\alpha$/Fe] versus [M/H] (from \textit{GSP-Spec}) for ellipsoidal candidates. Top panel: The points are color-coded according to their orbital period. Solar values are marked with dotted grey lines. Bottom panel: Same sample but the color bar corresponds to the spectroscopic mass of the primary. Young $\alpha$-rich stars (YAR) from \cite{Jofre23} based on APOGEE observations are included as squares, those with a red cross mark are spectroscopic binaries.}
 \label{fig:alpha_mh_pvr}
\end{figure}

Having the astrophysical parameters derived as part of the \textit{Gaia} \textit{GSP-Spec} module, the properties of ellipsoidal and rotational binaries can be analyzed in the context of Galactic stellar populations. Figure~\ref{fig:alpha_mh_pvr} shows the [$\alpha$/Fe] versus [M/H] plane for ellipsoidal variables, color-coded according to the orbital period P$_{V_R}$ (top panel) and mass of the primary (bottom panel). The same restriction in metallicity error and vbroadM and vradM quality flags as in the previous Section were applied. The sample is even reduced compared to previous plots as only the stars with \texttt{KMgiantPar = 0} are included, since for stars with \texttt{KMgiantPar = 1} or \texttt{2}, there is no [$\alpha$/Fe] estimate available in \textit{Gaia} DR3. 

The distribution of periods and metallicities in the top panel of Figure~\ref{fig:alpha_mh_pvr} shows that most of the shortest period ellipsoidal binaries are metal-rich, non-$\alpha$ enhanced while those with the largest periods are mostly metal-poor ([M/H]$\sim -$0.5 dex) and $\alpha$-enhanced with respect to the solar values. Such an $\alpha$-enrichment level, at a given metallicity, is consistent with the disk population, although, at the low metallicity tail, some stars could be part of the halo population. To the best of our knowledge, this is the first time that the [$\alpha$/Fe]-[M/H] diagram is presented for ellipsoidal binary stars.

It is also worth mentioning that some of the $\alpha$-enhanced stars have masses larger than one solar mass, resembling the population of young $\alpha$-rich (YAR) stars in the thick disk. These stars were originally considered young systems due to their intermediate mass having enhanced [$\alpha$/Fe], consistent with relatively old ages. Most of these systems have been found to be part of binary systems \citep[][]{Jofre23} and their chemical enrichment is potentially due to mass transfer. The bottom panel in Figure~\ref{fig:alpha_mh_pvr} shows the spectroscopic mass estimate for the primary and the sample of YAR stars in \cite{Jofre23}, based on APOGEE measurements for the abundances, including those spectroscopically confirmed to be binaries. Our population of ellipsoidal binaries, particularly those with the shortest periods, have [$\alpha$/Fe] abundances consistent with YAR stars, which may imply that some of these systems have already experienced mass transfer events which enriched the chemistry of the current primary star.

\begin{figure}
   \centering
    \includegraphics[width=0.5\textwidth]{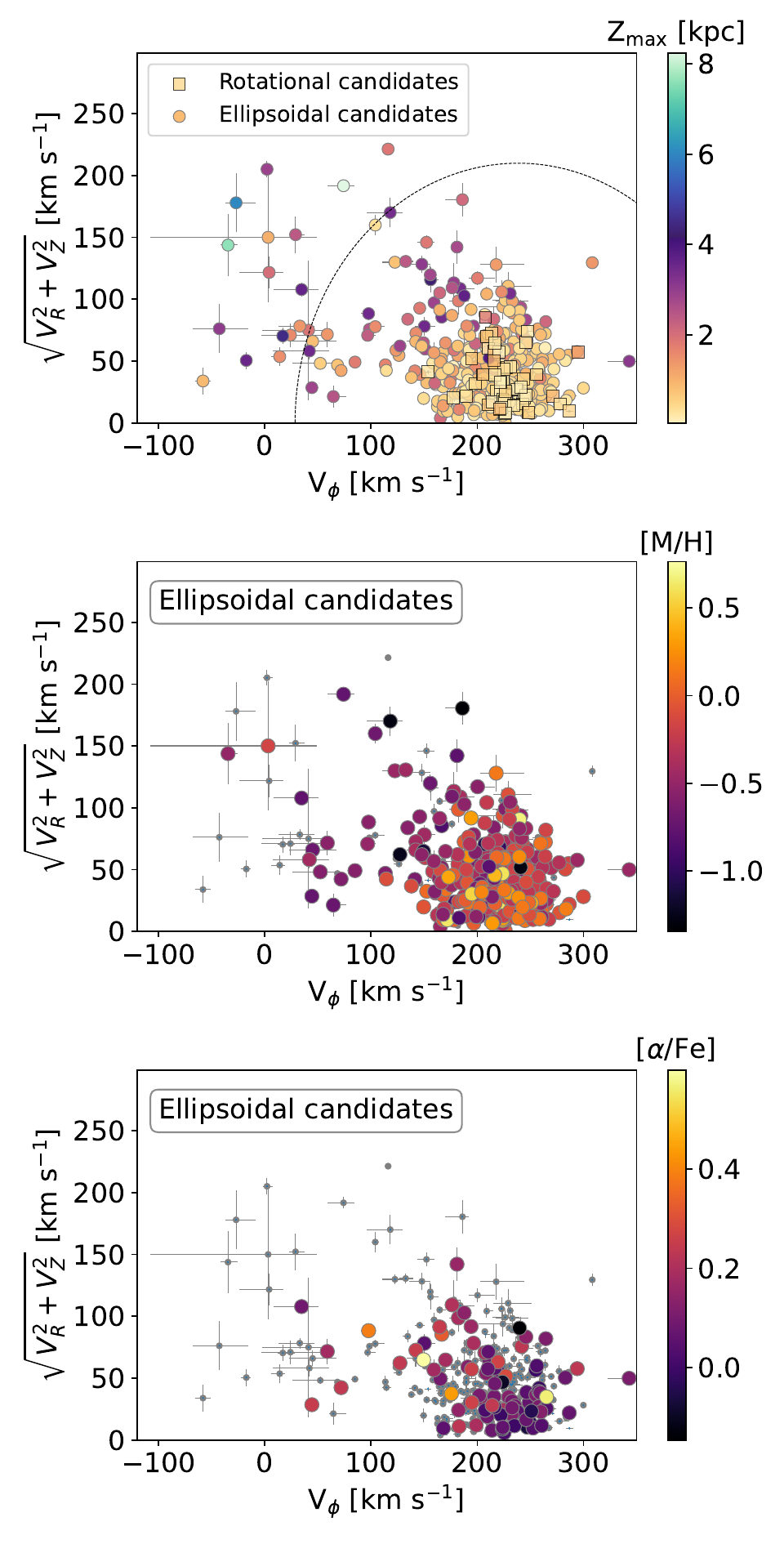} 
 \caption{Toomre diagram for ellipsoidal and rotational binary candidates, color-coded according to the maximum height from the Galactic plane Z$_{\rm max}$ (top panel), ellipsoidal candidates only, color-coded based on their spectroscopic metallicity [M/H] (middle panel) and [$\alpha$/Fe]-abundance (bottom panel). The dashed circle marks the kinematics of the Galactic disk at $\sqrt{V^2_R+V^2_Z+(V_{\phi}-V_0)^2} < 210$ km s$^{-1}$.} \label{fig:chemokine}
\end{figure}

Combining the chemical information with the kinematics, Figure~\ref{fig:chemokine} shows the Galactic spherical velocities, metallicities and $\alpha$-element abundances for ellipsoidal variable candidates. The Galactic velocities and Zmax of the orbit, as well as their low and upper limits, were retrieved from the \texttt{gaiadr3.chemical\_cartography} table, part of \textit{Gaia} DR3 \citep[see the description of the catalogue in][]{Gaia_cartography}\footnote{The method assumed a solar position and velocities of R = 8.249 kpc, Z = 20.8 pc, (V$_{R}$, V$_{\phi}$, V$_{Z}$) = (-9.5, 250.7, 8.56) km s$^{-1}$ \citep{Gravity}.}. The top panel shows the azimuthal velocity V$_{\phi}$ versus the tangential velocity component $V_{\bot} = \sqrt{V^2_{R}+V^2_{Z}}$, color-coded by the maximum $Z$ above the plane, for ellipsoidal (circles) and rotational binaries (squares), without applying any cut in the quality of the atmospheric parameters. The middle and bottom panels present only ellipsoidal candidates having a metallicity error $<$0.15 dex from the \textit{GSP-Spec} module, color-coded based on their [M/H] (middle panel) and [$\alpha$/Fe] (bottom panel) calibrated abundances. The dashed circular line corresponds to the boundary inside which resides the disk population, i.e., $\sqrt{V^2_R+V^2_Z+(V_{\phi}-V_0)^2} < 210$ km s$^{-1}$, being V$_0$ = 238.5 km s$^{-1}$ as in \cite{Gaia_cartography}. From the kinematics, it is clear that all the rotational binaries are part of the disk population, having all prograde circular V$_{\phi}$ velocities, centered around $\sim$220 km s$^{-1}$. In the literature, a sample of eight rotational RS CVn systems having chemo-dynamical information has also been found to be all disk population by \cite{Morel04} while there is only one known RS CVn system in the halo \citep{Bubar11}. The effect of the induced activity can severely affect the metallicities and [$\alpha$/Fe] abundances of rotational, close-binaries and therefore we do not aim to see any trends in chemo-dynamical space for these stars. 

The chemo-dynamical properties of ellipsoidal variables are much more varied: the kinematics are consistent with thin/thick disk as well as halo populations. As expected, the stars having halo-like kinematics are the most metal-poor of the sample, having also relatively high [$\alpha$/Fe] values. Nonetheless, some systems with kinematics of the disk have the highest $\alpha$-enhancement. These are the stars with the shortest periods (see Fig.~\ref{fig:alpha_mh_pvr}) which may indicate that the interaction with the companion star is affecting the measurements of chemical abundances in the primary, as it happens in the case of rotational binaries. 

\begin{figure}
\centering
\includegraphics[width=0.5\textwidth]{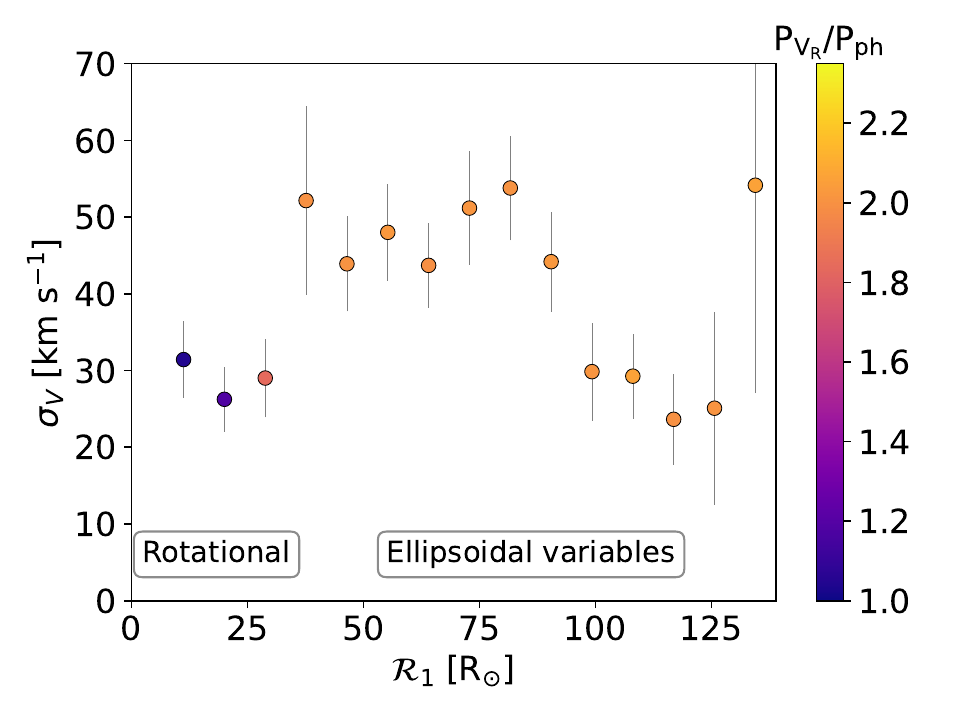}
 \caption{Velocity dispersion as a function of the primary radius, in equal bins of width $\sim$ 9.0 R$_{\odot}$. The color-bar corresponds to the ratio between the radial velocity and photometric period.} \label{fig:sigmaV}
\end{figure}

Given that the primary star in the case of rotational variables is less evolved than the one in the ellipsoidal systems (see their location in the Kiel and HR diagram in Fig.~\ref{CMD_period} and ~\ref{HR_Radius_Masses}, respectively) which, combined with their kinematics and metallicities (mostly consistent with a disk population, see Fig.~\ref{fig:chemokine}) may indicate a different stellar population also in terms of age. To test this, the velocity dispersion of the sample could be used as an indirect proxy for their relative ages. Figure~\ref{fig:sigmaV} shows the velocity dispersion, based on the quadratic sum of the velocity dispersion in $R$, $Z$ and $\phi$ directions in bins of primary radius, $\mathcal{R}_1$, which is the physical parameter with the lowest associated error, compared to, for example, the mass of the primary. There is a significant difference in the velocity dispersion of rotational candidates (having P$_{\rm V_R}$/P$_{\rm phot} \sim$ 1.0 and radii smaller than 30 R$_{\odot}$), being no larger than 30 km s$^{-1}$. Ellipsoidal variables tend to have an almost flat, much larger velocity dispersion, of the order of 50 km s$^{-1}$, for radii between 40 to 80 R$_{\odot}$. This marked difference in velocity dispersion may indicate that ellipsoidal variables, at least in the current sample, represent an older population than rotational candidates.

Summarizing, both samples appear to be intrinsically different in terms of radii, Galactic velocities, metallicities, and not only in orbital separations/orbital periods. Even though some outliers could be present in both samples, mostly in the case of poorly sampled light curves, their main stellar properties are markedly different. Nonetheless, we cannot exclude that they are two evolutionary stages of the same phenomenon: a binary in a close orbit in which, depending on the mass and evolution of the primary, can get tidally circularized rapidly, getting a synchronized rotation (induced by the tides) and orbital motion, leading to an enhanced chromospheric activity; or if massive enough and having a larger orbital separation, the system can evolve up to point in which the primary star increases its radius, being able to substantially fill its Roche lobe and get tidally deformed.

\subsection{Chemical enrichment: Cerium abundance}

From \textit{Gaia} RVS spectra, chemical abundances for up to 12 elements were derived\footnote{Depending on the spectral type of the source and its signal-to-noise ratio, not all the abundances are available.} in \textit{Gaia} DR3, including, for example, $\alpha$-elements ([Mg/Fe], [Si/Fe], [S/Fe], [Ca/Fe], [Ti/Fe]), iron-peak elements ([Cr/Fe], [Ni/Fe]), and heavy-elements \citep[{[Zr/Fe]}, {[Ce/Fe]}, {[Nd/Fe]}; see][for a detailed analysis of the abundance distribution of these elements in the Milky Way]{Gaia_cartography}. Among them, the abundance of Cerium and Neodymium, $s$-process elements of the second peak, are of great importance during the stellar evolution of single AGB stars as their abundances change depending on the stellar mass, metallicity, mass-loss and the number of thermal pulses and thermal dredge-up episodes \citep{Contursi23, Contursi24}. In the context of binaries, red giant branch or early-AGB stars in binary systems having experienced mass-transfer, can also present an overabundance of $s$-process elements, as is the case of S-stars \citep{Shetye18} and Ba-stars \citep{Baornot, Escorza17, Jorissen2019}. Therefore, the analysis of the Ce abundances, as representative of $s$-process elements, can shed light on the secondary component; in the case of binaries that went through mass transfer, the former primary (currently the secondary, probably a stellar remnant) was the donor which transferred the mass to the secondary (currently the primary, more luminous component); or could indicate that the current primary star has evolved without any influence from the companion, following the expected evolution of a single AGB star. 

\begin{figure*}
 \centering
 \includegraphics[width=\textwidth]{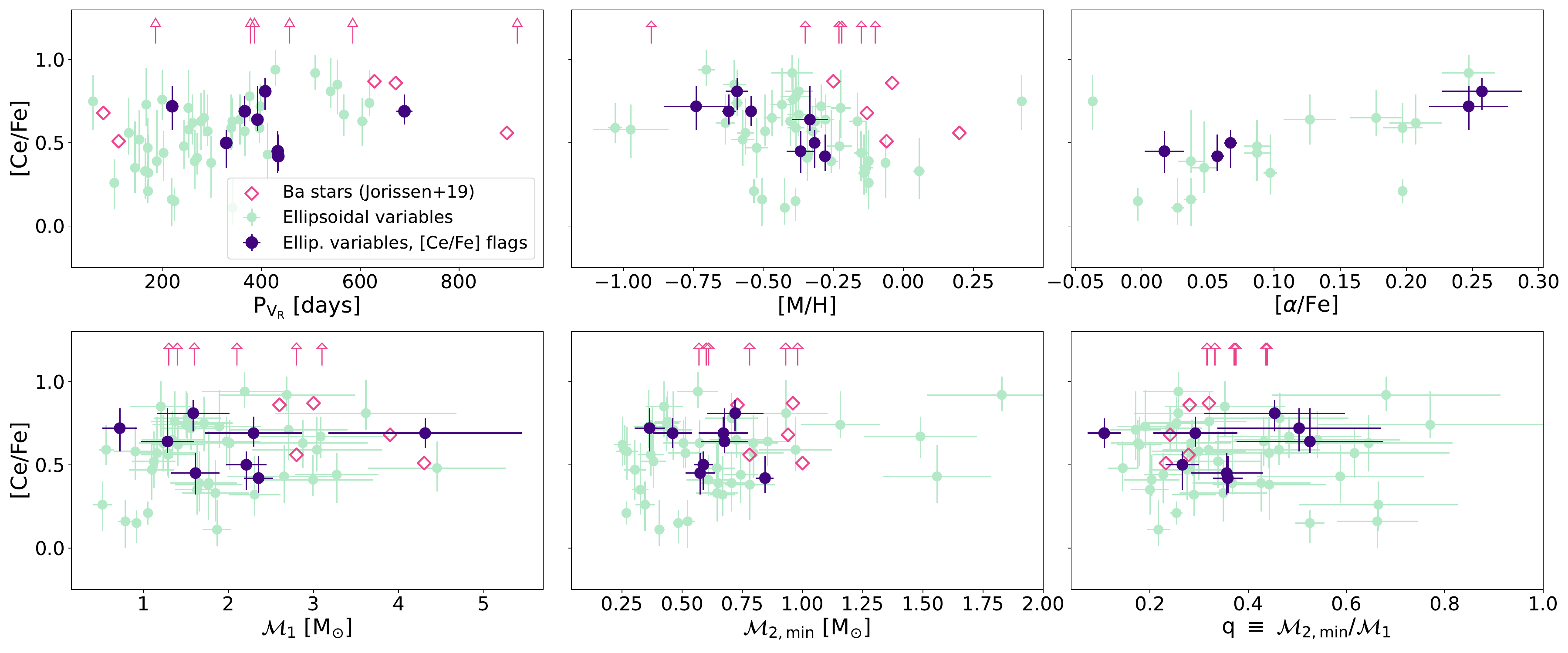}
 \caption{[Ce/Fe] abundances as a function of different orbital and stellar properties for ellipsoidal candidates with available measurements. Top row shows [Ce/Fe] against the orbital period (P$_{V_R}$ from \textit{Gaia} FPDR, left panel), \textit{GSP-Spec} calibrated [M/H] and [$\alpha$/Fe] (middle and right panel, respectively). The bottom row shows the abundance of Cerium as a function of the mass of the primary star $\mathcal{M}_1$ (left panel), the minimum mass of the secondary $\mathcal{M}_{2, {\rm min}}$ (middle panel) and the mass ratio $q$ (right panel). In all the panels, only ellipsoidal binaries with good quality flags for the extinction are included (mint circles). Stars following a cut in the [Ce/Fe] and [M/H] quality flags (see the main text) are shown as purple circles. For comparison, Ba stars from \cite{Jorissen2019} having orbital periods of $\leq$1000 days are included as magenta diamonds and arrows (when their [Ce/Fe] abundance is larger than 1.2 dex).} \label{fig:m2_ce}
 \end{figure*}

Figure~\ref{fig:m2_ce} shows [Ce/Fe] abundances for the ellipsoidal candidates as a function of different parameters associated to the orbit of the binary (orbital period), chemical abundances of the primary star (metallicity [M/H] and $\alpha$-element abundances [$\alpha$/Fe]) and the masses of the system (mass of the primary $\mathcal{M}_1$, minimum mass of the secondary $\mathcal{M}_{2, {\rm min}}$, and the mass ratio $q \equiv \mathcal{M}_{2, {\rm min}}/\mathcal{M}_1$). Stars having a \textit{GSP-Spec} measurement of [Ce/Fe] abundances without any extra quality cuts are shown as mint circles, in all panels, while those that additionally have good quality flags associated with the determination of the [Ce/Fe] abundances \citep[\texttt{CeUpLim}$<$3, \texttt{flagCeUncer}$<$2; see Table 2 in][]{GaiaRVS}, \texttt{vbroad} $\leq$ 13 km s$^{-1}$, \texttt{extrapol} $\leq$1 and [Ce/Fe] abundance uncertainty $\leq$ 0.2 dex \citep[see][]{Contursi23, Contursi24} as well as those sources with good metallicity flags (\texttt{vbroadM, vradM} $<$2, see App.~\ref{app:mh}) are included as purple circles. Ba stars from \cite{Jorissen2019} are also included as empty magenta diamonds and arrows, for those having [Ce/Fe]$>$1.2 dex. Overall, the sample of ellipsoidal variables having Ce abundances covered similar ranges of period, primary mass and secondary mass as the sample of \cite{Jorissen2019}, although having lower [Ce/Fe] values. This could indicate that our sample is not representative of binaries having experienced $s$-process enhancement due to mass-transfer.

In terms of the metallicity content, our sample tends to be slightly more metal-poor compared to the measurements in \cite{Jorissen2019}. An anti-correlation in the Ce abundance as a function of the metallicity can be seen, being those more metal-rich the ones with the lower values of [Ce/Fe]. This is expected from the AGB evolution, as the production of s-process elements is less efficient in more metal-rich stars \citep{Jorissen2019, Cristallo15, Contursi24}. However, single AGB chemical evolution models predict different abundances of Ce based on the metallicity, and mass of the AGB star. In particular, the less massive ($\mathcal{M}_1 \sim$ 0.7 M$_{\odot}$) star in our sample has an elevated Ce abundance [Ce/Fe]$\sim$0.8, while models predict a much lower Ce overabundance, of the order of $\sim$0.2 dex, for any metallicity \citep[see Fig. 9 in][]{Contursi24}. This overabundance of Ce could be the result of a previous episode of mass-transfer, in which the secondary star was in the AGB phase and was the donor of material. Only a handful of sources with [Ce/Fe] measurements have also [$\alpha$/Fe] abundances determined from \textit{GSP-Spec}. Despite being limited in number, a tentative positive correlation seems to be present, as the more metal-poor stars are also the ones with higher $\alpha$-overabundance.

For single AGB star evolution, \cite{Contursi24} found discrepant results using two different AGB yield models. For solar-mass stars, having [Fe/H]$\sim$-0.7, the maximum Ce abundance based on the two models predicted are [Ce/Fe] = $\simeq$0.2 and $\simeq$0.7 dex. Overall, the abundance of [Ce/Fe] in the sample of ellipsoidal candidates is lower than the predictions for metal-poor, intermediate-mass AGB in the Monash models \citep[and references therein]{Karakas18} but higher than the predictions of the FRUITY \citep[][and references therein]{Cristallo15} models \citep[see Fig.~9 of][]{Contursi24}. For stars having primary masses of 2.0 to 4.0 M$_{\odot}$, the [Ce/Fe] abundances observed are within the values predicted by the models \citep[see Fig. 9 and 10 in][]{Contursi24}. 

\cite{Contursi23} found a potential offset of up to 0.2 dex between the mean Ce abundance derived by \textit{Gaia} \textit{GSP-Spec} and other spectroscopic surveys (see their Fig. 7), although such an offset is not seen compared to some surveys. Therefore, other chemical abundances, such as Y, Ba and Pb would be useful to determine if the systems have s-process abundances consistent with the evolution of an AGB star, irrespectively of its companion or if the systems have gone through mass-transfer events.

\section{Period-Luminosity relations} \label{sec:PL_relations}

Ellipsoidal variables are known to define a linear sequence in the period-luminosity (absolute magnitude) diagram, first detected in the LMC as the sequence E \citep{Soszynski04, Soszynski07}. In the Milky Way, this relation has not been explored beyond the Galactic bulge region \citep[see e.g.,][]{Wrona22}, as this requires the knowledge of a large sample of ellipsoidal variables with known distances. \textit{Gaia} DR3 provides photo-geometric distances for hundreds of ellipsoidal variables, allowing us to study this relation for the Galactic population. In this Section, we aim to probe the effects of different parameters that are available thanks to the \textit{Gaia} observations into the period-luminosity relation of ellipsoidal variables and its intrinsic dispersion.

\begin{figure}
 \centering
 \includegraphics[width=0.5\textwidth]{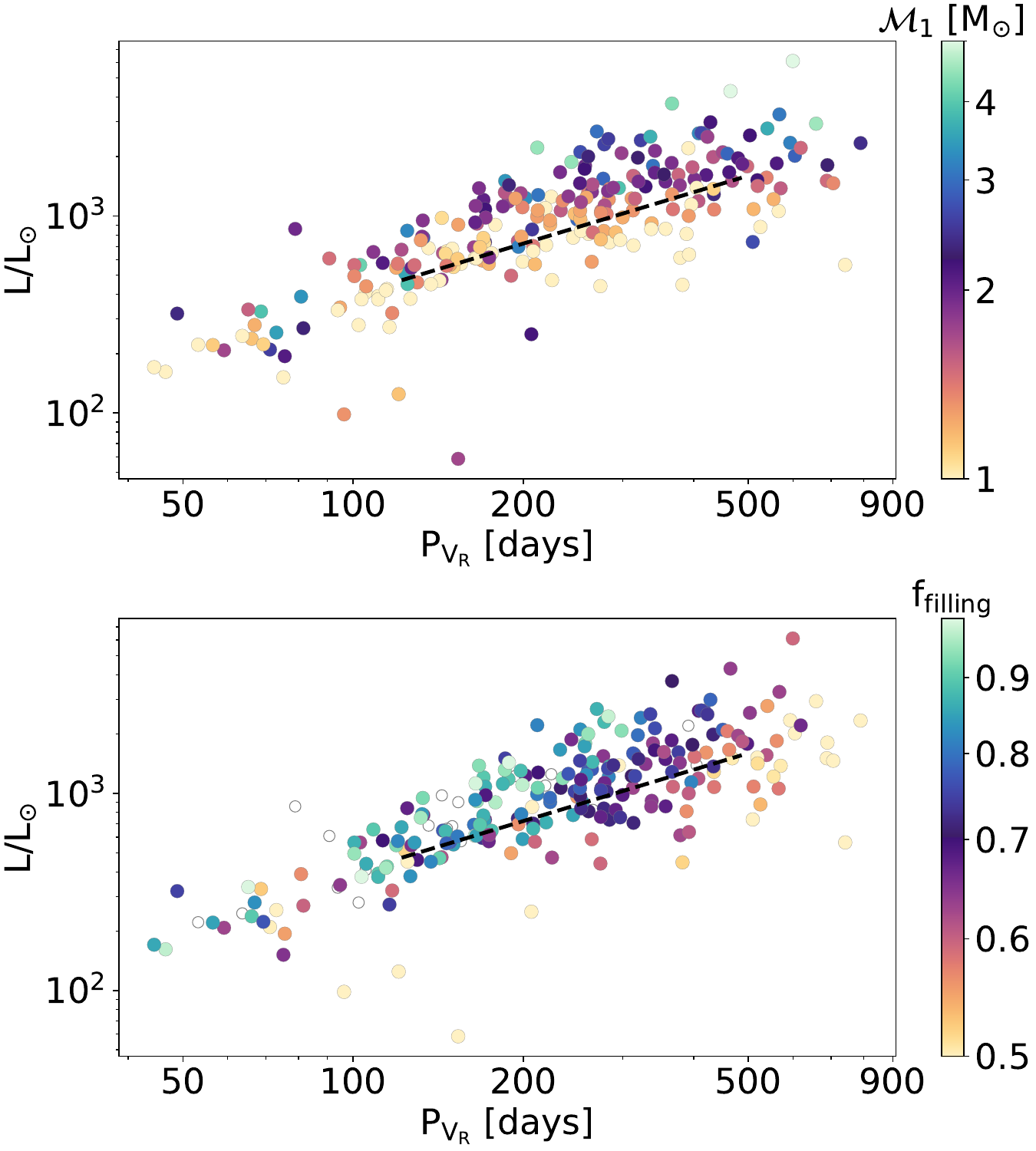}
 \caption{Period-luminosity diagram for the ellipsoidal candidates, colored based on the primary mass (top panel) and filling factor (bottom panel. Note that both color bars are shown in logarithmic scale. Systems having a filling factor larger than 1 are shown as unfilled circles. In both panels, the black dashed line corresponds to the fit to the sample of low-mass ellipsoidal variables in \cite{Nie17}.} \label{fig:PL_3p}
 \end{figure}

Figure~\ref{fig:PL_3p} shows the orbital period against the luminosity for ellipsoidal binaries. The top panel shows the ellipsoidal stars in this plane, colored according to the mass of the primary star. It reveals that the faintest systems have mostly lower masses, while the brighter systems, for a given period, are those with the largest masses. A few outliers at L/L$_{\odot} \leq 10^{2}$ could have an erroneous period determination. Excluding those, the trend in mass is striking. This is an expected behaviour as more massive systems have larger orbital separations $a$ for a given period, making only the larger, more luminous primary stars detectable as ellipsoidal variables due to their ellipsoidal distortion. The same trend, although restricted to $\sim$80 stars, was observed in the sample of ellipsoidal variables in the LMC in \cite{Nie17}. For reference, we fitted a straight line to the low-mass ($\leq$ 1.85 M$_{\odot}$) ellipsoidal binaries in \cite{Nie17}, similar to their Figure 7 but the fit was done directly into luminosity instead of $I$ magnitude. Our sample covers a larger range in period than the sample in the LMC but overall, for the low-mass systems, the slope of the period-luminosity sequence agrees, having a non-negligible population of low-mass binaries having lower luminosities than the fit, as in \cite{Nie17}.

Another parameter that could impact the period-luminosity sequence of ellipsoidal variables is the filling factor. For a given period and mass, there is a range of possible values for the filling factor under which the systems will undergo ellipsoidal variations. The bottom panel of Figure~\ref{fig:PL_3p} shows the Galactic ellipsoidal binaries color-coded according to the filling factor (see Section~\ref{sec:q_f}). The same trend as in \cite{Nie17} is recovered: the brighter systems tend to have larger filling factors, at a given period. This effect introduces a mass dispersion that produces an intrinsic scatter in the PL sequence of ellipsoidal stars. Systems with larger filling factors will also have circular binary orbits, while the fainter binaries with lower filling factors should have non-circular orbits. 

\begin{figure}
 \centering
 \includegraphics[width=0.5\textwidth]{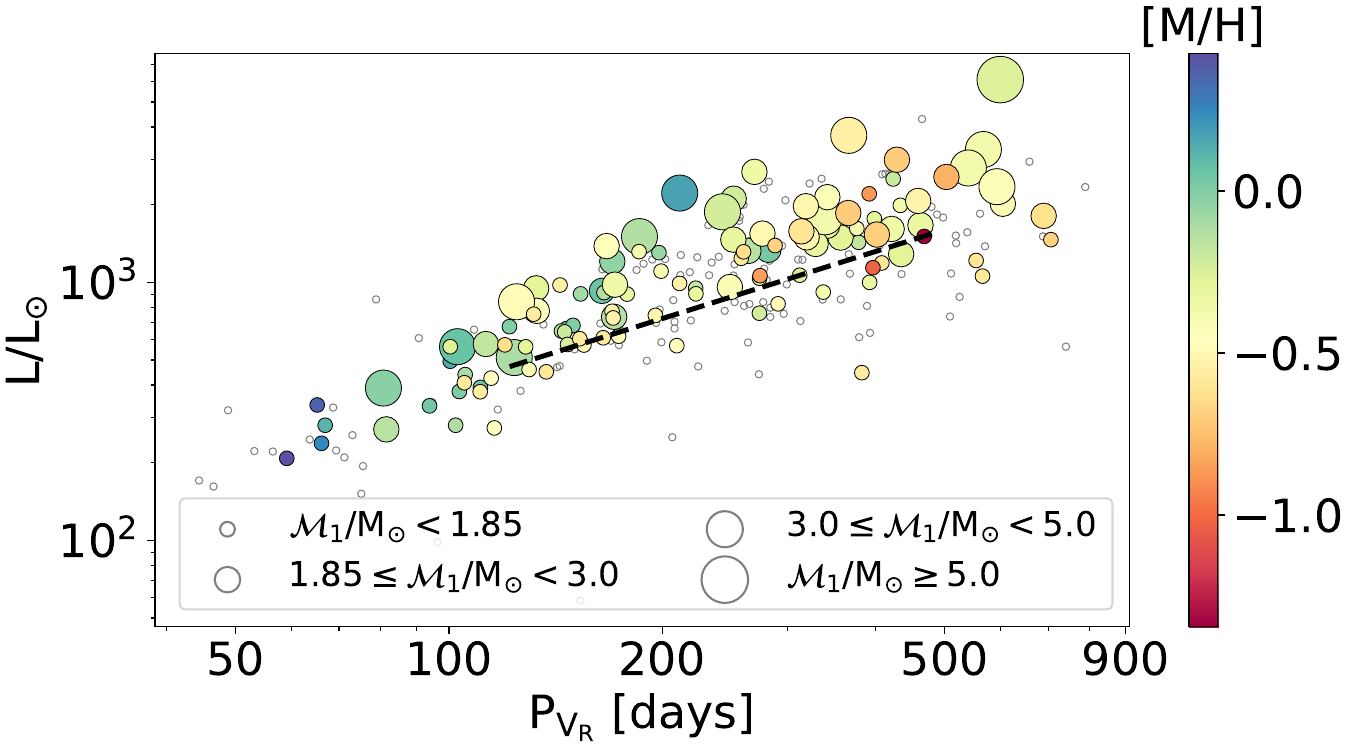}
 \caption{Period-luminosity diagram for ellipsoidal candidates having a reliable metallicity estimate based on \textit{Gaia} RVS spectra (derived by the \textit{Gaia} \textit{GSP-Spec} module). The size of the symbol corresponds to the mass of the primary star. Ellipsoidal variables without a reliable metallicity estimate (see Sect.~\ref{subsec:chemodynamics}) are shown as grey dots, of equal size.} \label{fig:PL_MH}
 \end{figure}

In the study of \cite{Nie17}, some low-mass systems ($\mathcal{M}_1$ $\leq$ 1.85 M$_{\odot}$) were found at the higher luminosity side of the period-luminosity sequence, possibly a consequence of being more metal-poor. More metal-poor stars, at a given period, will be smaller in radius and, therefore, only the bigger, more luminous low-mass stars, will be detected as ellipsoidal variables. Figure~\ref{fig:PL_MH} shows the systems in the period-luminosity diagram coloured according to their \textit{Gaia} \textit{GSP-Spec} metallicity, while the size of the symbol is proportional to the mass of the primary star. Unfilled circles are for those systems that do not pass the metallicity quality cuts (the same as the ones applied in the Fig.~\ref{fig:mdf}). There are some indications of the predictions in \cite{Nie17}, as for a fixed period and low mass, the brightest systems are more metal-poor. Nonetheless, the number of metal-poor, low-mass ellipsoidal variables among the high luminosity end of the period-luminosity sequence is quite limited and the strongest dependence for the luminosity is on the mass of the primary star.

In the case of rotational candidates, we do not find a relation between the mass and the dispersion of the visible linear PL sequence, as it is visible in Figure~\ref{PL_NSS}. This means that as much larger, in radius, and therefore, more luminous, a rotational binary is, it will be rotating more slowly. Hints of this relation are possibly found in \cite{Muraveva2014}, without presenting a discussion about the origin of this sequence, as well as in \cite{Martinez22}, where a sample of RS CVn systems were analysed in terms of their activity cycles and rotational periods.

\section{Summary} \label{sec:conclusions}

In this work, the physical parameters (luminosities, radii and masses) as well as orbital and chemo-dynamical properties of long-period red giant binary stars were derived. Our input sample consisted of more than 400 red giant binaries for which both photometric and radial velocity time series were available as part of the \textit{Gaia} Focused Product Release \citep{GaiaFPR}.

The input sample comprises two types of binaries: ellipsoidal red giants and rotational binaries. The separation boundary of the subtypes was performed considering the ratio between the orbital and photometric periods, both values derived in the \textit{Gaia} FPR \citep{GaiaFPR}, as well as through a visual inspection of the phase-folded $G$-band light curves. Even though we have shown that most of the periods are in good agreement with the literature, we cannot rule out having ellipsoidal binaries with erroneous photometric or orbital periods among the rotational sample. Considering the similar distribution in the period-radius-eccentricity plane (see Fig.~\ref{P_Radius_Masses}), both samples appear to be the manifestation of the same physical configuration, under two different regimes: a subgiant or a giant primary star,  for rotational or ellipsoidal binaries respectively, with different orbital separations. The variability of the rotational sample could be, therefore, not due to an enhanced rotation but instead produced by the geometrical effect of a less pronounced ellipsoidal deformation (see Fig.~\ref{fig:filling}). Nonetheless, the enhanced activity index in some of the rotational candidates (see Fig.~\ref{fig:logg_activity}) supports the scenario in which these binaries are chromospherically active due to enhanced rotation in a close binary, a different variability phenomenon than in the case of ellipsoidal variables. Moreover, we confirmed that their location on the 2MASS near-IR CMD coincides with the group of RS CVn stars identified in \cite{Phillips24}, based on a sample of 50\,000 rotational variables. The position in the CMD as well as their period range are consistent with the compilation of RS CVn variables in \cite{Leiner22}.

Our main findings for rotational and ellipsoidal binary candidates are summarized below.

\textit{Rotational variables}: Although the sample contains less than 50 stars, they appear to differ from ellipsoidal variables considering the period range, independent period-luminosity sequence and, most importantly, enhanced stellar activity. We have found that the enhanced activity prevents the use of atmospheric parameters from the \textit{GSP-Spec} module and, instead, the parameters from the XGBoost catalogue, based on spectro-photometric XP spectra were favoured. We found that the sample of rotational variables has stellar radii between 5 to $\lesssim$ 30 R$_{\odot}$, and primary masses from less than one solar mass up to no more than 3 M$_{\odot}$. The minimum masses for the secondary components are below 1.0 M$_{\odot}$. In terms of the orbital properties, almost all these variables have circularized orbits. Their kinematics reveal that they are concentrated in the Galactic disk, having preferentially a low velocity dispersion. This can be associated with a relatively younger population, at least when compared to the ellipsoidal binaries. 

\textit{Ellipsoidal binaries:} Physical parameters were derived for 243 ellipsoidal binaries, the largest collection with spectroscopically derived luminosities and masses. The previous largest collection, with photometric and spectroscopic data, comprised 81 ellipsoidal binaries in the LMC \citep{Nie17}. The properties of our sample confirm their nature as ellipsoidal binaries, since the period-radius distribution is at the edge of reaching the Roche lobe radius, having filling factors as large as unity. As expected, the larger the orbital period, the larger the radius. The eccentricities are close to zero, in contrast with the measurements for ellipsoidal binaries in the LMC which reach much larger values. This could imply different tidal circularization rate efficiencies \citep[see Section 5 in][]{Nie17}. The primary stars are mostly low- and intermediate-mass giants, covering a wide range from 0.4 up to 6.0 M$_{\odot}$. The minimum mass of the secondary component was found to be $\gtrsim$1.0 M$_{\odot}$, consistent with a possible compact remnant star. The chemo-dynamics of the sample are consistent with both Galactic disk and halo stars. Some stars having enhanced [Ce/Fe] abundances suggest that they may be the result of a previous mass transfer, similar to the scenario of enrichment in Ba-stars. The dispersion in the period-luminosity relation of ellipsoidal binaries was found to be a mass and filling-factor dependence, rather than dominated by the metallicity dispersion.

Despite the separation of the input sample into rotational and ellipsoidal binaries, their mean stellar properties (masses, radii, luminosities) and chemo-dynamical parameters lead us to propose them as two manifestations of a close binary, in which the primary star is a red giant (or subgiant stars) and a low-mass companion where, depending on the mass of the primary and orbital separation, the tidal synchronization or the ellipsoidal deformation takes place. Instead of considering both types of binaries as separated and unrelated classifications, this analysis has shown their connection as two evolutionary stages of the same phenomenon. 

Together with this article, we release the sample of 243 ellipsoidal binaries and 39 rotational binary candidates including extinction A$_{\rm G}$, luminosity, radius, and the spectroscopic mass of the primary component, as well as estimated for the minimum mass of the secondary component, mass-ratio and filling factors. This large sample would significantly contribute to the population studies of these two binary types as well as serve for future spectroscopic follow-up campaigns to fully characterize these binaries.

\section{Data availability}
Tables~\ref{tab:parameters} and ~\ref{tab:binary_ratios} are available in electronic form at the CDS via anonymous ftp to cdsarc.u-strasbg.fr (130.79.128.5) or via http://cdsweb.u-strasbg.fr/cgi-bin/qcat?J/A+A/.

\begin{acknowledgements} The authors would like to thank Dr M\'arcio Catelan and the anonymous referee for their constructive comments that helped to improve this manuscript. This work has made use of data from the European Space Agency (ESA) mission \textit{Gaia} (https://www.cosmos.esa.int/gaia), processed by the \textit{Gaia} Data Processing and Analysis Consortium (DPAC, https://www.cosmos. esa.int/web/gaia/dpac/consortium). Funding for the DPAC has been provided by national institutions, in particular the institutions participating in the \textit{Gaia} Multilateral Agreement. C.N. would like to thank Henry Lung, for the development and update of the mw-plot package \url{https://milkyway-plot.readthedocs.io/en/stable/} used to produce the Figure~\ref{fig:all_sky}. This work made use of the \texttt{cmastro} package \url{https://github.com/adrn/cmastro}. C.N acknowledges financial support from the Centre National d'Études Spatiales (CNES) fellowship. A.E. received the support of a fellowship from the La Caixa Foundation (ID 100010434) with fellowship code LCF/BQ/PI23/11970031. 
\end{acknowledgements}

\bibliographystyle{aa}
\bibliography{biblio}

\begin{appendix}

\section{Photometric periods in the literature}\label{app:periods}

\begin{figure}[h!]
   \centering
    \includegraphics[width=0.5\textwidth]{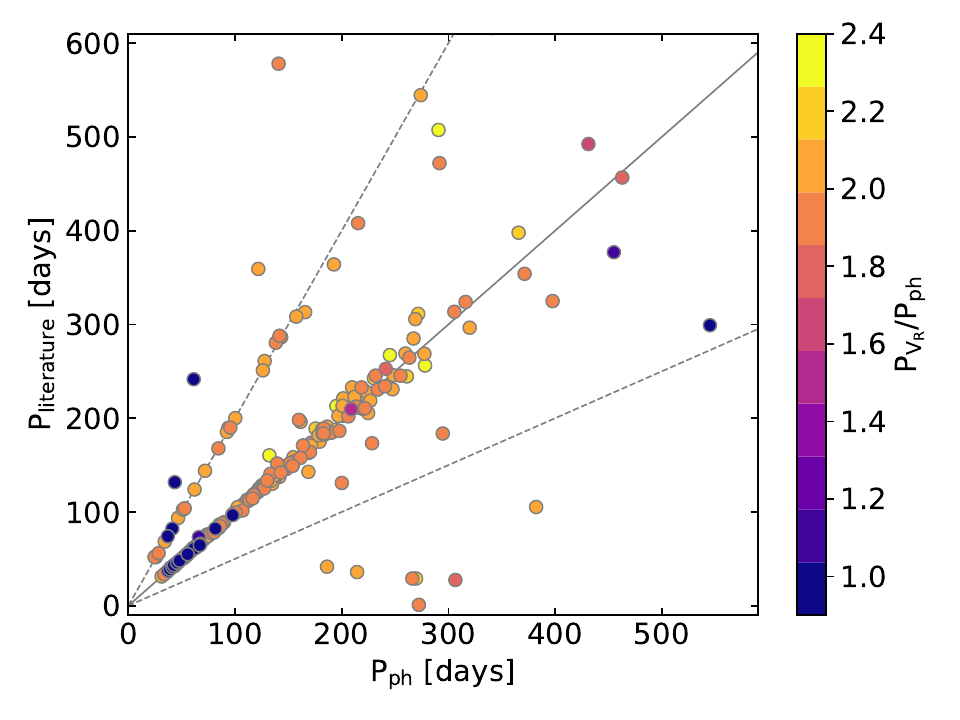} 
 \caption{\textit{Gaia} FPR photometric period, P$_{\rm ph}$, versus literature periods compiled in \cite{Gaia_var}. The stars are color-coded according to the ratio between the RV period and the photometric period. The dashed lines correspond to P$_{\rm lit}$/P$_{\rm ph}$ = 2 and 0.5, while the solid line corresponds to the 1:1 relation.} \label{fig:plit}
\end{figure}

Figure~\ref{fig:plit} shows the photometric periods from \textit{Gaia} FPR against the literature periods from the catalogue of \cite{Gaia_var} for all the binary candidate stars in common (239 out of 462), showing the good agreement of the \textit{Gaia} periods, even at periods longer than 300 days. There is a fraction of stars that have twice the period reported in \textit{Gaia} and only a handful of them have periods in the literature that are almost half of the \textit{Gaia} period, which could place them in the sequence for ellipsoidal variables (P$_{V_R}$/P$_{\rm ph}$ $\simeq$ 2). 

From the stars with P$_{V_R}$/P$_{\rm ph}$ $\simeq$ 1, there are literature periods for 22 out of 42 stars. Only four stars have periods that are either 2, 3 or 4 times the \textit{Gaia} P$_{\rm ph}$, and one has half the \textit{Gaia} period\footnote{\textit{Gaia} DR3 201703728789835904, rejected later based on the visual inspection of its light curve.}. A second search was made in the databases of different variability surveys, either through the \textit{Gaia} source ID and/or the equatorial coordinates. Literature periods were recovered for seven stars not included in \cite{Gaia_var}, all being the same or double the \textit{Gaia} P$_{\rm ph}$ value, as listed in Table~\ref{tab:plit}. 

There is a good agreement in the \textit{Gaia} periods from FPR and those reported in the literature, although in a few cases, the phase coverage and number of epochs could prevent the recovery of the true period, obtaining instead twice or half of it. The modelling of the individual light curves and period determination is out of the scope of this paper, but we visually inspect all the sources to identify clear misclassified sources, and remove highly uncertain cases from the analysis.

\begin{table*}
    \centering
    \caption{Literature periods for sources not included in the \cite{Gaia_var} catalogue. Variability types are either dubious, Semi-Regular variable (SR), long-period variable (LPV), Rotational variable (ROT) and RS Canum Venaticorum (RS CVn).}
    \begin{tabular}{cccccc}
   \hline
   \textit{Gaia} DR3 source id & $P_{\rm ph}$ & Other ID & Lit. period & Variability type & Reference \\
                                &          (d) &          &         (d) &                  &           \\
   \hline
1989111887224781056 & 36.89034 & ATO J346.0317+51.5494         &  77.013120   & dubious  & (1) \\
2037641061998679552 & 49.05138 & ATO J286.5005+28.5830         &  97.588819   & dubious  & (1)  \\
629379460969722496  & 48.67723 & CSS J100138.5+204857          &  50.3620289  &          & (2) \\
1869696952997313664 & 52.81437 & ZTF J205411.65+353803.3        &  53.8513758 & SR       & (3)\\
                    &          & ATO J313.5485+35.6342         &  106.784031  & LPV      & (1)\\
1756282633518972544 & 42.97533 & ASASSN-V J203640.72+133017.9  &  42.8644419  & ROT      & (4)\\
                    &          & ATO J313.5485+35.6342         &  85.883021   & LPV      & (1)\\
2028753576416486912 & 63.55899 & ASASSN-V J195140.08+285735.1  &  63.1252006  & SR       & (4)\\
2230939471064416128 & 39.94571 & GCVS                          &  39.64       & RS CVn   & (5)\\
\hline
    \end{tabular}
\tablefoot{[1] ATLAS survey \citep{ATLAS}; [2] Catalina Schmidt Survey \citep{CSS_var}; [3] Zwicky Transient Facility \citep{ZTF_var}; [4] All-Sky Automated Survey for Supernovae \citep[ASAS-SN;][]{ASASSN}; [5] Variable Star IndeX, from \cite{PASP}.}
    \label{tab:plit}
\end{table*}

\vfill

\section{Surface gravities} \label{app:logg}

\begin{figure}[h!]
   \centering
    \includegraphics[width=0.5\textwidth]{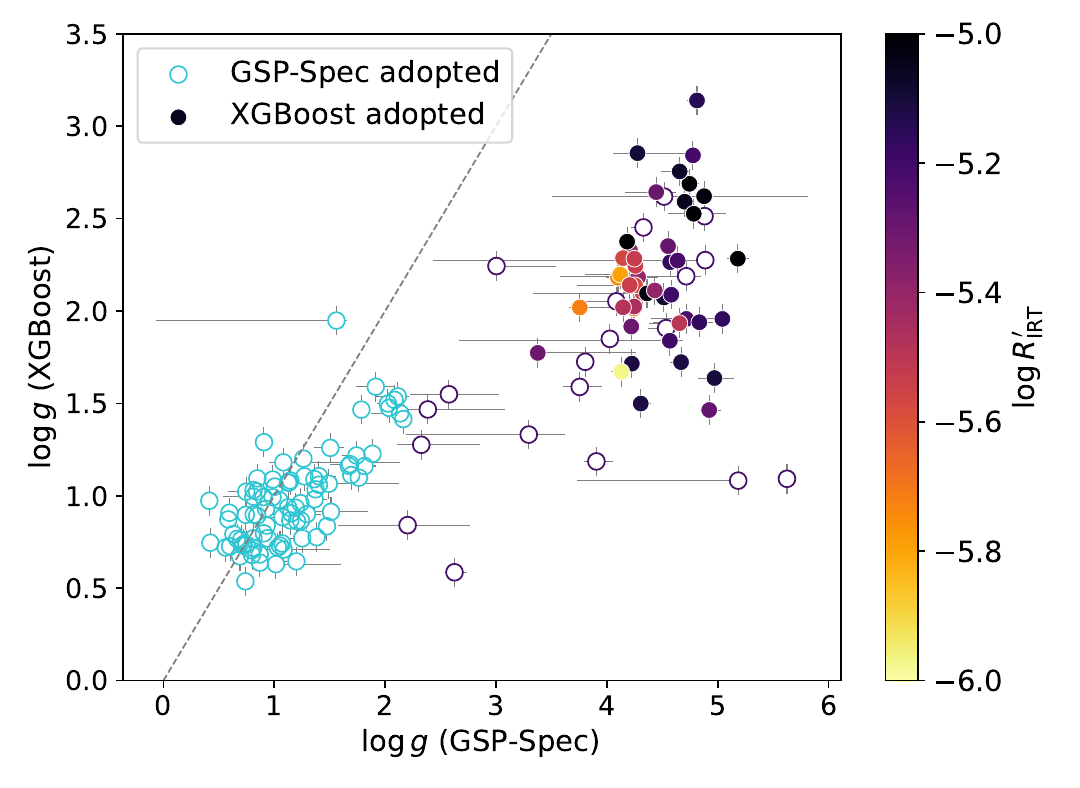} 
 \caption{Comparison between the $\log{g}$ values for stars with \texttt{KMgiantPar = 0} as derived in \textit{GSP-Spec} (calibrated based on the T$_{\rm eff}$) and in the XGBoost catalogue. Cyan and purple circles correspond to stars with \textit{GSP-Spec} calibrated surface gravities smaller and larger than 2.2 dex, respectively. In the later, the XGBoost parameters were adopted. The activity index (available for a subsample of sources) is shown in the color bar. Stars having overestimated \textit{GSP-Spec} surface gravities are precisely those sources with enhanced activity.} \label{fig:logg_comp}
\end{figure}

In Section~\ref{sec:atm}, stars with shorter periods and enhanced chromospheric activity were found to have overestimated spectroscopic $\log{g}$ from \textit{Gaia} RVS spectra. Figure~\ref{fig:logg_comp} shows the (calibrated) $\log{g}$ estimates from \textit{GSP-Spec} compared to the inferred values from XGBoost, for stars with good quality flags from \textit{GSP-Spec} (\texttt{KMgiantPar = 0}), both for the sources having $\log{g} \leq$ 2.2 (cyan points) and those with potentially overestimated $\log{g}$ values (purple points). In the figure, the stars having an entry for the activity index in Gaia DR3 are coloured based on $\log{R_{\rm IRT}}$. The good agreement between the \textit{GSP-Spec} and XGBoost values at $\log{g} \leq$ 2.2 makes us use this boundary to select atmospheric parameters from \textit{GSP-Spec}. In the rest of the cases, the atmospheric parameters from XGBoost were adopted.

\vfill

\section{Notes on individual systems} \label{app:comments}

\begin{figure}[h!]
   \centering
    \includegraphics[width=0.5\textwidth]{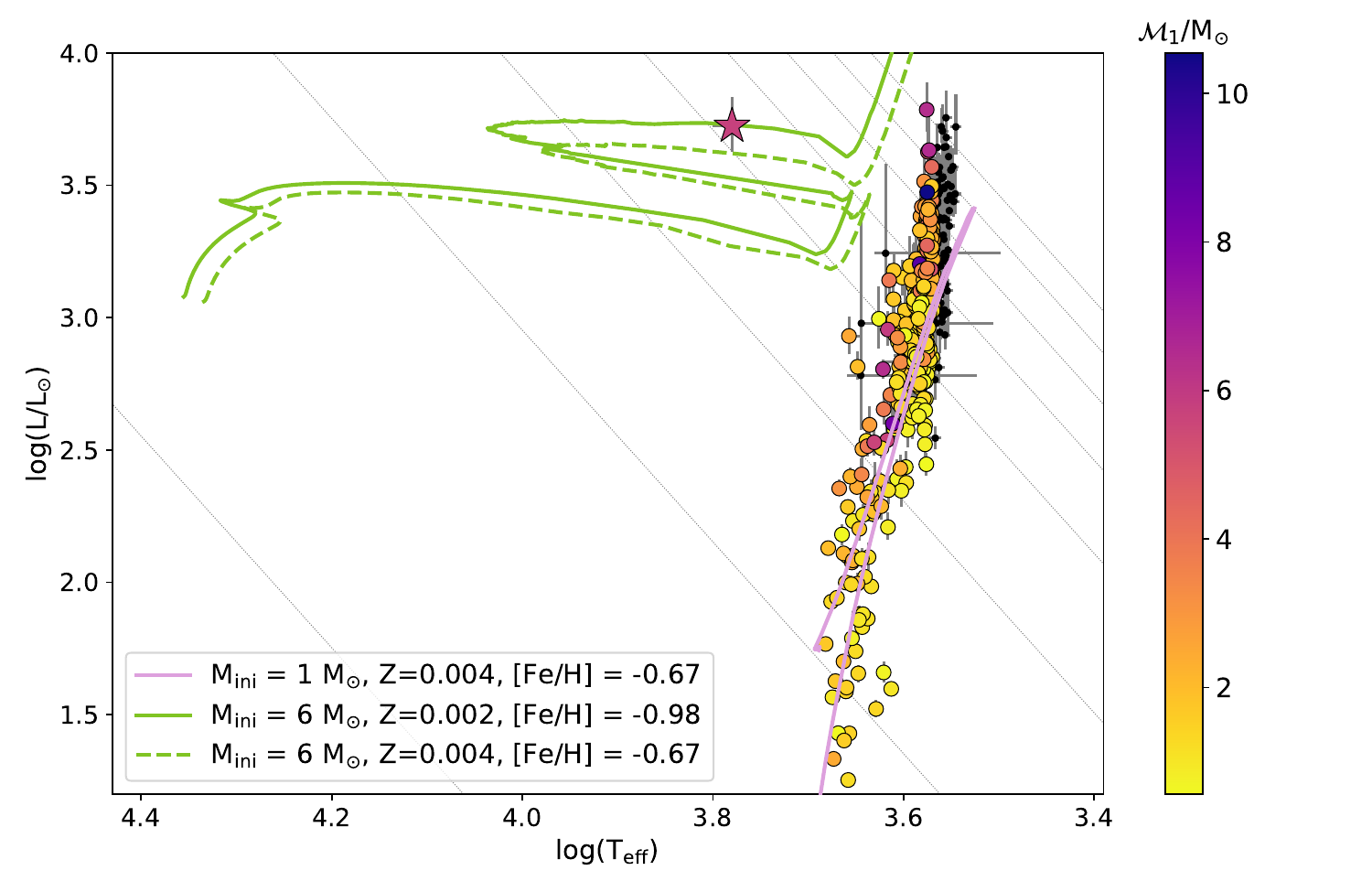} 
 \caption{HR diagram for \textit{Gaia} DR3 517611939345861120 (star symbol) along with the rest of the sample of red giant binaries, color-coded by their spectroscopic mass. The grey lines correspond to radii of 1, 10, 30, 60, 90, 120, 150, 180 R$_{\odot}$. Different evolutionary tracks from STAREVOL \citep{Lagarde12} have been included.} \label{fig:outlier}
\end{figure}

\textit{\textit{Gaia} DR3 517611939345861120:} This star has a period ratio of P$_{V_R}$/P$_{\rm ph}$ $\simeq$1.0 and therefore can be discarded as an ellipsoidal binary. However, its absolute magnitude M$_{G}$ = $-$4.63 mag and color (BP-RP)$_0$ = 0.66 mag are not consistent with a red giant. Their atmospheric parameters from \textit{GSP-Spec} are (T$_{\rm eff}$, $\log{\rm g}$, [M/H]) = (6025 $\pm$ 169 K, 1.56 $\pm$ 0.9 dex, $-$2.28 $\pm$0.4 dex) having \texttt{KMgiantPar = 0} \citep[i.e., good parametrization of the RVS spectrum,][]{GaiaRVS}. Its low metallicity is consistent with the reported value based on BP/RP spectra from XGBoost, [M/H] = $-$2.12 $\pm$0.1 dex (T$_{\rm eff}$ = 5258.7 K; $\log{\rm g}$ = 1.948 dex). Based on the \textit{GSP-Spec} atmospheric parameters, its spectroscopic mass should be of $\sim$5.8 M$_{\odot}$, overly large for its metallicity, but appropriate for its luminosity. Nonetheless, the error on the mass is quite large, having a lower value (16th-percentile) of 0.8 M$_{\sun}$ and upper value (84th percentile) of 21.2 M$_{\odot}$. If the atmospheric parameters from XGBoost are used, a larger mean mass is recovered: 16$_{-4.1}^{+5.7}$ M$_{\odot}$. Based on its position in the PL diagram (see Fig.~\ref{PL_NSS}), its absolute magnitude is $\sim$4 mag brighter than the rest of non-ellipsoidal, RS CVn candidates at the same period. We believe this star is a genuine massive giant, much hotter than the rest of the sample, with an erroneous metallicity estimate. Figure~\ref{fig:outlier} shows the HR diagram position for this star compared to the rest of the sample. Evolutionary tracks from the grid computed by \cite{Lagarde12} for stars with initial masses of 1 M$_{\odot}$ and 6 M$_{\odot}$ and different metallicities are included. The position of this star in the HR diagram is consistent with the evolutionary track of a 6 M$_{\odot}$ star with [Fe/H] $\simeq$ -1, in the prescription that includes both thermohaline convection and rotation-induced mixing. Further spectroscopic observations of this system are required to establish its nature.

\vfill
\clearpage
\section{Overestimated spectroscopic masses} \label{cleaned_masses}

The spectroscopic masses derived in this work strongly depend on the surface gravity values' accuracy. To confirm that the values for the masses are within the range expected given the effective temperature and luminosity measured, we compare the HR diagram of those sources having the largest masses $\mathcal{M}_1 \geq$ 5.0 M$_{\odot}$ with the expected luminosity for its given mass based on evolutionary tracks. Stars with lower masses are well within the expected position in the HR diagram (see the middle panel in Fig.~\ref{HR_Radius_Masses}). 

\begin{figure}
   \centering
    \includegraphics[width=0.45\textwidth]{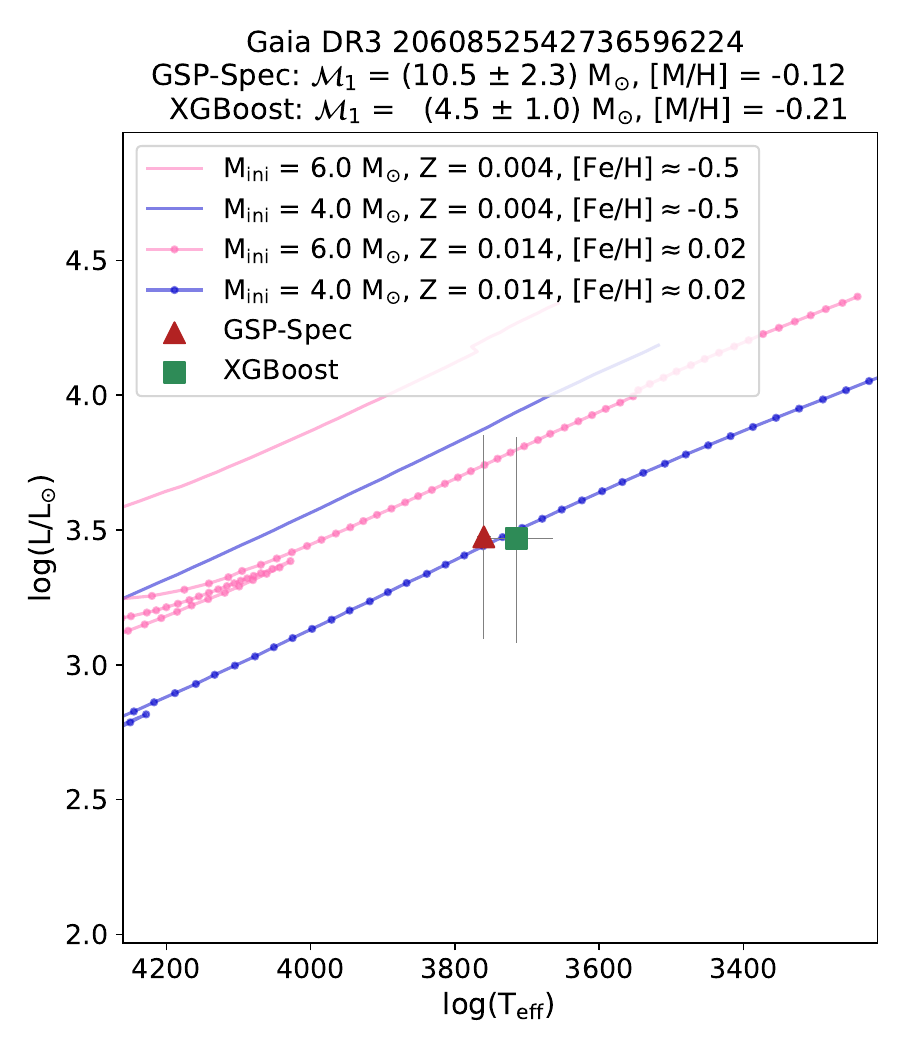} \\
    \includegraphics[width=0.45\textwidth]{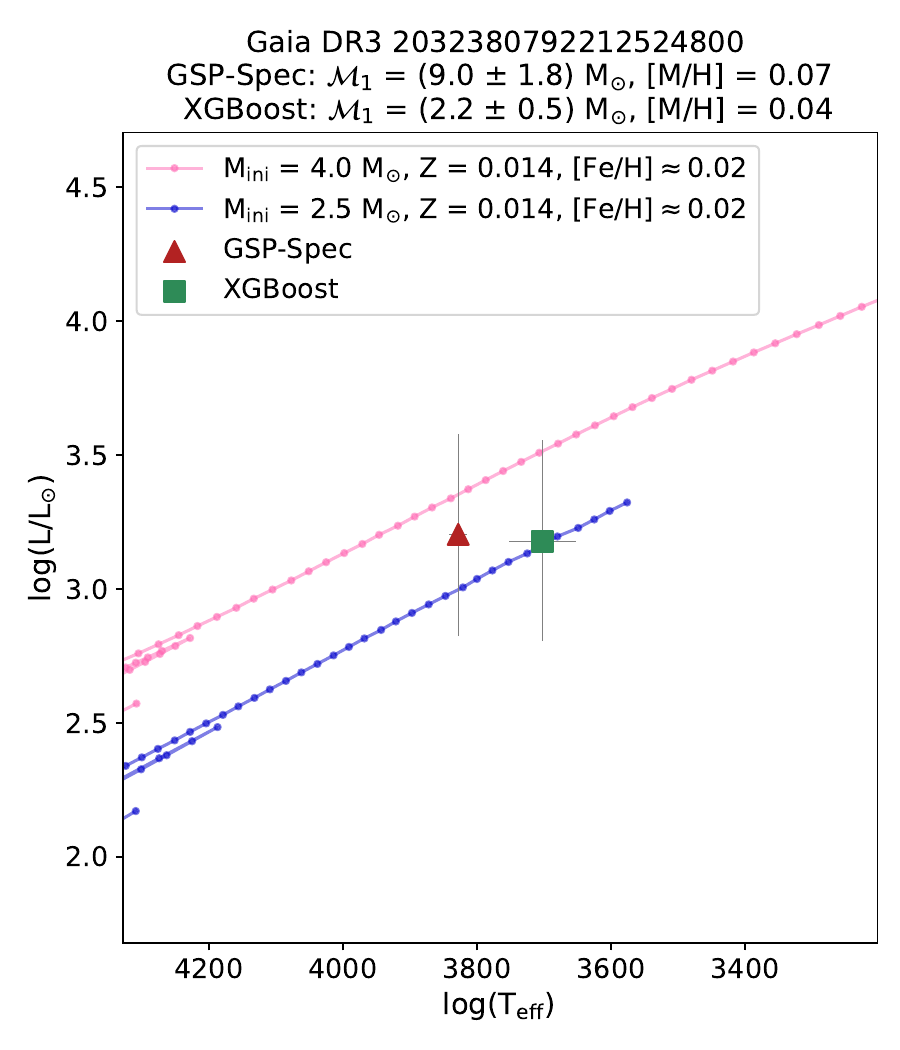} \\
\caption{HR diagram for sources having spectroscopic masses $\mathcal{M}_1 \geq$ 5.0 M$_{\odot}$ and the closest STAREVOL \citep{Siess2000} evolutionary tracks from \cite{Lagarde12}. The spectroscopic masses and metallicities from \textit{GSP-Spec} and XGBoost (if available) are in the title of each panel. The parameters from XGBoost and the associated mass were adopted. } \label{fig:s0}
\end{figure}

\begin{figure*}
   \centering
    \includegraphics[width=0.45\textwidth]{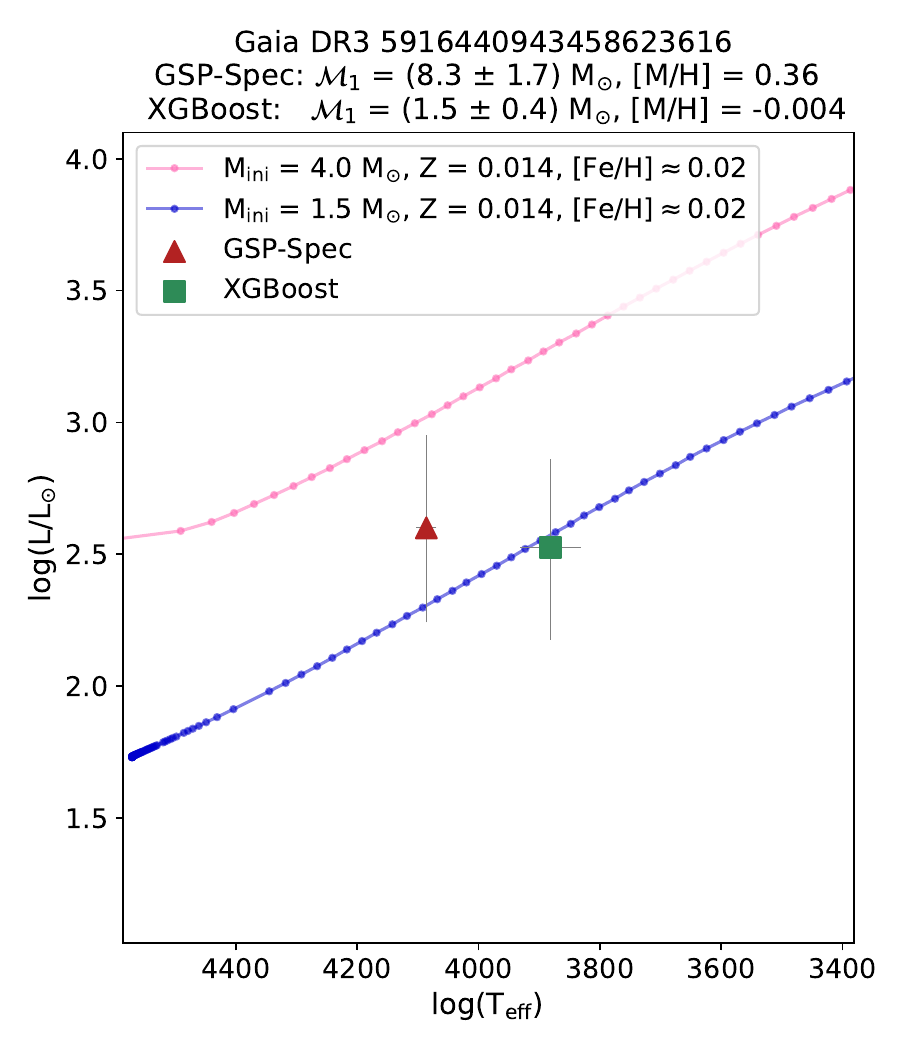} 
    \includegraphics[width=0.45\textwidth]{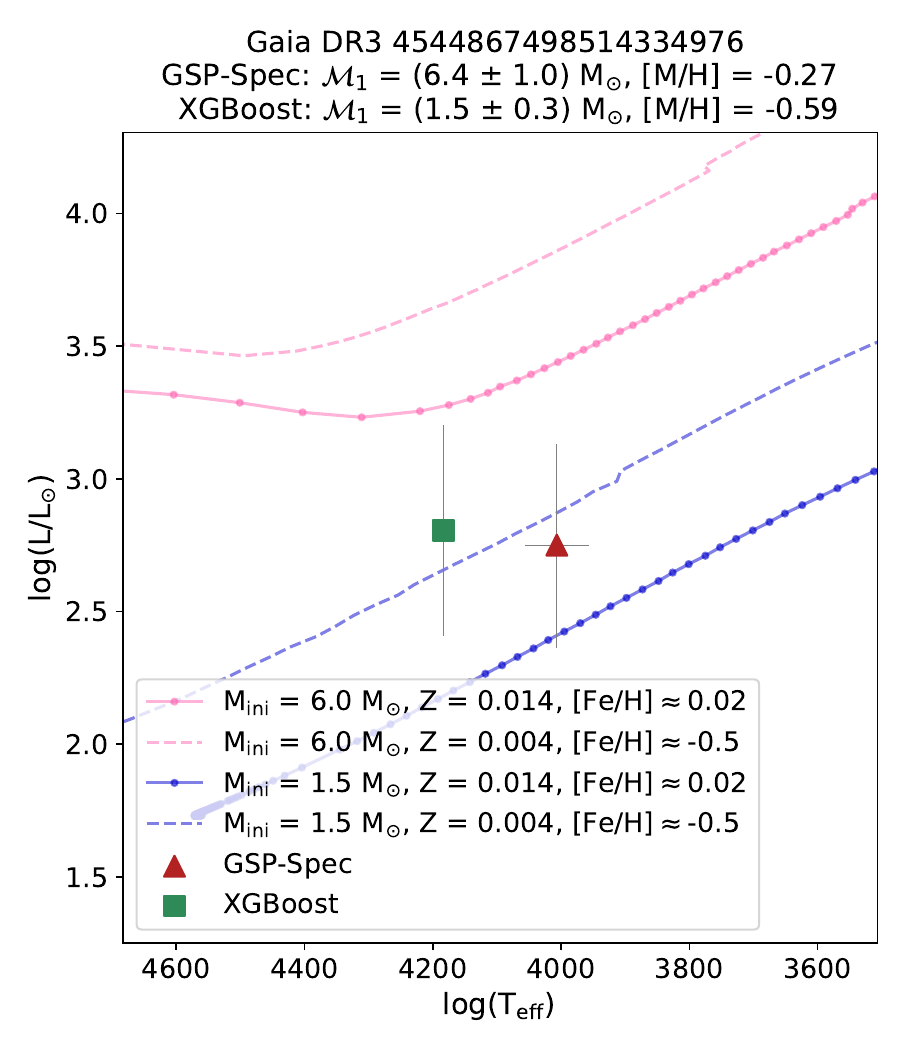}     
    \includegraphics[width=0.45\textwidth]{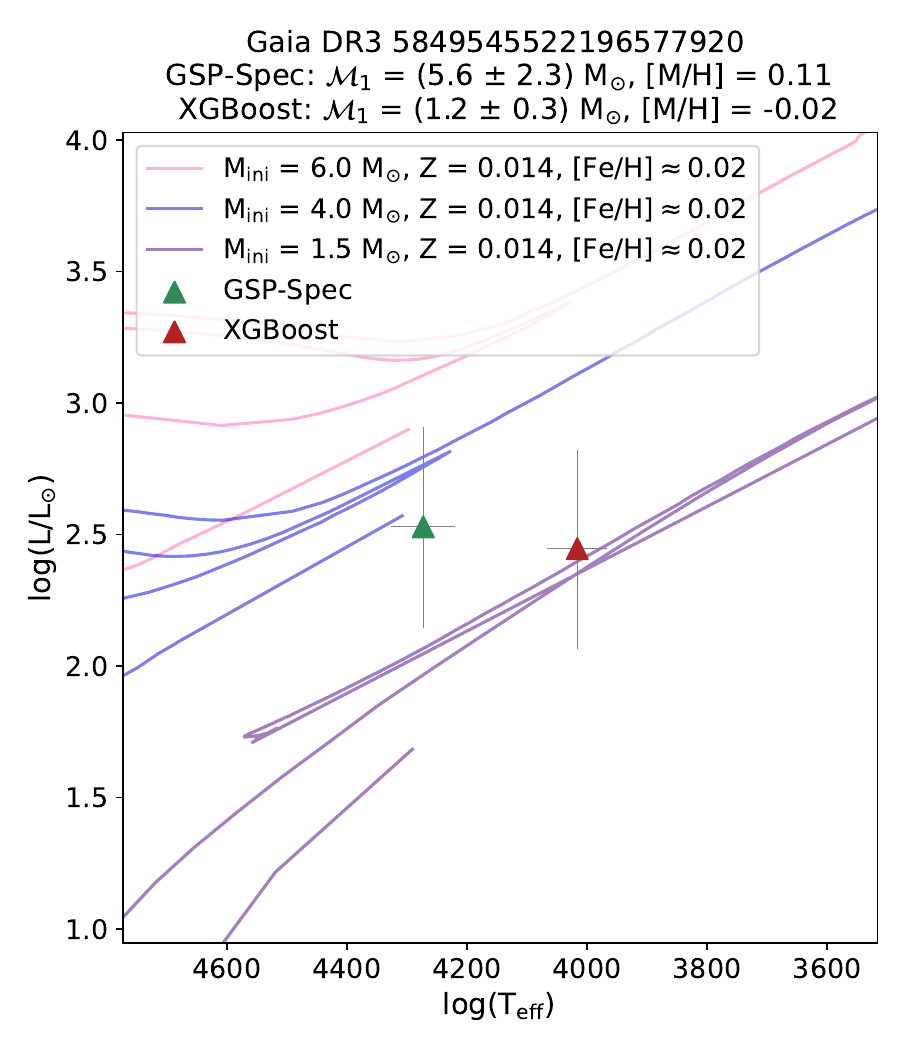} 
    \includegraphics[width=0.45\textwidth]{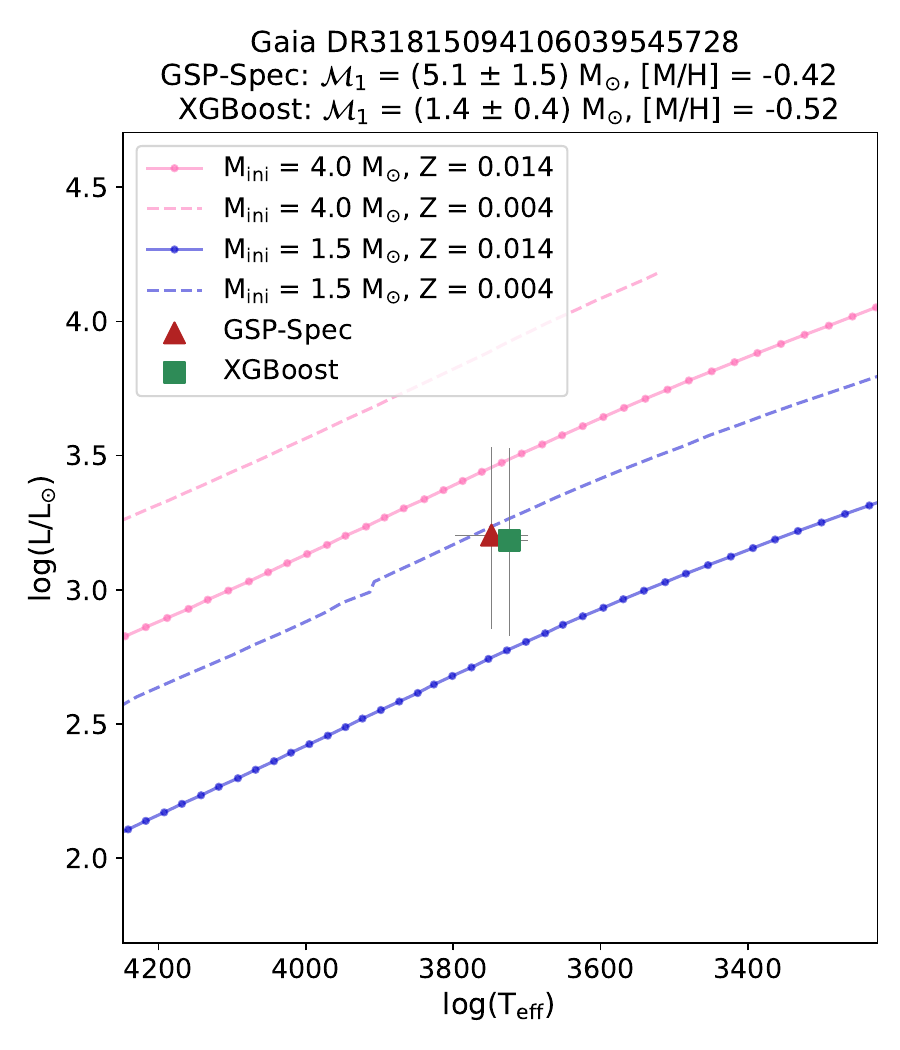}         
\caption{Same as Figure~\ref{fig:s0}. The parameters from XGBoost and the associated mass were adopted.}\label{fig:s1}
\end{figure*}

\begin{figure*}
\centering
    \includegraphics[width=0.45\textwidth]{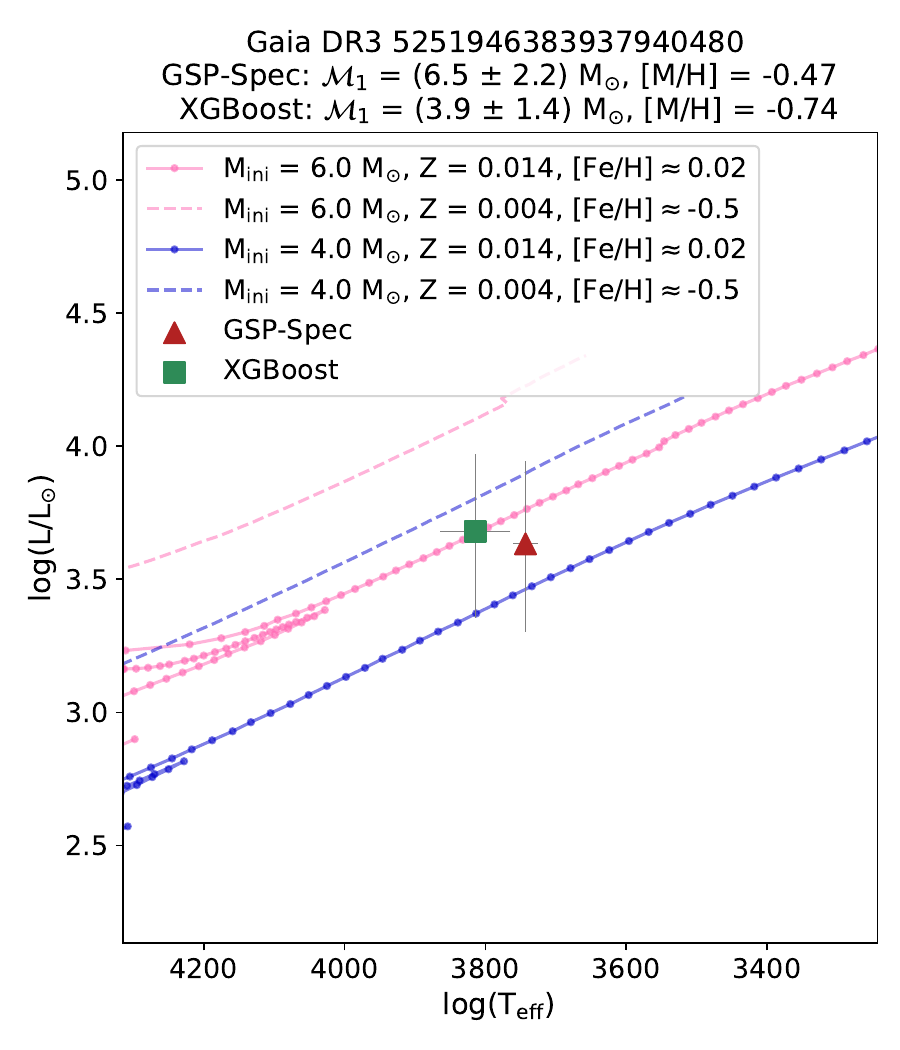}     
    \includegraphics[width=0.45\textwidth]{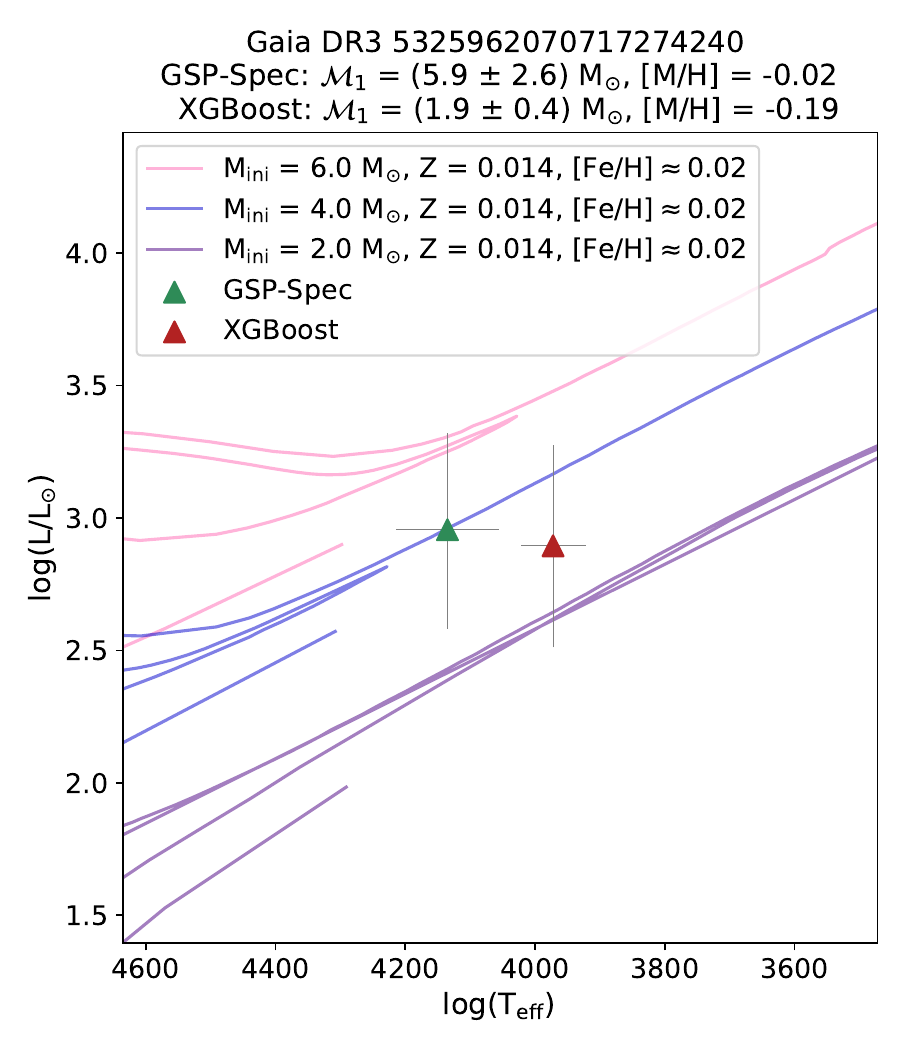} 
    \includegraphics[width=0.45\textwidth]{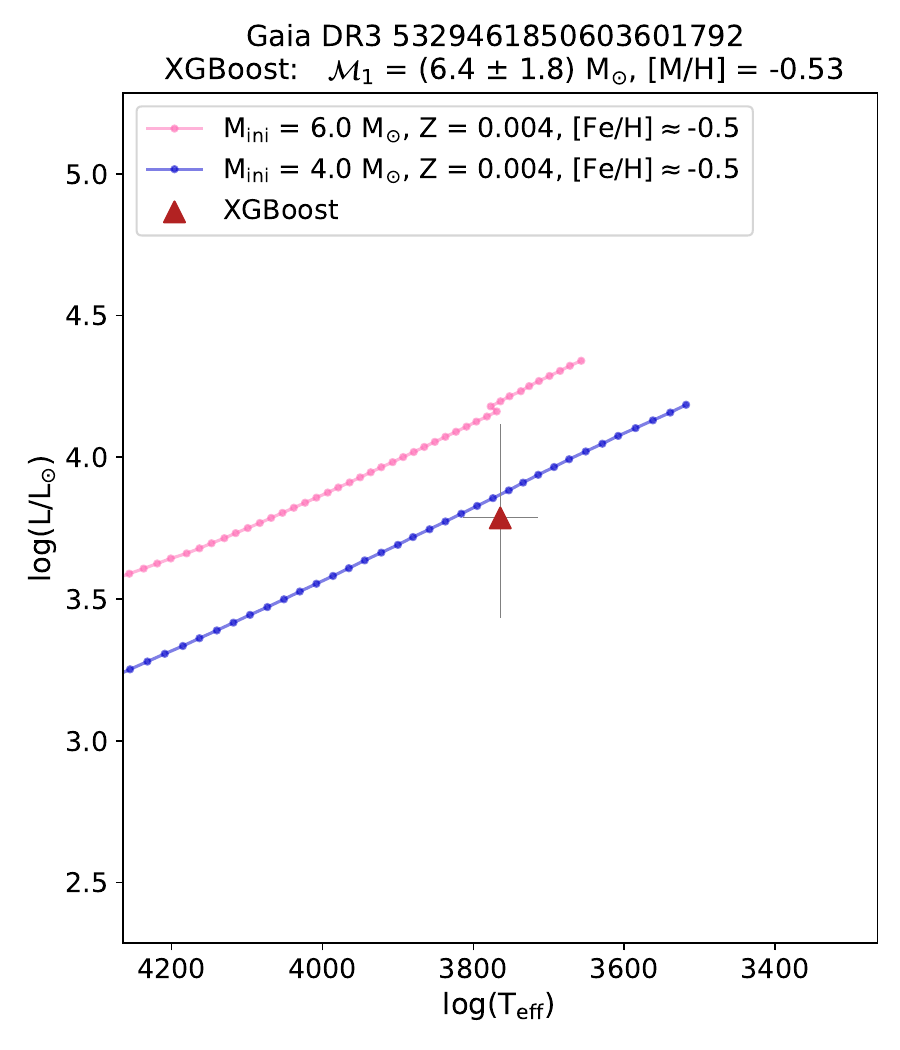}     
\caption{Same as Figure~\ref{fig:s0}. In this case, we adopted the XGBoost parameters for all the sources, but \textit{Gaia} DR3 5251946383937940480, based on the good agreement with respect to the evolutionary tracks. For \textit{Gaia} DR3 5329461850603601792, only atmospheric parameters from the XGBoost catalogue are available and their high mass is validated based on the luminosity and temperature.} \label{fig:s2}
\end{figure*}

Figures~\ref{fig:s0}, \ref{fig:s1} and ~\ref{fig:s2} show the HR diagram for those massive sources, showing both the \textit{GSP-Spec} and XGBoost temperatures and luminosities (if available), STAREVOL \citep{Siess2000} evolutionary tracks derived by \cite{Lagarde12}, for the closest metallicity and mass to our measured values and, indicating, in the title of each panel, the corresponding spectroscopic mass according to the parameters adopted, and its metallicity. In most of the cases, given the good match between the mass derived from XGBoost with respect to the evolutionary track, this spectroscopic mass was adopted even though there are atmospheric parameters measured from RVS spectra. The main culprit of the different masses is large differences in spectroscopic surface gravities.

In the case of \textit{Gaia} DR3 2032380792212524800, the mass obtained based on the XGBoost parameters is in much better agreement with the evolutionary tracks than the corresponding to the \textit{GSP-Spec} parameters. However, the determination of the reddening was flagged as inaccurate and therefore, the source was removed from the final sample.

From the ellipsoidal binary sample, two sources were discarded based on the poor precision associated with their spectroscopic masses (mass uncertainty larger than the median mass value). These two sources are \textit{Gaia} DR3 2216128156091604352 and 5215777399156162304. Their large errors in mass are due to the large error ($\gtrsim$0.4 dex) in the surface gravity measurement from \textit{GSP-Spec} compared to the rest of the sample having mean $\log{\rm g}$ uncertainty of 0.07 dex.

\FloatBarrier 

\clearpage

\onecolumn

\begin{landscape}
\section{Supplementary material}

\FloatBarrier

\begin{longtable}{cccccccccc}
    \caption{Spectroscopic atmospheric parameters and derived physical parameters for the primary star of Ellipsoidal and Rotational variables.} \\
   \hline
   \textit{Gaia} DR3 Source ID & Period & T$_{\rm eff}$ & $\log{\rm g}$ & A$_{\rm G}$ & $L_1$ & $R_1$ & $\mathcal{M}_{\rm 1}$ & Source & Flag \\
     & (day) & (K) & & (mag) & (L$_{\odot}$) & (R$_{\odot}$) & (M$_{\odot}$) & \\ 
   \hline
   \multicolumn{10}{@{}c}{Rotational candidates}\\
   \hline
 429660629859797248 & 51.5 & 4775 $\pm$ 50 & 2.26 $\pm$ 0.08 & 1.29 $\pm$ 0.05 & 134.8 $\pm$ 8.5 & 17.0 $\pm$ 0.6 & 1.9 $\pm$ 0.4 & 1 & G \\
 508673425214610560 & 43.8 & 4303 $\pm$ 50 & 2.01 $\pm$ 0.08 & 1.11 $\pm$ 0.06 &  96.6 $\pm$ 6.3 & 17.7 $\pm$ 0.7 & 1.2 $\pm$ 0.2 & 1 & R \\
 629379460969722496 & 48.7 & 4692 $\pm$ 50 & 2.69 $\pm$ 0.08 & 0.11 $\pm$ 0.05 &  42.4 $\pm$ 2.5 &  9.8 $\pm$ 0.3 & 1.7 $\pm$ 0.3 & 1 & R \\
 793060733341021056 & 39.6 & 4469 $\pm$ 50 & 2.35 $\pm$ 0.08 & 0.00 $\pm$ 0.03 &  55.0 $\pm$ 2.4 & 12.4 $\pm$ 0.4 & 1.2 $\pm$ 0.2 & 1 & U \\
1756282633518972544 & 41.8 & 4726 $\pm$ 50 & 2.53 $\pm$ 0.08 & 0.16 $\pm$ 0.05 &  36.8 $\pm$ 2.4 &  9.1 $\pm$ 0.3 & 1.0 $\pm$ 0.2 & 1 & G \\
\hline
  \multicolumn{10}{@{}c}{Ellipsoidal variable candidates}\\
\hline   
178734862161885440 & 337.2 & 3748 $\pm$ 15 & 0.81 $\pm$ 0.11 & 2.22 $\pm$ 0.05 &  917.1 $\pm$ 166.4 &  72.1 $\pm$  6.2 & 1.2 $\pm$ 0.4 & 0 & R \\
186039124063799936 & 153.3 & 4805 $\pm$ 50 & 2.59 $\pm$ 0.08 & 1.31 $\pm$ 0.04 &   58.5 $\pm$   3.3 &  11.0 $\pm$  0.4 & 1.7 $\pm$ 0.3 & 1 & G \\
201703728789835904 & 560.1 & 3753 $\pm$ 50 & 0.76 $\pm$ 0.08 & 1.39 $\pm$ 0.09 & 4235.4 $\pm$ 830.6 & 153.1 $\pm$ 14.3 & 5.0 $\pm$ 1.3 & 2 & U \\
201831508357553536 & 421.0 & 3753 $\pm$ 50 & 0.81 $\pm$ 0.08 & 1.34 $\pm$ 0.08 & 1611.3 $\pm$ 194.8 &  94.3 $\pm$  6.3 & 2.1 $\pm$ 0.5 & 2 & R \\
203635841300473984 & 278.0 & 3774 $\pm$ 50 & 0.76 $\pm$ 0.08 & 1.80 $\pm$ 0.11 & 2305.0 $\pm$ 640.6 & 112.8 $\pm$ 17.5 & 2.6 $\pm$ 0.9 & 2 & R \\
   \hline
\hline
\hline
    \label{tab:parameters}
\end{longtable}
\tablefoot{\textit{Gaia} DR3 source ID, RV (orbital) period, adopted effective temperature and surface gravity, the derived G-band extinction, A$_{\rm G}$, luminosities, masses and radii for rotational and ellipsoidal candidates. Only sources having conservative flags associated with the extinction estimate and validated masses are listed. The errors correspond to half the difference between the 84th- and 16th-percentile values. The Source column indicates the origin of the atmospheric parameters used to derive the extinction, luminosity, mass and radius, being equal to 0 when the atmospheric parameters used are those from the \textit{GSP-Spec} module, 1 when the atmospheric parameters are those from the XGBoost sample (\texttt{KMgiantPar=0}, $\log{g} \geq$2.2 or in the case of an overestimated mass for its luminosity, see App.~\ref{cleaned_masses}), and 2 when the atmospheric parameters are only available from XGBoost (\texttt{KMgiantPar=1,2}). The last column dubbed Flag indicates the quality of the light curve and, therefore, the certainty in the classification of the variables (G = Good, R = Regular, U = Uncertain). Only the first five rows for each binary type are shown. The full table will be available at the CDS.}    

\FloatBarrier

 \begin{table*}
    \centering
    \caption{Estimates for the companion mass, mass ratio and filling factors ef Ellipsoidal and Rotational variables.}
    \begin{tabular}{ccccc}
   \hline
   \textit{Gaia} DR3 Source ID & $\mathcal{M}_{\rm 2, min}$ [M$_{\odot}$] & $\mathcal{M}_{\rm 2, max}$ [M$_{\odot}$] & q ($\mathcal{M}_{\rm 2, min}$/$\mathcal{M}_{\rm 1}$) & f$_{\rm filling}$ \\
   \hline
   \multicolumn{5}{@{}c}{Rotational candidates}\\
   \hline
429660629859797248 & 0.320 $\pm$ 0.020 & & 0.165 $\pm$ 0.020 &  0.414 $\pm$ 0.019 \\
508673425214610560 & 0.844 $\pm$ 0.075 & & 0.718 $\pm$ 0.137 & 0.657 $\pm$ 0.036 \\
629379460969722496 & 0.681 $\pm$ 0.044 & & 0.393 $\pm$ 0.050 & 0.283 $\pm$ 0.012 \\
793060733341021056 & 0.208 $\pm$ 0.021 & & 0.169 $\pm$ 0.032 & 0.417 $\pm$ 0.023 \\
1756282633518972544	& 0.461 $\pm$ 0.056 &  & 0.467 $\pm$ 0.112 & 0.354 $\pm$ 0.022 \\
\hline
   \multicolumn{5}{@{}c}{Ellipsoidal variable candidates}\\
\hline   
178734862161885440 & 0.482 $\pm$ 0.096 & 2.182 $\pm$ 0.301 &	0.401 $\pm$ 0.148 & 0.646 $\pm$ 0.074 \\
186039124063799936 & 1.020 $\pm$ 0.115 & & 0.595 $\pm$ 0.135 & 0.154 $\pm$ 0.008 \\
201831508357553536 & 0.412 $\pm$ 0.0579 & 2.504 $\pm$ 0.280	& 0.196 $\pm$ 0.052 & 0.561 $\pm$ 0.052 \\
203635841300473984 & 0.924 $\pm$ 0.218 & 2.950 $\pm$ 0.536 & 0.356 $\pm$ 0.153 & 0.878 $\pm$ 0.158 \\
271966748051655424 & 0.228 $\pm$ 0.037 & 0.913 $\pm$ 0.129 & 0.435 $\pm$ 0.053 & 0.630 $\pm$ 0.069 \\
   \hline
\hline
    \end{tabular}
    \label{tab:binary_ratios}
   \tablefoot{Companion minimum and maximum mass, minimum mass-ratio and filling factor (adopting a zero eccentricity) for rotational and ellipsoidal variable candidates. The errors correspond to the standard deviation for each parameter after sampling from the errors of the primary mass, orbital periods and radii. The full table will be available at the CDS.} 
\end{table*}

\end{landscape}

\FloatBarrier

\twocolumn
\section{\textit{GSP-Spec} [M/H] validation}\label{app:mh}

\begin{figure}[h!]
   \centering
    \includegraphics[width=0.5\textwidth]{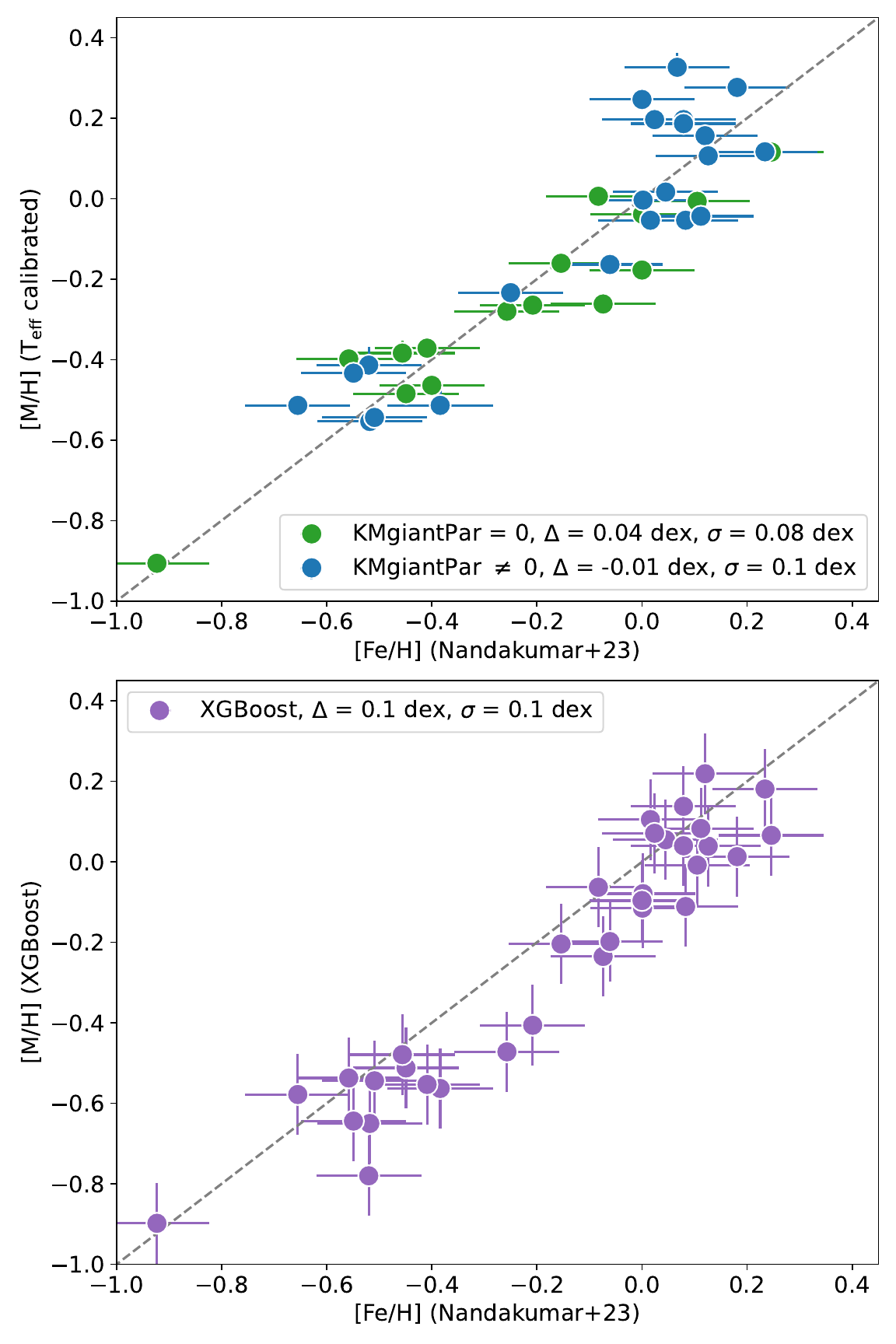} 
 \caption{Metallicity comparison between [Fe/H] values from \cite{Govind23}, based on high-resolution spectra, and [M/H] from \textit{Gaia} RVS by the \textit{GSP-Spec} module (top panel), for stars having good quality flags for the metallicity estimate, and \textit{Gaia} XP spectra (bottom panel, from XGBoost). Stars having \texttt{KMgiantPar=0} are shown in green circles while those having \texttt{KMgiantPar=1} are shown as blue circles. The median difference and dispersion of each group are also reported.}\label{fig:mh_govind}
\end{figure}

Atmospheric parametrization for cool giants from \textit{Gaia} RVS spectra can be challenging and in some cases, not successful \citep[see][]{GaiaRVS}. In those cases, the flag \texttt{KMgiantPar} was set to 1 or 2, and the reported \textit{Gaia} DR3 temperature and surface gravity were fixed to 4250 K and 1.5 dex, respectively. Nonetheless, the metallicity determination, as it is based on the Ca triplet lines, can be reliably measured for those sources. To validate the metallicities obtained in this way, we cross-match \textit{Gaia} DR3 \texttt{astrophysical\_parameters} table with the reported metallicities for $\sim$45 cool M-giants from \cite{Govind23}. The latter were derived using near-IR high-resolution spectra. We applied the calibration to the \textit{GSP-Spec} metallicities as in \cite{ARB24} and considering only those sources with \texttt{vbroadM} and \texttt{vradM} flags equal to zero. Figure~\ref{fig:mh_govind} shows the 37 stars obtained in this way, separated based on the value of the \texttt{KMgiantPar} flag, being 0 for the stars with successful M-giant parameterization (green circles) and $\neq$0 for those stars without reliable T$_{\rm eff}$ and $\log{g}$ (blue circles). For the latter, the metallicity was calibrated based on the fixed T$_{\rm eff}$ = 4250 K, adopted by the \textit{Gaia} \textit{GSP-Spec} module. At the bottom left corner, the median difference $\Delta$ in the sense $\Delta$ = [Fe/H] (Nandakumar+23) $-$ [M/H] \textit{Gaia}, as well as the dispersion between both samples, is shown. The bottom panel shows the metallicity derived by XGBoost for a subsample of the same stars.

The median difference is of the order of 0.1 dex for XGBoost metallicities, ten times larger than the median difference between the \textit{GSP-Spec} metallicities for stars with \texttt{KMgiantPar = 1, 2}. Figure~\ref{fig:mh_govind} shows that, even though the atmospheric and physical parameters such as mass and radius, cannot be derived for these sources using \textit{GSP-Spec} parameters, the metallicity estimates are reliable and can be used to study the metallicity content of our sample.
 
\end{appendix}
\end{document}